\documentclass[12pt,letterpaper]{article}
\pdfoutput=1
\usepackage[includeheadfoot,
            marginratio={1:1,2:3}, 
            width=412pt, 
            height=688pt,]{geometry}

\usepackage{amsmath}
\usepackage{amsfonts}
\usepackage{amssymb}
\usepackage{graphicx}
\usepackage{cite}
\usepackage{ulem}
\usepackage{dsfont}
\usepackage{longtable}
\usepackage{afterpage}
\usepackage{booktabs}
\usepackage{hyperref}
\usepackage{tikz} 
\usetikzlibrary{decorations.pathreplacing}
\usepackage{subfigure}
\usepackage{verbatim}
\usepackage{multirow}

\newcommand{\eq}[1]{\begin{equation}
                     \begin{split} #1 \end{split}
                     \end{equation}}

\newcommand{\ov}{\overline}

\newcommand{\kahler}{K\"ahler}

\allowdisplaybreaks[2]
\numberwithin{equation}{section}

\begin{document}

\normalem
\vspace*{-1.5cm}
\begin{flushright}
  {\small
  MPP-2018-34 \\
  }
\end{flushright}

\vspace{1.5cm}

\begin{center}
  {\LARGE
The  Refined Swampland Distance Conjecture \\[0.3cm]
     in Calabi-Yau Moduli Spaces 
}
\vspace{0.4cm}

\end{center}

\vspace{0.35cm}
\begin{center}
  Ralph Blumenhagen$^{1}$, Daniel Klaewer$^{1}$,
Lorenz Schlechter$^{2,1}$, Florian Wolf$^{1}$
\end{center}

\vspace{0.1cm}
\begin{center} 
\emph{$^{1}$ Max-Planck-Institut f\"ur Physik (Werner-Heisenberg-Institut), \\ 
   F\"ohringer Ring 6,  80805 M\"unchen, Germany } \\[0.1cm] 
\vspace{0.25cm}
\emph{$^{2}$ Technische Universit\"at M\"unchen, Fakult\"at f\"ur
  Physik, \\ 85748 Garching, Germany} \\[0.1cm]

\vspace{0.2cm}

\end{center} 

\vspace{1cm}


\begin{abstract}
\noindent
The Swampland Distance Conjecture claims that effective theories
derived from a consistent theory of quantum gravity only have a finite
range of validity. This will imply drastic consequences for
string theory  model building. 
The refined version of this conjecture says that this range is of the order of the naturally built in scale,
namely the Planck scale. It is investigated whether the 
Refined Swampland Distance Conjecture  is consistent with proper field distances arising in the
well understood moduli spaces of Calabi-Yau compactification.
Investigating in particular the non-geometric phases of 
K\"ahler moduli spaces of dimension $h^{11}\in\{1,2,101\}$, we always find proper field distances
that are smaller than the Planck-length.
\end{abstract}


\clearpage
\tableofcontents


\section{Introduction}

For relating a theory of quantum gravity, like string theory,  to the real world it is of utmost
importance to understand which properties of the low-energy 
effective action can be realized in the string theory landscape and
which are outside  its reach.  The set of theories that cannot be
consistently UV completed by a quantum gravity theory has been 
termed the swampland \cite{Vafa:2005ui}. If one could make a clear statement that
a certain property of the low-energy effective action cannot be
realized in the landscape, this would open up the possibility of in principle
falsifying string theory.

During the last  years progress in this direction has been made in
particular in the realm of string cosmology, where the question arose
whether string theory admits controllable models of large field
inflation. In such scenarios the inflaton transverses trans-Planckian
field ranges and, due to the Lyth bound,  gives a tensor-to-scalar
ratio $r>10^{-3}$, i.e. values that can be measured with current and
future experiments. Well motivated candidates for such models
are based on axions, whose shift symmetry can protect 
the scalar potential against quantum gravity corrections.

However, it was argued  that the Weak Gravity Conjecture (WGC)
\cite{ArkaniHamed:2006dz} provides strong constraints 
on models where the axion potential is generated by non-perturbative
effects
\cite{Rudelius:2014wla,Rudelius:2015xta,Montero:2015ofa,Brown:2015iha}. 
In effect, the WGC implies that for models with one \cite{Freese:1990rb}
or more axions  \cite{Kim:2004rp,Dimopoulos:2005ac}, 
increasing the field range to the trans-Planckian regime
means that one moves outside the regime  of a controlled low-energy
effective action (see also the early work \cite{Svrcek:2006yi}).

A second approach towards realizing axion inflation is to impose  a controlled spontaneous breaking of the
axionic shift symmetry \cite{Kaloper:2008fb} by adding branes or
fluxes.  This ansatz is called axion monodromy inflation and was introduced in
the stringy context in \cite{Silverstein:2008sg}. One at first sight
promising mechanism  to
generate  a polynomial potential for  axion monodromy inflation 
is to turn on background fluxes generating a tree-level F-term scalar
potential  \cite{Marchesano:2014mla,Blumenhagen:2014gta,Hebecker:2014eua}.
However, concrete attempts to realize such a scenario
in a consistent scheme of string theory moduli stabilization 
 \cite{Blumenhagen:2014nba,Blumenhagen:2015kja} showed
that one encounters major obstacles. In fact,  the prior assumptions
about the range of validity of the  low-energy effective action were
violated. For instance, often one could not keep the moduli masses
smaller than the Kaluza-Klein masses.
A more general observation \cite{Blumenhagen:2015qda,Baume:2016psm} was that 
taking the backreaction of the rolling axion onto the other moduli
into account,  the backreacted proper field distance showed a logarithmic behavior 
$\Theta\sim \lambda^{-1} \log\theta$.
Here $\lambda^{-1}$ can be considered as the scale in field distance where the backreaction
becomes substantial.

It was realized in \cite{Baume:2016psm,Klaewer:2016kiy} that this logarithmic scaling of the proper
field distance is very generic and that it  precisely reflects one of  the
conjectured behaviors  by  Ooguri/Vafa \cite{Ooguri:2006in} to distinguish 
effective field theory models that can be realized in string theory
(the landscape) from those that cannot be coupled in a UV complete way
to gravity (the swampland) \cite{Vafa:2005ui}. 

This now called `Swampland Distance Conjecture (SDC)' \cite{Grimm:2018ohb} was
abstracted from the very simple example  of string theory compactified
on a circle, where it is the Kaluza-Klein tower that shows this
behavior in terms of the proper field distance.
It says that if one moves over very large distances in the moduli
space  of  an effective  quantum gravity theory, there appears an 
infinite tower of states whose mass scales as $m\sim m_0 \exp(-\lambda
\Delta\Theta)$. This means that for $\Delta\Theta> \lambda^{-1}$ 
the effective theory breaks down. 
A new field theoretic perspective on the relation between the logarithmic scaling 
and the infinitely many exponentially light  states was recently given
in \cite{Grimm:2018ohb,Heidenreich:2018kpg}.

This paper is concerned with the question, at which scale
$\lambda^{-1}$ this happens, i.e. what the size of the region where
the effective theory can be trusted is.
In fact, the concrete models of axion backreaction discussed in \cite{Baume:2016psm} always had
$\lambda=O(1)$, that is the cut-off in the field distance was close to the Planck-scale.
This motivated  to formulate the {\it Refined Swampland Distance
  Conjecture} (RSDC) \cite{Klaewer:2016kiy}, that extends the  original
one of Ooguri/Vafa by the statement that
$\lambda=O(1)$. Clearly, if this conjecture holds also for axions
with a potential, it will provide a generic quantum gravity obstacle
to realize axion-monodromy inflation in a controlled way using
a low energy effective action. Recent discussions of the Swampland Distance
Conjecture in the context of axion monodromy inflation
are
\cite{Valenzuela:2016yny,Bielleman:2016olv,Blumenhagen:2017cxt,Hebecker:2017lxm}
(see also \cite{Cicoli:2018tcq} for a geometric
reason for finite distances in moduli spaces).

It is the purpose of this paper to go a step back to the original
work of Ooguri/Vafa and analyze the RSDC not only for moduli spaces
of toroidal compactifications but for genuine Calabi-Yau (CY)
compactifications of  type II superstring theory. This means that the
axions do not receive a potential and remain as periodic flat
directions. More concretely, we will challenge the RSDC  by
computing proper field distances along geodesics (and sometimes
non-geodesic trajectories)  in the K\"ahler moduli
spaces of Calabi-Yau manifolds. For that purpose we will use the
available two methods. First one can compute the periods of the mirror
dual CY manifold and second one can extract the K\"ahler potential
from  the 2-sphere partition function of  the gauge linear sigma model
(GLSM).

It is well known that these K\"ahler moduli spaces do not
only contain (geometric) regions/phases,  where points at infinite distance
exist, but also non-geometric phases of stringy nature, like the
Landau-Ginzburg (LG) phase. These often do only have a finite radius and
therefore do not admit points at infinite distance. To reach the latter one
first has to cross the non-geometric phase and move to another geometric
phase where such a point do exist.
If the distances one can travel along 
geodesics in such  non-geometric phases were  already larger than $M_{\rm pl}$, it would directly
falsify the RSDC. Therefore, our main objective is to  compute geodesic distances in the  moduli space of Calabi-Yau
compactifications and compare them  with the expectation from the
RSDC.  We emphasize that very similar in spirit, in \cite{Conlon:2016aea}
axion decay constants
have been evaluated in the geometric phases for $h^{11}\in \{1,2\}$ CY
threefolds with the result that they are bounded  from above by 
an order one parameter times the Planck-scale.

In section 2, in more detail we explain  the RSDC and our approach to
check it in Calabi-Yau moduli spaces. For self-consistency, section 3 contains a summary of
the  known techniques we employed to gain the necessary information on the
moduli spaces. The reader familiar with these techniques can safely
skip this section. The sections 4 and 5 are the main sections of this
paper. For a couple of one and two moduli examples we analyze 
(geodesic) proper field distances between points in their moduli
spaces
and compare them with the expectation from the RSDC. 
In section 6, based on a recent
advance \cite{Aleshkin:2017fuz,Aleshkin:2017oak}, we also investigate
the full 101 dimensional K\"ahler moduli space of the mirror quintic.
The result of this endeavor can be summarized by saying that in all
instances our results are consistent with the RSDC, thus providing
non-trivial evidence for it. We finally will make the compelling
observation
that our data seem to suggest that the finite proper field distance one can traverse in a single
non-geometric phase scales inversely with the number of
phases.

\section{Prerequisites and objectives}
\label{sec:RSC}

Before we dwell into the more formal analysis, let us
explain the objective of this paper. This is an analysis of the 
Refined Swampland Distance Conjecture in the moduli space of 
Calabi-Yau compactifications.

\subsection{Basics on Calabi-Yau moduli spaces}

Let us first introduce some very basics of type II compactifications
on CY manifolds.   Consider the compactification of  type IIA superstring
theory on a Calabi-Yau manifold ${\cal M}$ with Hodge numbers
$(h^{21},h^{11})$. The four-dimensional effective theory has ${\cal
  N}=2$ space-time supersymmetry.
The dimension of complex structure
moduli space gives rise to $h^{21}$ hypermultiplets  and the dimension
of the K\"ahler moduli space to $h^{11}$ vectormultiplets.
Such a  compactification is dual to type IIB compactified on the mirror
CY manifold ${\cal W}$ with Hodge numbers $h^{21}({\cal
  W})=h^{11}({\cal M})$ and $h^{11}({\cal W})=h^{21}({\cal M})$, which
gives the same massless spectrum.

Since hyper- and vectormultiplets decouple in ${\cal N}=2$
supergravity, the vectormultiplet  moduli space of the type IIB model on the mirror CY
does not receive world-sheet instanton corrections and can be reliably
computed at string tree-level (in $\alpha'$). This complex structure
moduli space is mapped via mirror
symmetry to the K\"ahler moduli space of the type IIA model. The
latter indeed receives $\alpha'$ corrections which can be read-off
once one knows the precise mirror map between the moduli.

In this paper we will be concerned with field distances in  these vectormultiplet moduli spaces.
In section 3, we will review  methods to compute their K\"ahler
potential and consequently their  metric. The classic method
of Candelas et al. \cite{CANDELAS199121} applies to the complex structure moduli space
of the mirror Calabi-Yau manifold  ${\cal W}$ where techniques  were
developed  to  compute the
periods of the holomorphic three-form and from them all the
information
on the (classical) complex structure moduli space. More recently,
methods were developed \cite{Benini:2012ui,Jockers:2012dk,Gomis:2012wy} that work directly for the K\"ahler moduli
space of the original CY ${\cal M}$ and allow
to compute the 2-sphere partition function utilizing the description
in terms of gauged linear sigma models (GLSMs). From the partition 
function one can deduce the K\"ahler potential.

Of course,  the two descriptions have to agree so that we are
free to use either the language of the complex structure or the K\"ahler
moduli space. Throughout this paper, we will use the K\"ahler moduli
space, i.e. we will speak of large and small radius regime. When we
say conifold locus (that is actually defined in the complex structure
moduli space), we mean  the mirror dual of it. 
As we will see in the next section, the advantage is that K\"ahler
moduli more directly  set  the mass scale of the Kaluza-Klein modes.

\subsection{The Refined Swampland Distance Conjecture}

In this section, we briefly  explain the Swampland Distance Conjecture and recall one
of its motivations. In the original paper by Ooguri/Vafa \cite{Ooguri:2006in}, what 
later has been called the `Swampland Distance Conjecture',  was just one of the
proposed criteria to discriminate effective field theories
arising in the string landscape from those that do not admit a UV-completion,
i.e. lie in the swampland. Certainly, this criterium was the most
quantitative one and, as the name suggests,  is about distances in field space. Let us explain
its origin for the K\"ahler moduli space of the type IIA model on
${\cal M}$.

In the large volume regime the K\"ahler potential for the K\"ahler
moduli of ${\cal M}$ 
is given in terms of the triple intersection numbers as
\eq{\label{Kpotlcs}
           K=-\log \Big( -{i\over 6}\,{\kappa^{ijk}}\, (t_i - \ov t_i)(t_j - \ov t_j)(t_k - \ov t_k)\Big)
}
where the $t_i$, with  $i=1,\ldots,h^{11}$, denote the complexified
K\"ahler moduli 
\eq{
t_i=\int_{\Sigma_i}  B +i \int_{\Sigma_i} J
}
 where the two-cycles $\Sigma_i$ are a basis of $H_2({\cal M})$.
Following a generic trajectory inside the K\"ahler cone ${\rm Im}(t_i)\sim
\alpha_i \,r$, $\alpha_i\in\mathbb R$, for very  large $r$  the effective K\"ahler potential  behaves as
$K=-3 \log r$. Hence,  the K\"ahler metric along the trajectory becomes
\eq{
                      G(r)={3\over 4\, r^2}  \, .
}
There might exist special trajectories for which the numerator is
smaller than three but it will never be larger.
The proper field distance $\Theta$ along a trajectory $x^\alpha(\tau)$
is defined as 
\begin{equation}
  \Theta=\int_{\tau_0}^{\tau_*}d\tau\,\sqrt{G_{\alpha\overline{\beta}}
        \frac{dx^\alpha}{d\tau}\frac{d\overline{x}^{\overline{\beta}}}{d\tau}}\;,
\end{equation}
which in this case becomes
\eq{
\label{properdistancer}
    \Theta=\int_{r_0}^{r_*}  dr\, \sqrt{G(r)} = {1\over
      \lambda}\log\Big(   {r_*\over r_0}\Big)\,
}
with $\lambda={2\over \sqrt{3}}\approx 1.15$.
This logarithmic scaling has a dramatic consequence for the validity
of the effective field theory in which the Kaluza-Klein (KK) modes are
assumed to be integrated out. The generic  KK mass-scale can be
estimated as
\eq{
              M_{\rm KK}\sim {M_s\over \sqrt{r}} \sim {M_{\rm pl}\over
                r^2} \sim M_{{\rm KK},0}\, \exp(-2 \lambda \Theta)
}
which implies that for trans-Planckian field excursions $\Theta>\lambda^{-1}\approx
0.87$ an infinite tower of KK-states becomes light, thus spoiling the
validity of the effective field theory.

In \cite{Ooguri:2006in} Ooguri and Vafa formulated this in a  more concise manner.
They provided a couple of conjectured criteria that an effective theory in the
landscape necessarily should satisfy.
The most quantitative criterion was termed the Swampland Conjecture in
\cite{Klaewer:2016kiy} and later was renamed as Swampland Distance
Conjecture \cite{Grimm:2018ohb}.
It says:
\begin{quotation}
\noindent
{\bf Swampland Distance Conjecture:}

{\it For any point $p_0$ in the continuous scalar moduli space of a
consistent quantum gravity theory (the landscape), 
there exist other points $p$ at arbitrarily large distance. As the
geodesic distance $\Theta=d(p_0, p)$ diverges,  an infinite tower of states
exponentially light in the distance appears, 
meaning that the mass scale of the tower varies as
\eq{
\label{swamp_mass}
M \sim  M_0\, e^{-\lambda \,\Theta}\,.
} 
Thus, the number of states in the tower which are below any finite mass
scale diverges as $\Theta\to\infty$.
The geodesic distance  is measured with the metric on the moduli space.}
\end{quotation}
In the initial version of the conjecture,  $\lambda$ is still an undetermined  parameter that specifies when 
the exponential drop-off becomes significant. 
As we have seen in \eqref{properdistancer}, for trajectories (not
necessarily geodesic) 
in the large volume regime of CY
compactifications one has $\Theta_\lambda \sim \lambda^{-1}\sim O(1)$.

Infinitely many states becoming exponentially light  in field space indicates that
the effective quantum gravity theory  at the point $p_0$ only has
a finite range 
of validity in the scalar moduli space. 
To determine the exact value of the displacement where the effective theory breaks down,
all relevant mass scales have to be taken into account.
Therefore, the exact upper bound on the displacement is highly
model-dependent, but it is sure that  in the presence of the
exponential drop-off, any physics that 
we might derive for larger values
$\Theta>\Theta_\lambda = \lambda^{-1}$ cannot be trusted.

However, it is not a priori clear at which finite value of the distance $\Theta_0$ along a geodesic this exponential drop-off first sets in.
 Indeed, the scale $\Theta_\lambda$ was derived
already in the large radius regime without any reference to the point
$p_0$ where the  geodesic started. 
In \cite{Baume:2016psm,Blumenhagen:2017cxt}, by analyzing a couple of string theory models,  evidence was
provided that $\Theta_0$ is equal to the natural 
mass scale in quantum gravity, namely $M_{\rm pl}$\footnote{There, the displaced field was actually an axion, hence the consequences of the SDC were only visible through backreaction effects induced by moduli stabilization.}.
For the simple examples chosen,  all scales turned out to be related, i.e.
$\Theta_0 \simeq \Theta_\lambda = \lambda^{-1}$.
Let us point out that these papers focused on stabilized moduli in contrast to the original work of \cite{Ooguri:2006in}.
In the case of unstabilized moduli, general arguments were given in
\cite{Klaewer:2016kiy} in favor of a \emph{Refined Swampland Distance Conjecture} 
which states that not only $\Theta_\lambda\lesssim \mathcal{O} (1)$ but also
$\Theta_0\lesssim \mathcal{O} (1)$ (in Planck units). In other words,
for any starting point of a geodesic, one cannot follow  it for a  longer
distance than $M_{\rm pl}$ before the validity of the effective theory
breaks down.

The objective of this paper is to test this highly non-trivial conjecture 
in the K\"ahler moduli space
of type IIA string compactications on CY manifolds. 
For that purpose we head out to compute the critical values $\Theta_0$
and $\Theta_\lambda$  for a plenitude of different
trajectories/geodesics  in various CY manifolds. Let us 
explain this idea in more detail.

\subsection{Objective: Testing the  RSDC in CY moduli spaces}

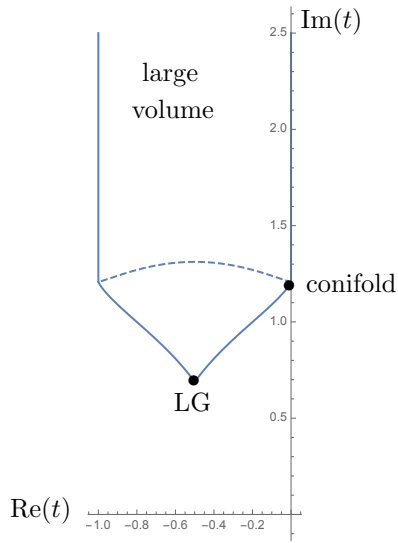
\begin{figure}[ht]
  \centering
  \includegraphics[width=0.2\textwidth]{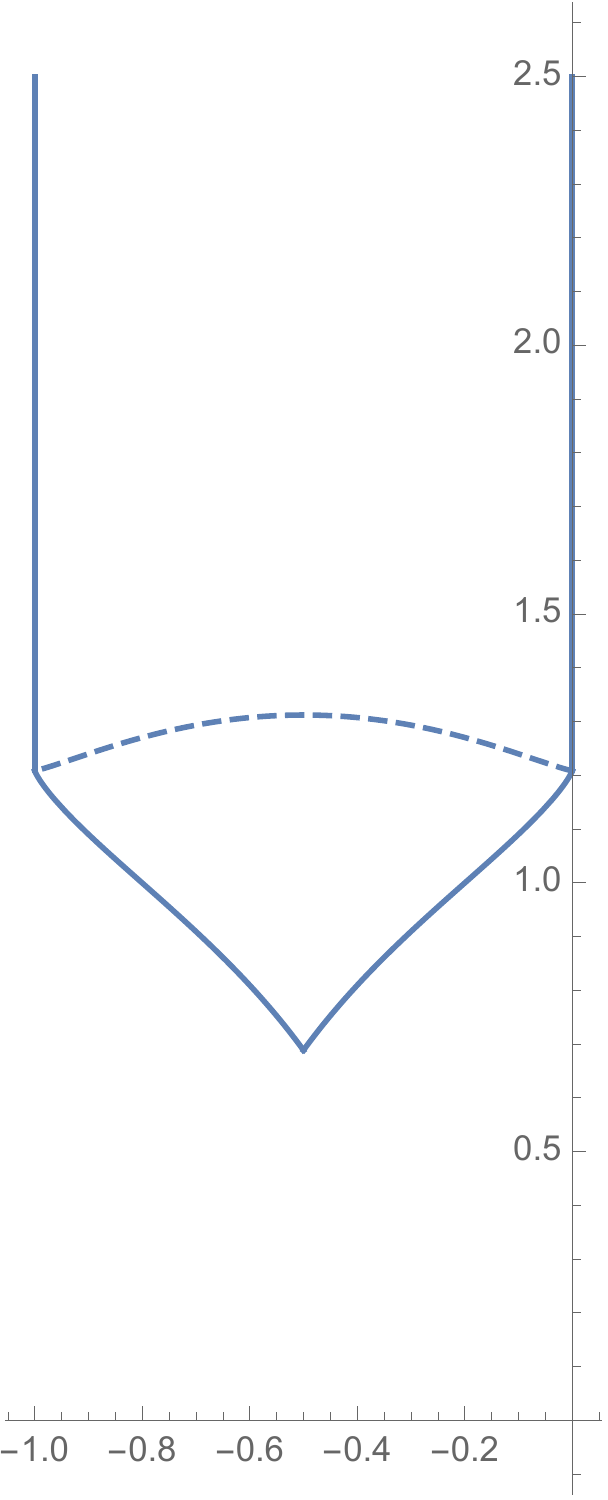}
 \begin{picture}(0,0)
   \put(-115,10){\footnotesize${\rm Re}(t)$}
   \put(-5,195){\footnotesize${\rm Im}(t)$}
    \put(-65,175){\footnotesize large}
     \put(-69,161){\footnotesize volume}
   \put(-53,50){\footnotesize LG}
   \put(-48,59){\footnotesize${\bullet}$}
   \put(-3,95){\footnotesize${\rm conifold}$}
    \put(-12,95){\footnotesize${\bullet}$}
  \end{picture}
   \caption{Sketch of the K\"ahler moduli space of the quintic.}
  \label{fig:toymodulispace}
\end{figure}

As we have recalled, the intuition for the Swampland Distance Conjecture derives from the form of
the K\"ahler potential in the large radius regime\footnote{Other examples were presented already in \cite{Ooguri:2006in}, for example the Planck-normalized mass of the oscillator modes of the F- and D-strings as a function of the string coupling.}. It is known that 
the K\"ahler moduli space of a Calabi-Yau manifold also contains non-geometric
regions where $\alpha'$ corrections become important. For instance,
for the quintic the form of the K\"ahler moduli space has the structure shown
in  figure \ref{fig:toymodulispace}. 

\vspace{0.2cm}

\noindent
As indicated, there are three distinguished points: the large volume point, the
conifold and the Landau-Ginzburg (LG) point. The LG or Gepner point is the
one of minimal radius. To cover the whole moduli space, one needs at
least two charts, whose radii of convergence are shown by the dashed
arc in figure \ref{fig:toymodulispace}.

\begin{figure}[ht!]
\centering
\begin{tikzpicture}[xscale=0.8,yscale=0.8]
\node[inner sep=0pt] at (0,0)
{\includegraphics[scale=0.8]{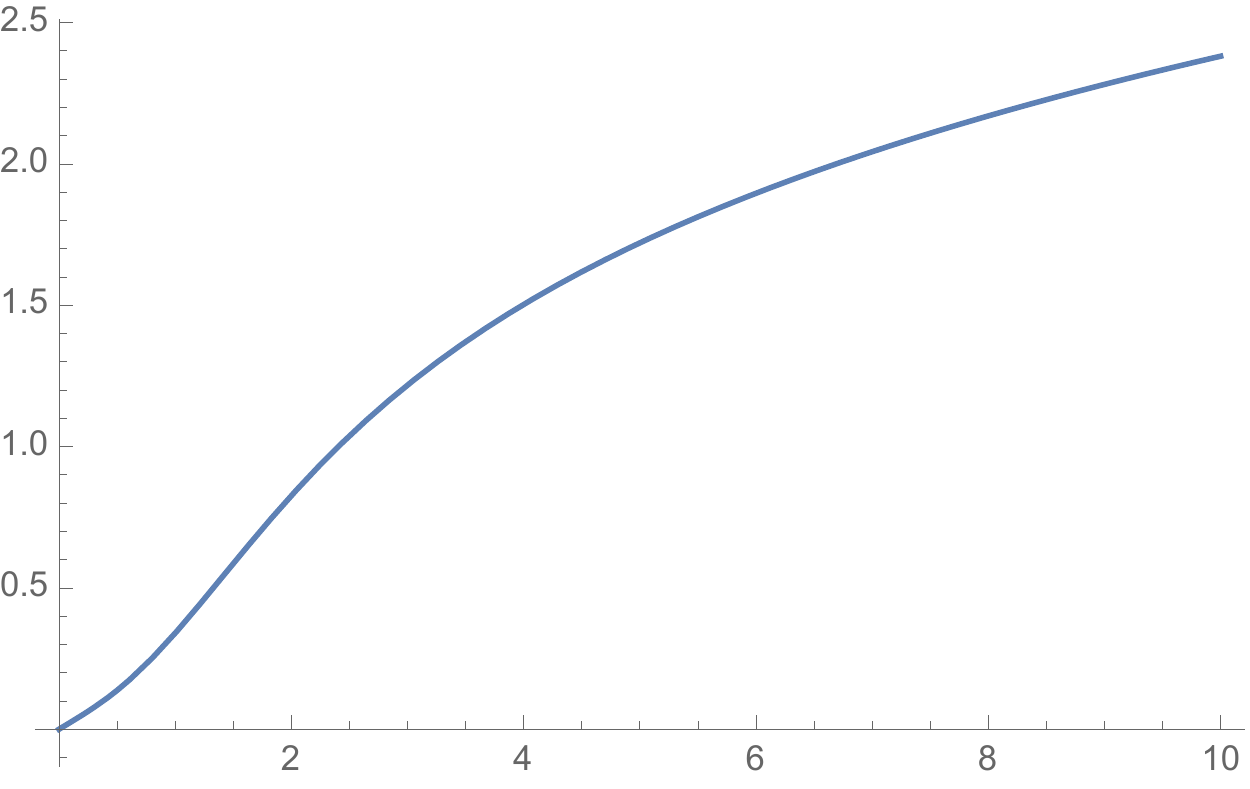}};
\node at (-5.2,4) {$\Theta$};
\node at (6,-4.1) {${\rm Im}\, t$};
\node at (-7.2,-0.5) {$\Theta_0$};
\node at (-7.2,2.3) {$\Theta_c$};
\node at (-2.8,-4.1) {${\rm Im}\, t_0$};
\node at (2,-4.1) {${\rm Im}\, t_c$};
%
\node[align=center,below] at (-1.5,-0.4) {\footnotesize logarithmic\\ \footnotesize behavior\\ \footnotesize relevant};
\node[align=center,below] at (3.2,2.3) {\footnotesize significant\\ \footnotesize decrease of\\ \footnotesize mass scale};
\node[red] at (-2.9,-0.5) {\textbullet};
\node[red] at (1.9,2.25) {\textbullet};
\draw[dashed,red] (-5.7,-0.5) -> (-2.9,-0.5);
\draw[dashed,red] (-2.9,-0.5) -> (-2.9,-3.425);
\draw[dashed,red] (-5.7,2.25) -> (1.9,2.25);
\draw[dashed,red] (1.9,2.25) -> (1.9,-3.4);
\draw [decorate,decoration={brace,amplitude=10pt},xshift=-4pt,yshift=0pt]
(-6.25,-0.5) -- (-6.25,2.25) node [black,midway,xshift=-0.6cm] 
{$\Theta_\lambda$ \ };
\end{tikzpicture}
\caption{\small Expected relation between proper field distance $\Theta$ and ${\rm Im} \, t$.}
\label{nicefigure}
\end{figure}
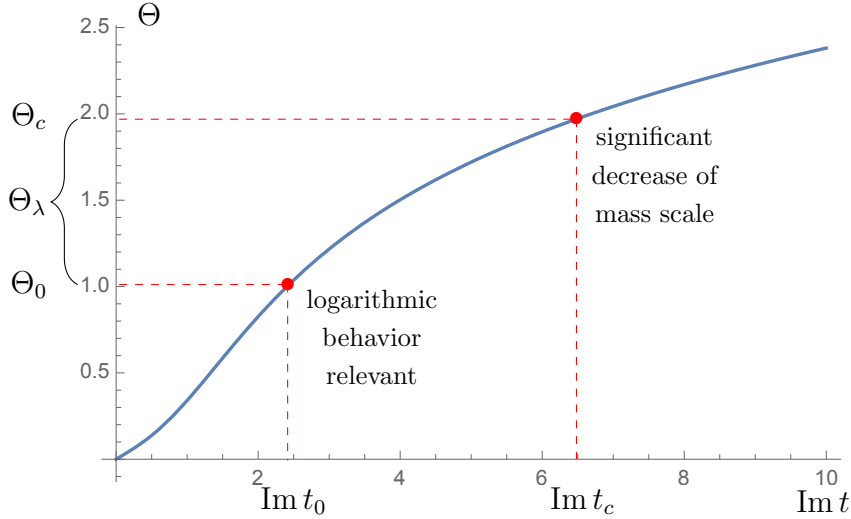

The concrete description of this moduli space
will be recalled in section \ref{sec_quintic_pheno}. At this stage,  we just want to mention that one can
ask the question whether the RSDC  still
holds for points $p_0$ in the small volume regime. We will see  that the conifold and LG points are at {\it finite}
distance in the moduli space so that the only region featuring infinite
distances is the large volume regime. 
Following a geodesic from the LG point to the large volume
regime, one expects that the proper field distance depends on ${\rm Im} \, t$
 like shown in figure \ref{nicefigure}.

\vspace{0.3cm}
\vspace{0.1cm}

\noindent
As long as one stays in the small volume regime the proper field
distance scales polynomially with ${\rm Im} \, t$ and at some point $({\rm Im} \, t_0,\Theta_0)$ the logarithmic scaling becomes dominant.
As a consequence, we define the critical field distance as the sum
\eq{
               \Theta_c=\Theta_0+\lambda^{-1}\,,
}
which includes the distance $\Theta_0$. This makes sense as we would
like to know how much distance can one travel along a geodesic before
the effective field theory breaks down.
For the quintic example, displacing the \kahler{} modulus from the LG
point towards the large volume phase, $\Theta_0$ would be at least the proper distance to the edge of the convergence region of the LG phase, which is approximately the same as the distance between the LG and conifold point.

Clearly, if $\Theta_0$ determined in this way was already larger
than the Planck-scale, the RSDC  would  be
falsified. Said the other way around, if the RSDC is correct, the
proper field  distance that can be traveled
in the small volume regime must be smaller than $M_{\rm pl}$.
Therefore, the (proper) radius of convergence for any chart that does not
contain a region of infinite distance should be sub-Planckian.

Now, let us confirm that the identification of $\Theta_0$ with the
radius of convergence is really a good approximation.
In the case of a single modulus, the general asymptotic form of the prepotential is \cite{CANDELAS199121}
\begin{equation}
  \mathcal{F}=-\frac{5}{6}t^3+i\underbrace{\frac{\zeta(3)}{2(2\pi)^3}\chi(\mathcal{M})}_{\equiv -c}+\ldots\;,
\end{equation}
where the constant is the one-loop contribution and the dots denote non-perturba\-tive corrections. From this we find that the asymptotic form of the proper distance is
\begin{equation}\label{eq:Theta_expansion}
    \Delta\Theta= \frac{\sqrt{3}}{2}\log(t)+\frac{\sqrt{3}c}{5}\frac{1}{t^3}+\mathcal{O}\left(\frac{1}{t^6}\right)\;.
\end{equation}
The point where the perturbative $1/t^3$ contribution is of the same order as the asymptotic logarithm is found to be at
\begin{equation}
    t_\text{eq}\simeq e^{\frac13 W(2 c/5)}\;,
\end{equation}
where $W(x)$ is the Lambert $W$-function. For the quintic with Euler
characteristic $\chi=-200$, this is approximately at
$t_\text{eq}\simeq 1.06$, a value well below the minimal value of $t$
in the large radius phase $t_\text{min}\simeq1.21$. Indeed, at the phase boundary, the logarithm is approximately twice the size of the $1/t^3$ term, which can be considered a small correction for all $t$ larger than this. This is the case for all models that we analyzed and of course consistent with the fact that the $1/t$ expansion is a perturbative expansion which breaks down in the non-geometric regions of the moduli space.

As a result, the logarithm is a good approximation for the behavior of
the proper distance over the whole large volume phase and only breaks
down at the boundary to non-geometric regions. 
Thus,  the only relevant contribution to $\Theta_0$ comes from inside the non-geometric phases and the behavior predicted by the SDC sets in immediately after crossing the phase boundary.

It is the goal of this paper  to check two very concrete
predictions of the RSDC,
namely that  both $\Theta_\lambda$  and $\Theta_0$  should be bounded
by $\mathcal{O}(1)$ in Planck units.


\section{The K\"ahler metric on moduli spaces}
\label{sec_KahlerMetric}

To be able to test the RSDC,  for concrete examples of CY
manifolds   one needs the  explicit form of the metric on the K\"ahler moduli space. In this rather
technical section we describe the two employed methods  to calculate this metric.

First we use the conjecture of \cite{Jockers:2012dk} which was proven
by \cite{Gomis:2012wy}. There the \kahler{} potential and thereby the
metric can be read-off from the partition function of a corresponding
gauged linear sigma model (GLSM). 
Second one calculates the periods of the holomorphic three-form for
the mirror CY and transforms them into an integral symplectic basis. 
Using two different methods allows us to also compare the numerical
results. 

Moreover the two methods have different advantages and
disadvantages. While the GLSM construction is very efficient and
mathematically easier, it is hard to extract the periods and thereby
the mirror map from the partition function in the non-geometric
phases. The period formalism on the other hand becomes quite involved,
but from  the periods one also obtains the mirror  map directly.

\subsection{Sphere partition function for GLSM}
\label{glsm1}

In the GLSM approach the \kahler{} potential is given in terms of the 2-sphere partition function of the gauged linear sigma model

\begin{equation}
\label{local}
e^{-K}=Z_{S^2}\;.
\end{equation}
The partition function for a general $\mathcal{N}=(2,2)$ GLSM was
calculated in \cite{Benini:2012ui,Doroud:2012xw}. In the Coulomb
branch the relevant fields in the path integral are the scalar parts
$a_j$ of the vector multiplets and the field strengths $F_j$ of the
vector part of the vector multiplet. The index $j$ labels the $s$
vector multiplets of the theory, while the index $i$ labels the $M$
chiral fields. As the sphere is a compact space, the field strengths
are quantized, ${1\over 2\pi}\int_{S^2}F_j=m_j$ with the $m_j$ being
integers. 
Thus for an abelian gauge group $G=U(1)^s$ it can be written as
\begin{equation}
\label{partition}
Z_{S^2}(\xi,\bar{\xi},Q,R) =\sum\limits_{m_1\in\mathbb{Z}}\dotsc\sum\limits_{m_s\in\mathbb{Z}}\int\limits_{-i\infty}^{i\infty}da_1\;\dotsc \int\limits_{-i\infty}^{i\infty}da_s\; Z_\text{class}\;Z_\text{gauge}\;Z_\text{chiral}
\end{equation}
where the purely imaginary integration contour for the scalar fields $a_i$ is chosen to simplify the expressions. For an abelian gauge group $Z_\text{gauge}=1$ is trivial.
The classical contribution is
\begin{equation}
Z_\text{class}=\prod\limits_{j=1}^s\; e^{-4\pi i r_j a_j+i\theta_j m_j}\;,
\end{equation}
and the contribution from the 1-loop determinants of the chiral fields are given by
\begin{equation}
Z_\text{chiral}=\prod\limits_{i=1}^M{\Gamma\left(R_i/2+\sum\limits_{j=1}^s Q_{i,j}\cdot(a_j-m_j/2)\right)\over \Gamma\left(1-R_i/2-\sum\limits_{j=1}^s Q_{i,j}\cdot(a_j+m_j/2)\right)}
\;.
\end{equation}

After performing the integration, the partition function only depends
on the complexified Fayet-Iliopoulos (FI) parameters $\xi_j=2\pi r_j+i\theta_j$, and the
gauge- and $R$-charges $Q_{i,j}$ and $R_i$ of the chiral fields. In the
next section we describe how to identify these parameters for a given
Calabi-Yau.

\subsubsection*{Data of GLSM}

The GLSM is a two-dimensional  $\mathcal{N}=2$ supersymmetric gauge theory build out of
$M$ chiral multiplets $\Phi_m$ with $m=1,\ldots,M-N$ and $\Sigma_n$ with
$n=1,\ldots,N$. In addition there   are $s$ vector multiplets $V_j$.
The chiral multiplets carry the $U(1)$ charges shown in table
\ref{table_charge}.
  
\begin{equation}
\label{table_charge}
\begin{aligned}
\begin{array}{c|cccc|ccc}
&  \Phi_1 & \Phi_2  & \ldots & \Phi_{M-N} & \Sigma_1 & \ldots & \Sigma_N \\
 \noalign{\hrule height 1pt}
 U(1)_1  & Q_{1,1}   & Q_{2,1}  & \ldots\ldots & Q_{M-N,1} & Q_{M-N+1,1} &
 \ldots &  Q_{M,1}  \\
 U(1)_2  & Q_{1,2}  & Q_{2,2}  & \ldots\ldots & Q_{M-N,2} & Q_{M-N+1,2} &
 \ldots &  Q_{M,2}\\
 \vdots & \vdots  &    \vdots  &  \vdots  &  \vdots & \vdots & \vdots& \vdots \\  
U(1)_s  & Q_{1,s}   &  Q_{2,s}  & \ldots\ldots & Q_{M-N,s} & Q_{M-N+1,s} &
 \ldots &  Q_{M,s}
\end{array}
\end{aligned}
\end{equation}

\vspace{0.2cm}
\noindent
The $M-N$ fields $\Phi_m$ can be considered as the homogeneous coordinates
of an ambient toric variety that has $s$ identifications. The $N$
fields $\Sigma_n$ provide the constraints. The dimension of the
resulting space is $D=M-2N-s$.
The CY condition is related to the vanishing of the mixed
gravitational - abelian gauge  anomalies,  which are proportional to the sum of the gauge charges
\eq{
                  \sum_{i=1}^{M} Q_{i,j}=0\,.
}
In addition there exist a superpotential 
\eq{
                     W=\sum_{n=1}^N   P_n(\Phi) \,  \Sigma_n
}
that is linear in the $\Sigma_n$ and the polynomials are such that
$W$ carries   vanishing charges.
Moreover, one associates   $R$-charges $R_i$ to  the chiral fields such
that the superpotential has $R$-charge $2$.
The superpotential only depends on the complex structure, therefore
the A-model partition function is independent 
of the superpotential itself. The superpotential only restricts the value of the $R$-charges.
The $R$-charges of the fields are determined by the condition that
$R(W)=2$ with the  $R$-charges of all chiral field positive. This does not uniquely fix
the $R$-charges,  but the remaining freedom corresponds to a rescaling of the partition function or equivalently a \kahler{} transformation of the \kahler{} potential. 

In this paper we will be concerned with for instance non-singular hypersurfaces in
weighted projective spaces
\eq{
W\mathbb{CP}^D_{k_1,\dots,k_{D+1}}[d_1,\dots,d_N]\,.
}
Since $s=h^{11}=1$ there is  only a single gauge symmetry and the chiral fields
carry charge $Q_{m,1}=k_m$, $m=1,\ldots,D+1$ and $Q_{n,1}=-d_n$ with
$n=1,\ldots,N$.

If the space is singular, the singularities have to be resolved. If
this is possible by toric methods, additional $U(1)$ gauge symmetries are
added. The gauge charges of the chiral fields are in this case
determined by the generators of the Mori cone, which can be calculated
using the PALP package \cite{Kreuzer:2002uu}.

\subsubsection*{Coordinates on K\"ahler moduli space}

The partition function \eqref{partition} is given in terms of
the FI-parameters $\xi_j$ of the GLSM used to describe the CY. We
define $z_j=\exp(2\pi i \xi_j)$ and $q_j=\exp(2\pi i t_j)$, where in
the second definition $t_i$ are the complexified \kahler{} moduli,
$t_i=\int_{\Sigma_i}(B+iJ)$. While the calculations are simpler in the $z$ coordinate
system, most results in the literature are stated in terms of the
``algebraic'' coordinates, denoted by $\phi$ or $\psi$. These  are the
coefficients of the deformations of the defining polynomial in the
mirror dual. To relate
these coordinates to each other, one determines special points in the
moduli space. These special points are the Landau-Ginzburg point at
$\phi=0$, the large complex structure point $\phi=\infty$ and the
singular loci. In the GLSM coordinates, the singular loci are located
at the solutions to the 
equations \cite{Morrison:1994fr}
\begin{equation}
\label{coni1}
\prod_{i=0}^M\big\langle\delta_i\big\rangle ^{Q_{i,j}}=z_j\;.
\end{equation}
and
\begin{equation}
\label{coni2}
\prod_{i=0}^M\Big\langle\delta_i|_H\Big\rangle ^{Q_{i,j}}=z_j\;,
\end{equation}
where $\langle\delta_i\rangle$ denotes the VEV of the operator
$\delta_i=\sum_{j=1}^{s}Q_{i,j}\,a_j$. Here $a_j$ is the
scalar part of the vector multiplet. The $\delta_i|_H$ are obtained
from $\delta_i$ by setting all $a$ which are charged under a subgroup
$H$ of $G$ to 0 and by only taking the uncharged chiral fields into
account. The equation has to be fulfilled for all subgroups of $G$
whose charges span $\mathbb{R}^{M-D-{\rm rank}(H)}$ with positive coefficients.

In the algebraic coordinates, the manifold becomes singular at the
points where the transversality condition is violated, i.e.
\begin{equation}
\label{coni3}
P(x) = 0 \, , \qquad 
{\partial P(x)\over\partial x_i}=0 \, , \qquad\forall x_i\;.
\end{equation}
Matching these two loci allows to fix the coordinate transformation $z\rightarrow \phi$. 

The procedure is the same for all models, so we will demonstrate it in the case of the two parameter model $\mathbb{P}^4_{11222}[8]$ with gauge group $G=U(1)\times U(1)$. The gauge charges are
\begin{equation}
  Q=\begin{pmatrix}
0 & 0 & 1 & 1 & 1 & 1 & -4 \\
1 & 1 & 0 & 0 & 0 & -2 & 0 
\end{pmatrix}\;.
\end{equation}
Inserting this into \eqref{coni1} and expressing the $\delta_i$ in terms of $a_j$ gives
\begin{equation}
\begin{aligned}
z_1&=(-4s_1)^{-4}s_1^3(s_1-2s_2)\\[0.1cm]
z_2&=s_2^2(s_1-2s_2)^{-2}
\end{aligned}
\end{equation}
where for ease of notation $\langle a_j\rangle=s_j$. These equations have solutions if 
\begin{equation}
2^{18}z_1^2z_2-(1-2^8z_1)^2=0\;.
\end{equation}
Moreover, the only subgroup of $G$ whose charges span $\mathbb{R}$ is the first $U(1)$. The remaining charges are $Q_H=(1\; 1\; -2)$. Therefore \eqref{coni2} produces the additional relation
\begin{equation}
z_2=s_2^2(-2s_2)^{-2}=1/4\;.
\end{equation}

Turning to the algebraic coordinates, the defining polynomial is
\begin{equation}
  P(x)=x_1^{12}+x_2^{12}+x_3^6+x_4^6+x_5^2-12\psi\, x_1 x_2 x_3 x_4 x_5-2\phi\, x_1^6 x_2^6\;.
\end{equation}
Solving the system of equations \eqref{coni3} with this polynomial gives the conditions $\phi^2=1$ and $(\phi+8\psi)^2=1$.
The equations $\phi^2=1$ and $z_2=1/4$ fix the transformation 
\begin{equation}
z_2={1\over 4\phi^2}\; .
\end{equation}
After some algebraic manipulations the remaining equations fix
\begin{equation}
z_1={\phi \over 2^{11}\psi^4}\; .
\end{equation}
Note that these transformations agree with \cite{Aspinwall:1994ay} up to  an
interchange of the indices. 

\subsubsection*{Evaluating the partition function}
\label{glsm2}

The partition function (\ref{partition}) is of the Mellin-Barnes type
\begin{equation}
Z=\int\limits_{\gamma+i\mathbb{R}^n} d\tau \, h(t)\; e^{-p\cdot \tau }\prod_{i}{\Gamma[p_i(\tau )]\over\Gamma[g_i(\tau )]}
\end{equation}
where $h(\tau )$ is an entire function, $p_i$ and $g_i$ are
polynomials in $\tau $ and $\gamma$ and $p$ are constant vectors. In
the literature, numerical algorithms solving this type of integrals are described. We use the method of \cite{Gerhardus:2015sla} which applies a construction of \cite{Zhdanov1998}. 

Moreover, in some phases it is possible to directly rewrite the
integral as an infinite sum over residues, 
which will be demonstrated in the examples where it applies.

\subsection{Examples}

In this section we apply the methods of the sections \ref{glsm1} and
\ref{glsm2} to some CY spaces. We start with the quintic, which was
well studied in \cite{CANDELAS199121}. Second, we look at the one
parameter models of \cite{Berglund:1993ax}. Finally we study the
two-parameter models of \cite{Candelas:1993dm}.

\subsubsection*{Quintic \texorpdfstring{$\mathbb{P}^4_{11111}[5]$}{Quintic}}

The quintic $\mathbb{P}^4_{11111}[5]$ has $h^{11}=1$ and is one of the simplest examples. The gauge group is $G=U(1)$. The defining polynomial is

\begin{equation}
P=x_1^5+x_2^5+x_3^5+x_4^5+x_5^5\;,
\end{equation}
so that one chooses the superpotential
\begin{equation}
W=(\Phi_1^5+\Phi_2^5+\Phi_3^5+\Phi_4^5+\Phi_5^5)\,\Sigma_1\;,
\end{equation}
and the charges $Q=(1,1,1,1,1,-5)$. The superpotential forces the
$R$-charges to be of the 
form $R=(2q,2q,2q,2q,2q,2-10q)$, up to the choice of a single parameter $q$. With this choice, $q$ is limited to be taken  from the open interval $(0,2/5)$, such that all $R_i>0$. The actual value of $q$ does not influence the results.

Inserting these values in the partition function (\ref{partition})  and choosing $q=1/5$ one obtains
\begin{equation}
Z=\sum_{m}\int\limits_{-i\infty}^{i\infty}da\; z^{-\frac{m}{2}-a}{\bar{z}}^{\frac{m}{2}-a}\frac{\Gamma \left(-5 \left(a-\frac{m}{2}\right)\right) \Gamma^5 \left(-\frac{m}{2}+a+\frac{1}{5}\right)}{\Gamma^5 \left(-\frac{m}{2}-a+\frac{4}{5}\right) \Gamma \left(5 \left(\frac{m}{2}+a\right)+1\right)}\;.
\end{equation}

In the case $a>0$, i.e. the geometric phase, one can close the contour
to the left, while for $a<0$, i.e. the LG phase, one closes the
contour to the right. The gamma functions in the denominator cancel
the poles such that only one type of gamma functions contribute in
each phase. In the LG phase this is only the gamma function
corresponding to the $\Sigma_1$ field, leading to  first order
poles. The partition function can be rewritten in terms of a sum over
the residues at these poles. Shifting the summation index $m$ and
using the 
identity
\begin{equation}
\Gamma[x+b]={(-1)^b\,\Gamma[1-x]\Gamma[1+x]\over b\,\Gamma[1-x-b]}
\end{equation}
for integer $b$, in terms of hypergeometric functions one  obtains\cite{Jockers:2012dk}
\begin{equation}
\begin{aligned}
Z_{\rm LG}=&{1\over 5} \sum_{l=0}^{3}(-1)^l\, (z\bar{z})^{-{l\over 5}}\,
{\Gamma^5\left({1+l\over 5}\right)\over \Gamma^2(l+1)\,\Gamma^5\left({4-l\over 5}\right)}\cdot \\
&{\textstyle \left| _{4}F_{3}\left({1+l\over 5},\dots ,{1+l\over 5};{2+l\over 5},\dots ,{\widehat{(5-l)+\delta\over 5}}\dots ,{5+l\over 5};-{1\over 5^5z}\right)\right|^2}\;,
\end{aligned}
\end{equation}
where the $\hat{f}$ term is omitted.

The partition function in this LG phase does not contain any
logarithms.
 In the large volume  phase,  the five $Q=1$ chiral fields contribute, leading to fourth order poles. These result in logarithms and polygamma functions in the partition function and a quite lengthy expression.
The metric following from these two partition functions is shown in
figure \ref{fig:metric}. It perfectly agrees with \cite{CANDELAS199121}.

\subsubsection*{Hypersurface in weighted projective space}
\label{sec_1paraquinticmodels}

There are four CYs constructed by the vanishing of a polynomial in a
single weighted projected spaces with $h^{11}=1$. These are
$\mathbb{P}^4_{11111}[5]$, $\mathbb{P}^4_{11112}[6]$,
$\mathbb{P}^4_{11114}[8]$ and $\mathbb{P}^4_{11125}[10]$. The gauge
charges are given by the weights of the projective space and the
degree of the polynomial, $Q_m=k_m$, $Q_6=-d$, and the $R$-charges are
given by $R_m={(2-q)\cdot k_m\over d}$, $R_6=q$. The value of $q$ can
be chosen in the open interval $(0,2)$, without changing the
result. To obtain the highest numerical stability one has to integrate
as far away from the poles as possible, making $q=1$ the most stable 
choice. All models in this category have a similar geometry as the quintic,
i.e. they consist of a geometric phase and a Landau-Ginzburg-orbifold phase
with a $\mathbb{Z}_d$-symmetry.

\subsubsection*{Two parameter models}

In addition  we look at CYs with $h^{11}=2$. This class of CYs
contains for example $\mathbb{P}^4_{11222}[8]$,
$\mathbb{P}^4_{11226}[12]$ or $\mathbb{P}^4_{11169}[18]$. In these
models the charge matrix $Q_{i,j}$ is given by the $l$-vectors, which
span the Mori cone of the given model. These are known for a great
variety of CYs or can be calculate using the PALP
package \cite{Kreuzer:2002uu}.  The gauge charges are listed in table \ref{2param}. 
\begin{table}[ht!]
\caption{Charges for the two-parameter models}
\label{2param}
\resizebox{\linewidth}{!}{
\begin{tabular}{|c|c|c|}
\hline
model&$Q$&$R$\\
\hline
$ \mathbb{P}^4_{11222}[8]$ & $
\begin{pmatrix}
0 & 0 & 1 & 1 & 1 & 1 & -4 \\
1 & 1 & 0 & 0 & 0 & -2 & 0 
\end{pmatrix}
$ &
$\Big(2\times(2 - q_1 - 4 q_2)/8,3\times(2 - q_1)/4, q_2 , q_1\Big)$ \\
\hline
$\mathbb{P}^4_{11226}[12]$& $
\begin{pmatrix}
0 & 0 & 3 & 1 & 1 & 1 & -6 \\
1 & 1 & 0 & 0 & 0 & -2 & 0 
\end{pmatrix}
$ &$\Big(2\times(2 - q_1 - 6 q_2)/12 ,2\times(2 - q_1)/6, (2 - q)/2, q_2 , q_1\Big)$\\
\hline
$\mathbb{P}^4_{11169}[18]$& $
\begin{pmatrix}
0 & 0 & 0 & 2 & 3 & 1 & -6\\
1 & 1 & 1 & 0 & 0 & -3 & 0 
\end{pmatrix}
$ &$\Big(3\times(2 - q_1 - 9 q_2)/18 , (2 - q_1)/9, (2 - q_1)/6, q_2 , q_1\Big)$\\
\hline
\end{tabular}}
\end{table}

\noindent
Here, the $R$-charges of the $\Sigma$ field is denoted $q_1$ and the 
$R$-charge of the field corresponding to the exceptional divisor $q_2$.
The  $R$ charges for the remaining five fields are generally
determined by the superpotential 
condition to be 
\begin{equation}
R_m= {1\over d} \Big(2-q_1-{d\over Q_{6,2}}  q_2\Big)
\end{equation}
for the fields with gauge charge $(0,1)$ and
\begin{equation}
R_m=(2-q_1) {Q_{6,2}\over d}
\end{equation}
for the other fields.
With this choice of parameterization $q_1$ and $q_2$ have to fulfil the conditions $0<q_1<2$ and $0<q_2<{2-q_1\over d/Q_{6,2}}$.

These parameters contain all information needed to evaluate the
partition function and consequently the metric. 
The result is given in terms of the FI parameters which, as described,
can be mapped to the algebraic complex structure moduli
of the  mirror dual CY. In order to compare with the large volume regime,
we finally need the map from the algebraic complex structure moduli
to the K\"ahler moduli $t_i=\int_{\Sigma_i} (B + i J)$ of the original CY.
For that purpose we need the mirror map
that is best computed via the periods of the holomorphic three-form on the mirror CY.
Let us review how this works.

\subsection{K\"ahler potential and mirror map via periods}
\label{app_metric}

The \kahler{} potential can also be determined from the periods of the mirror dual. This will  provide  us with a cross-check for the GLSM method, as
well as with a  way to compute the more detailed information of the
periods and the mirror map. These  are not
easily extracted from the 2-sphere partition  function.

For computing an independent set of periods all over the complex
structure moduli space, for  a simple set of mirror dual CYs,  one can
proceed as follows.
For CYs that are mirror dual to hypersurfaces
in a weighted projective space, one basic  period  can be computed by direct integration
of the holomorphic three-form in the large complex structure regime.
Then one can  analytically continue this expression  into the
Landau-Ginzburg region, in which one can  generate a linearly independent set of
periods by using the symmetries of the deformed polynomial.
A subsequent analytic continuation allows to cover the whole moduli
space.

If the Calabi-Yau threefold $\mathcal{M}$ is
defined by a vanishing polynomial $P$ in a weighted projective space
$W\mathbb{CP}$, then the mirror $\mathcal{W}$ can be constructed as a
quotient $\mathcal{M}/G$, where $G$ is a product of $\mathbb{Z}_n$
symmetries \cite{Greene:1990ud}. This is the so-called Greene-Plesser construction.
The polynomial $P$ is split into a defining (Fermat type) polynomial $P_0$, a
fundamental deformation
$\Phi_0\cdot \prod_{i=1}^5 x_i=\Phi_0\,e_0$ and all other 
possible deformations $\Phi_\alpha e_\alpha$. 
The $\Phi_i$ can be considered as variables on the complex structure
moduli space of  $\mathcal{W}$.

The holomorphic three-form is  given by the residue
\begin{equation}
\Omega(\Phi_\alpha)=\text{Res}_{\mathcal{W}}\left[{ \prod_{i=1}^{5} dx_i\over P(x_i,\Phi_\alpha)}\right]\;,
\end{equation}
and the fundamental period is defined as

\begin{equation}
\label{eq:fund1}
\omega_0(\Phi_\alpha)=-\Phi_0\oint_{B_0} \Omega(\Phi_\alpha)=-\Phi_0{C\over(2\pi i)^{5}}\int_\Gamma{\prod_{i=1}^{5}dx_i\over P(x_i,\Phi_\alpha)}\;,
\end{equation}
where $C$ is an arbitrary constant and the sign as well as factors of
$2\pi i$ can be reabsorbed into it. Moreover, $B_0$ is the fundamental cycle
which is a $T^3$ in the limit $\Phi_0\to\infty$, while $\Gamma$ is an
auxiliary contour  in $\mathbb{C}^{5}$, which allows for a rewriting as a residue integral.
In \cite{Berglund:1993ax},  in the large complex structure limit
$\Phi_0\to\infty$, the residue 
integral \eqref{eq:fund1} has been carried out perturbatively in $1/\Phi_0$ to all orders. 

Given the fundamental period in the large $\Phi_0$ region of the moduli space, one can analytically continue the fundamental period to small $\Phi_0$. In this region, one can obtain a complete set of periods by
\begin{equation}
\omega_j(\Phi)=\omega_0(A^j \Phi)\;,
\end{equation}
where $A$ is the symmetry group of the Fermat type polynomial $P_0$. 
Afterwards, these periods can be analytically continued back to the large $\Phi_0$ region.

\subsubsection*{Periods for one-parameter models}

The first class of CY threefolds are hypersurfaces with $h^{21}=1$, like the
mirror duals of non-singular hypersurfaces in weighted projective
spaces $W\mathbb{CP}_{k_1,\ldots,k_5}[d]$.
In this case \eqref{eq:fund1} evaluates to

\begin{equation}
\omega_0(\Phi_0)=\sum\limits_{r=0}^{\infty}{\Gamma(d\,r+1)\over \prod_{j=1}^5 \Gamma(k_j\,r+1)\,\Phi_0^{d\,r}}\;.
\end{equation}
A full basis of the periods for the one-parameter CY hypersurfaces  have been determined in \cite{Font:1992uk,Klemm:1992tx}.

For  the quintic, i.e. $\mathbb{P}^4_{1,1,1,1,1}[5]$ with
$\Phi_0=5\psi$, the fundamental period in the large complex structure/large volume regime $|\psi|>1$ reads
\begin{equation}
\label{eq:fundamental_period_quintic}
\omega_0(\psi)=\sum\limits_{r=0}^{\infty}{\Gamma(5r+1)\over\Gamma^5(r+1)(5\psi)^{5r}}=\sum\limits_{n=0}^{\infty}{(5n)!\over (n!)^5(5\psi)^{5n}}\;.
\end{equation}

\subsubsection*{Mirror map for one-parameter models}

\label{mirror}
To obtain the mirror map, from the literature
\cite{Berglund:1993ax,Hosono:1994ax} we take  the known form of the
fundamental period  around the large complex structure point 
\begin{equation}
  \omega_0(z)=\sum\limits_{n=0}^{\infty}c_nz^n\;.
\end{equation}
For CYs with $h^{11}$ given as smooth hypersurfaces in weighted
projective space, the coordinate $z$ is related to the deformation
parameter of the fundamental deformation via
\begin{equation}
	z=\frac{\prod_{j=1}^5 k_j^{k_j}}{d^d}\psi^{-d}\;.
\end{equation}
For complete intersections, $d$ is replaced by the sum
of hypersurface degrees. The Picard-Fuchs equation admits exactly one
solution linear in  logarithms
\begin{equation}
  \tilde{\omega}(z)=\frac{1}{2\pi i}\omega_0\log(z)+\sum\limits_{n=0}^{\infty}\tilde{c}_nz^n\;.
\end{equation}
The series coefficients $\tilde{c}_n$ can be determined algorithmically from $\omega_0$ \cite{Hosono:1994ax} as
\begin{equation}
  \tilde{c}_n=\left. \frac{1}{2\pi i}\frac{\partial}{\partial\rho}c_{n+\rho}\right|_{\rho\to 0}\;,
\end{equation}
where implicitly the coefficients $c_n$ have to be analytically
continued in $n$. The mirror map in the large complex structure/large
radius regime  is determined by its monodromy properties as
\begin{equation}
\label{mirrormapA}
  t(z)=\frac{\tilde{\omega}(z)}{\omega_0(z)}\;.
\end{equation}
The continuation to small $\psi$ is  achieved by finding a Mellin-Barnes integral representation for the power series over $c_n$ and $\tilde{c}_n$.

For degree $d$ hypersurfaces in $W\mathbb{CP}^4$, with projective weights $k_j$, we have
\begin{equation}
  \begin{aligned}
    c_n&=\frac{\Gamma(d n+1)}{\prod_{j=1}^5\Gamma(k_j n+1)}\;,\\
    \tilde{c}_n&=\frac{c_n}{2 \pi i}\left(\Psi(d n+1)-\sum_{j=1}^5\Psi(k_j n+1)\right)\;.
  \end{aligned}
\end{equation}
Here $\Psi$ denotes the polygamma function. From the coefficients one
can see that the series converge in the region $|z|<\prod_j k_j^{k_j}/d^d$. The corresponding Mellin-Barnes integrals are
\begin{equation}
  \begin{aligned}
    \omega_0(z)&=\int_{\gamma}\frac{d\nu}{2 i \sin(\pi\nu)}c_\nu(-z)^\nu\;,\\
    \sum\limits_{n=0}^{\infty}\tilde{c}_nz^n&=\int_{\gamma}\frac{d\nu}{2 i \sin(\pi\nu)}\tilde{c}_\nu(-z)^\nu\;.
  \end{aligned}
\end{equation}
For small $z$, corresponding to large $\psi$, i.e. the LCS phase, we pick up the residues of the sine at $\nu\in\mathbb{N}_0$. The residue integral for $\omega_0(z)$ only gets contributions from the simple poles of the gamma function at $\nu=-n/d,\; n\in\mathbb{N}$. The second integrand also has poles at the same values of $\nu$, but now up to second order from the combination of the gamma and polygamma functions.
For the analytic continuation of $\omega_0(z)$ into the Landau-Ginzburg regime, we find
\begin{equation}
  \omega_0(z)=-\frac{\pi}{d}\sum\limits_{n=1}^{\infty}\frac{(-1)^n}{\sin(\pi
    n/d)}\frac{(-z)^{-n/d}}{\Gamma(n)\prod_{j=1}^5 \Gamma(1-n
    k_j/d)}\;,\qquad |z|>\frac{\prod_j k_j^{k_j}}{d^d}\;.
\end{equation}
On the other hand, for the period containing  the logarithm in the LCS phase we obtain
\begin{equation}
	\tilde{\omega}(z)=\frac{\pi}{2i d}\sum\limits_{n=1}^\infty\frac{(-1)^n}{\sin(\pi n/d)}\frac{(-z)^{-n/d}}{\Gamma(n)\prod_{j=1}^5 \Gamma(1-n k_j/d)}\biggr(\cot(\pi n/d)+i\biggr)\;.
\end{equation}
From this information one can now determine the mirror map via
\eqref{mirrormapA}.

\subsubsection*{Periods for two-parameter models}

The second class of CYs is a set of five two-parameter ($h^{11}=2$) Fermat hypersurfaces in $W\mathbb{CP}^4$, for which the full set of periods in the Landau-Ginzburg phase was calculated in \cite{Berglund:1993ax}. These are the manifolds
\begin{equation}
\begin{gathered}
  \mathbb{P}^4_{(1,1,2,2,2)}[8]^{86,2}_{-168}\;,\quad \mathbb{P}^4_{(1,1,2,2,6)}[12]^{128,2}_{-252}\;,\quad \mathbb{P}^4_{(1,1,1,6,9)}[18]^{272,2}_{-540}\;,\\[0.2cm]
   \mathbb{P}^4_{(1,4,2,2,3)}[12]^{74,2}_{-144}\;,\quad \mathbb{P}^4_{(1,7,2,2,2)}[14]^{122,2}_{-240}\;.
\end{gathered}
\end{equation}
Their mirrors are given by the vanishing set of the polynomial
\begin{equation}
  P=\sum_{j=1}^{5}x_j^{d/k_j}-\psi\, x_1 x_2 x_3 x_4 x_5-\frac{d}{q_1}\phi \,x_1^{q_1} x_2^{q_2} x_3^{q_3} x_4^{q_4}x_5^{q_5}\;,
\end{equation}
after modding out an appropriate discrete symmetry group. Here $d$ denotes the degree of the polynomial, the $k_i$ are the projective weights, $D=d/q_1$ is always an integer and the $q_i$ ($i \neq 1$) can be computed from the projective weights and $D$ through

\begin{equation}
    \frac{q_i k_i}{q_1}=\begin{cases}
      0\;,\qquad i\geq D \\1\;,\qquad i<D
    \end{cases}\;.
\end{equation}
For the computation of the periods it suffices to know that $D=2$ for all of the above models, except for the case\footnote{Due to the differing value of $D$, the periods for this model behave in a slightly different way. The subsequent formulae are valid for the case $D=2$ but the method can be easily carried over to $D=3$, see appendix \ref{app_11169}.}
of $\mathbb{P}^4_{(1,1,1,6,9)}$, where $D=3$.
The fundamental period $\omega_0$ in the large complex structure/large volume regime has been computed in
\cite{Berglund:1993ax} to be given by

\begin{equation}
\label{eq:omega_LCS}
    \omega_0(\psi,\phi)=\sum\limits_{l=0}^{\infty}\frac{(q_1 l)!(d\psi)^{-q_1 l}(-1)^l}{l!\prod_{i=2}^5 \left(\frac{k_i}{d}(q_1-q_i)l\right)!}U_l(\phi)\;,
\end{equation}
where the function $U_\nu(\phi)$ can be written in terms of hypergeometric functions as
\begin{eqnarray}
\label{eq:ufunct}
  U_\nu(\phi)&\!\!\!=\!\!\!&\frac{e^{\frac{i\pi\nu}{2}}\Gamma\Big(1+\frac{\nu}{2}(k_2-1)\Big)}{2\Gamma(-\nu)}\left[2
    i \phi \frac{\Gamma(1-\nu/2)}{\Gamma\left(\frac{1+\nu
          k_2}{2}\right)}\enspace
    {}_2F_1\left(\frac{1-\nu}2,\frac{1-k_2\nu}2;\frac32;\phi^2\right)+\right. \nonumber\\
  &&\phantom{aaaaaaaaaaaaaaaaa}\left.+\frac{\Gamma(-\frac{\nu}{2})}{\Gamma\left(\frac{2+\nu k_2}{2}\right)}\enspace{}_2F_1\left(-\frac{\nu}2,-\frac{k_2\nu}2;\frac12;\phi^2\right)\right]\,.
\end{eqnarray}
For fixed values of $\phi$ the series converges for sufficiently large
$\psi$. The actual convergence criterion is model-dependent. 

In order to obtain a full set of periods, this expression has to be analytically continued to small $\psi$. The result is \cite{Berglund:1993ax}
\begin{equation}
\label{eq:omega_phi}
  \omega_0(\psi,\phi)=-\frac2d \sum_{n=1}^\infty \frac{\Gamma(\frac{2
      n}{d})\,(-d \psi)^n\,
    U_{-\frac{2n}{d}}(\phi)}{\Gamma(n)\,\Gamma\big(1-\frac{n}{d}(k_2-1)\big)\,
    \prod_{i=3}^{5}\Gamma\big(1-\frac{k_i n}{d}\big)}\;,
\end{equation}
which converges for sufficiently small $\psi$.
By acting with the phase symmetry of the polynomial one derives the
remaining periods
\begin{equation}
  \omega_j(\psi,\phi)=\omega_0(\alpha^j\psi,\alpha^{jq_1}\phi)\;,
\end{equation}
where $\alpha$ is a $d$-th root of unity.

As this set of periods is overcomplete, we have to choose a linearly
independent subset\footnote{In the cases of interest
  to us, one can take the first 6 periods.}. This form of the periods
is useful for performing the analytic continuation to large $\phi$,
since it can be done by standard techniques for the hypergeometric
function. In order to continue the periods back to the region where $\psi$
is large, we will find it useful to use an alternative form, where the
principal summation runs over powers of $\phi$ times a certain
generalized hypergeometric function in $\psi$. The result of a rather
lengthy computation is that $\omega_j(\psi,\phi)$ can be decomposed
into eigenfunctions $\eta_{j,r}(\psi,\phi)$ of the phase symmetry
$(\psi,\phi,j)\to(\alpha\psi,-\phi,j+1)$ 
as
\begin{equation}
\label{eq:omega_psi}
    \omega_j(\psi,\phi)=-\frac2d \sum_{r=1}^d (-1)^r \,e^{2\pi i j r/d}\;\eta_{j,r}(\psi,\phi)\;
\end{equation}
with
\begin{equation}
    \eta_{j,r}(\psi,\phi)=\frac12 \sum_{n=0}^\infty e^{i \pi n
      (j+1/2)}\, \frac{(2\phi)^n}{n!}\;V_{n,r}(\psi)\;.
\end{equation}
Now, the full  $\psi$-dependence is contained  in  the functions $V_{n,r}(\psi)$
\begin{equation}
    V_{n,r}(\psi)=N_{n,r}\,(d\,\psi)^r\,H_{n,r}(\psi)\,,
\end{equation}
that are  analogues of the function $U_\nu(\phi)$ in \eqref{eq:ufunct}.
They consist of a generalized hypergeometric function $H_{n,r}(\psi)$
and a numerical prefactor $N_{n,r}$.
The first is explicitly given by
\begin{equation}
    H_{n,r}(\psi)={}_{(d+1)}F_d\bigg(1,\frac{n}{2}+\frac{r}{d},\underbrace{1+\frac{r}{d}-\frac{l_2+1-\tfrac{n}{2}}{k_2},1+\frac{r}{d}-\frac{l_i+1}{k_i}}_{i=3,\dots,5\qquad l_i=0,\dots,k_{i}-1};\underbrace{\frac{r+l}{d}}_{l=0,\dots,d-1};\prod_{j=1}^5 k_j^{k_j}\, \psi^d\bigg)\;
\end{equation}
where the underbrackets indicate that for each value in the allowed index range for $i,l_i,l$ we have to insert the corresponding parameter in the hypergeometric function. For the relevant models $k_1=1$, so we indeed obtain a hypergeometric function with $(p,q)=(2+k_2+\cdots+k_5,d)=(d+1,d)$.

The numerical prefactor can explicitly be expressed as
\begin{equation}
  \begin{aligned}
    N_{n,r}&=\pi^{d-3}\, d^{\frac12 -r}\left(\prod_{j=2}^5k_j^{-\frac12 +\frac{k_j r}{d}}\right)k_2^{\frac{n}{2}}\\
    &\times\frac{\Gamma\left(\frac{n}{2}+\frac{r}{d}\right)}{\prod\limits_{l=0}^{d-1}\Gamma\left(\frac{l+r}{d}\right)\prod\limits_{l_2=0}^{k_2-1}\Gamma\left(\frac{l_2+1-n/2-k_2 r/d}{k_2}\right)\prod\limits_{i=3}^5\prod\limits_{l_i=0}^{k_i-1}\Gamma\left(\frac{l+1-k_i r/d}{k_i}\right)}\;.
  \end{aligned}
\end{equation}
This expression is valid for $\phi<1$ and arbitrary $\psi$, with
implicit analytic continuation of the hypergeometric function
understood. In practice the degree of the generalized hypergeometric
function ${}_pF_q$ is reduced in all these examples to at most $(p,q)=(13,12)$.

To compute the metric on moduli space, we first have to transform a
linearly independent set of periods into a symplectic basis
$\Pi=(F_\Lambda,X^\Lambda)$. Then one  can  calculate the \kahler{}
potential via
\begin{equation}
\label{kaehlerpotperiods}
    K=-\log\Big(\! -i\overline{\Pi}\,\Sigma\,\Pi\Big)=-\log\Big(\! -i(X^\Lambda \overline{F}_\Lambda-\overline{X}^\Lambda F_\Lambda)\Big)\;.
\end{equation}
Here 
\eq{\Sigma=\begin{pmatrix}
	0 & \mathbf{1}\\
	-\mathbf{1} & 0\\
\end{pmatrix}
}
is the symplectic scalar product.

The basis transformation can be found case by case by a monodromy calculation \cite{Candelas:1993dm,Candelas:1994hw} or by the algorithmic procedure of \cite{Aleshkin:2017eby}.
In our analysis of the RSDC for two parameter
models, we will focus on the three manifolds
$\mathbb{P}^4_{11222}[8],\mathbb{P}^4_{11169}[18]$ and
$\mathbb{P}^4_{11226}[12]$. For the first two, the mirror map and
change to symplectic basis can be found in
\cite{Candelas:1993dm,Candelas:1994hw}. For
$\mathbb{P}^4_{11226}[12]$, we will compute it in the following.

\subsubsection*{An Example: $\mathbb{P}^4_{11226}[12]$}

To illustrate the calculation of
the periods, the  mirror map and  the metric on the whole moduli
space, we consider the CY defined as the mirror of the
hypersurface $\mathbb{P}^4_{11226}[12]$.  A cousin of this model, $\mathbb{P}^4_{11222}[8]$, has first
been analyzed in great detail in \cite{Candelas:1993dm}, although the
emphasis has been mostly on the LCS region. The
defining polynomial 
is
\begin{equation}
  P(x)=x_1^{12}+x_2^{12}+x_3^6+x_4^6+x_5^2-12\psi\, x_1 x_2 x_3 x_4 x_5-2\phi\, x_1^6 x_2^6\;,
\end{equation}
where we have to mod out a $H=\mathbb{Z}_6^2\times\mathbb{Z}_2$ phase
symmetry \cite{Candelas:1993dm}. The transformation
$(\phi,\psi)\to(-\phi,\alpha\psi)$ can be absorbed into a coordinate
redefinition of the ambient space which leaves the hypersurface
constraint $P(x)$ invariant, so the actual (uncompactified) moduli
space becomes the corresponding $\mathbb{Z}_{12}$ quotient of
$\mathbb{C}^2$. The manifold develops a conifold singularity at 
\begin{equation}
  864\psi^6+\phi=\pm1\;.
\end{equation}
Before analytic continuation, a detailed analysis of the asymptotic
behavior of $U_\nu(\phi)$ and application of the Cauchy root test
shows that the periods \eqref{eq:omega_phi} converge in the region
$|\phi|<1$ and $|864\psi^6|<|\phi\pm1|$, where the ``$\pm$" indicates
the minimum of the two values. Upon analytical continuation one finds
four distinct regions of the moduli space as summarized in table
\ref{tab:P11226regions}. 

\begin{table}[ht]
\centering
  \begin{tabular}{l r}
    Region & Convergence Criterion\\
    \hline
    Landau Ginzburg & $|\phi|<1$ and $|864\psi^6|<|\phi\pm1|$\\
    Hybrid: $\mathbb{P}^1$ fibration & $|\phi|>1$ and $|864\psi^6|<|\phi\pm1|$\\
    Hybrid: orbifold & $|\phi|<1$ and $|864\psi^6|>|\phi\pm1|$\\
    LCS & $|\phi|>1$ and $|864\psi^6|>|\phi\pm1|$
  \end{tabular}
  \caption{Different physical regions in the complex structure moduli space of the mirror of $\mathbb{P}^4_{11226}$. The $\pm$ is to be interpreted as a logical ``and".}
  \label{tab:P11226regions}
\end{table}

\noindent
The LG and LCS  regions are familiar from the quintic.
 In addition we get
two hybrid regions, which share properties of the 
LG and LCS regions. The hybrid region where $\phi\to\infty$ and $\psi$ stays small will be called the $\mathbb{P}^1$(-fibration)-phase, whereas the region with $\psi\to\infty$ and $\phi$ small will be referred to as the orbifold(-hybrid)-phase, for reasons that will become clear in section \ref{sec:two_parameter}.

The reduced set of periods $(\omega_0,\dots,\omega_5)$ form a
basis. We can calculate these in the LG phase by expanding the
hypergeometric function in \eqref{eq:omega_phi} around $\phi=0$. The
result is polynomial in both $\phi$ and $\psi$. In the $\mathbb{P}^1$
fibration region we expand the hypergeometric function around
$i\infty$. We find that the even periods $\omega_{2j}$ now contain
simple logarithms in $\phi$ 
\begin{equation}
\frac{\vec{\omega}_{\mathbb{P}1}(\psi,\phi)}{\psi}=
  \begin{pmatrix}
    (4.39+0.00i) \\
    (5.61+0.70 i)+(1.21+0.70i)\log\phi\\
    (2.20+3.80i) \\
    (2.20+5.21i)+(0.00+1.40 i) \log\phi\\
    (-2.20+3.80i) \\
    (-3.41+4.51 i)-(1.21-0.70i)\log\phi\\
  \end{pmatrix}
  \phi^{-1/6}+\mathcal{O}\left(\phi^{-5/6}\right)\,.
\end{equation}
In a similar fashion, in order to obtain an expression for the periods in the orbifold hybrid phase, we expand the generalized hypergeometric function
in \eqref{eq:omega_psi} around $i\infty$, upon which all of the
periods except $\omega_0$ acquire logarithmic terms up to third order
$\log(\psi)^3$.

For the LCS region, standard tools are available to compute the metric and periods such as INSTANTON \cite{Hosono:1994ax}. For this reason we will not further pursue the analytic continuation of the periods into the LCS region.

\subsubsection*{Mirror map and intersection matrix for two-parameter models}
\label{app_mirror11226}

We now explain the computation of the mirror map and intersection
matrix. These can both be determined calculating the monodromy
matrices of the periods around certain boundary divisors of the
compactified moduli space. For the manifolds $\mathbb{P}^4_{11222}[8]$
and $\mathbb{P}^4_{11169}[18]$ this has been done in
\cite{Candelas:1993dm} and \cite{Candelas:1994hw}, respectively. Next,
we analyse the case  $\mathbb{P}^4_{11226}[12]$, for which the
procedure is analogous to $\mathbb{P}^4_{11222}[8]$ (see \cite{Candelas:1993dm}).

Using the just determined explicit form of the periods, it is straightforward to calculate the monodromy transformations. For the monodromy obtained by moving around $\phi=1$, we find $\vec{\omega}\to \mathrm{B}\,\vec{\omega}$ with
\begin{equation}
  \mathrm{B}=\begin{pmatrix}
    1 & 0 & 0 & 0 & 0 & 0 \\
    1 & 0 & 0 & 0 & -1 & 1 \\
    -1 & 1 & 1 & 0 & 1 & -1 \\
    0 & 0 & 1 & 0 & 2 & -2 \\
    0 & 0 & -1 & 1 & -1 & 2 \\
    0 & 0 & 0 & 0 & 0 & 1 \\
  \end{pmatrix}\;.
\end{equation}
The monodromy matrix $\mathrm{T}$ about the conifold is determined to be
\begin{equation}
  \mathrm{T}=\begin{pmatrix}
    2 & -1 & 0 & 0 & 0 & 0 \\
    1 & 0 & 0 & 0 & 0 & 0 \\
    -1 & 1 & 1 & 0 & 0 & 0 \\
    -2 & 2 & 0 & 1 & 0 & 0 \\
    2 & -2 & 0 & 0 & 1 & 0 \\
    1 & -1 & 0 & 0 & 0 & 1 \\
  \end{pmatrix}\;.
\end{equation}
Finally, we calculate the monodromy matrix $\mathrm{A}$ corresponding to the monodromy around $\psi=0$
\begin{equation}
  \mathrm{A}=\begin{pmatrix}
    0 & 1 & 0 & 0 & 0 & 0 \\
    0 & 0 & 1 & 0 & 0 & 0 \\
    0 & 0 & 0 & 1 & 0 & 0 \\
    0 & 0 & 0 & 0 & 1 & 0 \\
    0 & 0 & 0 & 0 & 0 & 1 \\
    -1 & 0 & 0 & 0 & 0 & 0 \\
  \end{pmatrix}\;.
\end{equation}
Following \cite{Candelas:1993dm} we define the matrix $\mathrm{T}_{\infty}=(\mathrm{A} \mathrm{T})^{-1}$. The monodromies around the boundary divisors whose intersection is the large complex structure point are then $\mathrm{S}_1=\mathrm{T}_{\infty}^2$ and $\mathrm{S}_2=\mathrm{B}^{-1}\mathrm{T}_{\infty}$. We also define $\mathrm{R}_i=\mathrm{S}_i-1$. We check that the triple products between the $\mathrm{R}_i$ reproduce the triple intersection numbers of $\mathbb{P}^4_{11226}[12]$. Using the ansatz
\begin{equation}
  t^i=\frac{\vec{A}^i\cdot\vec{\omega}}{\omega_0}\;,
\end{equation}
where $\vec{A}^i\;,i=1,2$ are row vectors and demanding that monodromies around the LCS boundary divisors correspond to shifts of the $B$-field, hence gauge transformations
\begin{equation}
  \vec{A}^i\cdot\mathrm{R}_j=\delta^i_j\, (1,0,0,0,0,0)\;,
\end{equation}
we can solve for the mirror map up to a constant shift. The result for the $A$-vectors is
\begin{equation}
  \vec{A}^1=\left(c_1,0,\frac12,0,\frac12,0\right)\;,\qquad \vec{A}^2=\left(c_2,\frac12,-\frac12,\frac12,-\frac12,\frac12\right)\;,
\end{equation}
where we fix the constants to be $c_1=-\frac12$ and $c_2=\frac12$.
By demanding that the monodromy in the symplectic basis of periods are in fact integral and symplectic, we determine the basis transformation to be
\begin{equation}
  \Pi=\begin{pmatrix}
    -1 & 1 & 0 & 0 & 0 & 0\\
    \frac12 & \frac12 & \frac32 & -\frac12 & \frac12 & -\frac32\\
    2 & 0 & 0 & 0 & -1 & 0\\
    1 & 0 & 0 & 0 & 0 & 0\\
    -\frac12 & 0 & \frac12 & 0 & \frac12 & 0\\
    \frac12 & \frac12 & -\frac12 & \frac12 & -\frac12 & \frac12\\
  \end{pmatrix}\omega\;.
\end{equation}
This can then be used to compute the K\"ahler potential \eqref{kaehlerpotperiods}.

\newpage

\section{RSDC for CY manifolds with $h^{11}=1$}

In this section we investigate the manifestation of the RSDC for
regions of one dimensional Calabi-Yau \kahler{} moduli spaces beyond
the large volume phase. As our prototype example, we will discuss
the quintic in very much detail.
Besides the large volume point, there also exist the conifold and Landau-Ginzburg orbifold points. Recall that in proper distance, the large volume point was infinitely far away from any other point in moduli space, but that for field distances larger than $\Theta_\lambda=\sqrt{3/4}M_\text{pl}<M_\text{pl}$ a logarithmic scaling sets in that renders infinitely many states exponentially light. Thus, at distances larger than $M_\text{pl}$ the effective field theory could not be trusted anymore.
The question is whether  proper distances $\Theta_0$,  accumulated
before by traversing  non-geometric phases, are also smaller than the
Planck-scale. 
Besides the quintic we also check the RSDC for the other three
one-parameter Calabi-Yau manifolds given as smooth hypersurfaces in
$W\mathbb{CP}$.

\subsection{An illustrative example: The quintic $\mathbb P_{11111}^4[5]$}
\label{sec_quintic_pheno}

After investigating the local properties of the metric on the quintic moduli space
at the LG and conifold point, we move on to discuss the global
properties of the moduli space and test the RSDC in this setting. This
requires finding the shortest geodesics between two points in moduli
space. Because of the complicated form of the metric, described in
terms of hypergeometric functions, we solve the geodesic equation
numerically. To obtain the metric for the \kahler{} moduli space of
the quintic we use known results for the periods of the complex
structure moduli space of the mirror quintic as well as the mirror
map.
Later, for the other  one-parameter Calabi-Yau threefolds,  we will
obtain the metric by the GLSM construction of section
\ref{sec_KahlerMetric}. The availability of these two separate
computational methods provides us with a useful crosscheck.

\subsubsection*{The Landau-Ginzburg point of the quintic}

We want to compute the metric on the \kahler{} moduli space of the
quintic threefold. For this purpose we consider the mirror dual of the 
quintic $\mathbb P_4[5]^{(101,1)}$,
whose single complex structure modulus is given by the complex
parameter $\psi$ appearing in the hypersurface constraint
\eq{   
P=\sum_{i=1}^5  x_i^5 - 5\psi \prod_{i=1}^5  x_i =0\,.
}
The transformation $\psi\to\alpha\psi$, where $\alpha^5=1$, can be
absorbed into a redefinition of the coordinates $x_i$. For this reason
the moduli space has the simple form of an angular wedge $0\leq\text{Arg}(\psi)<2
\pi/5$. By mirror symmetry the complex structure moduli space of this
Calabi-Yau is equivalent to the \kahler{} moduli space of the
quintic. After applying the mirror map, it has the structure  depicted
in figure \ref{fig:toymodulispace} . At $\psi=\infty$ there is  the
large complex structure point, which is mapped to the large volume
point $t=i\infty$ of the quintic.
For the co-dimension one conifold locus $\psi^5=1$, which is mapped to $t\simeq i 1.21$, this hypersurfaces becomes singular, i.e.
$P=\partial_i P=0$ for $i=1,\ldots,5$. Moreover, at $\psi=0$ one has
the Landau-Ginzburg  point which is the point of smallest (but still finite) volume in the mirror dual description where it is located at $t= 0.5+i\cot(\pi/5)/2$. Because of this property, it is also the point deepest inside the non-geometric regime of the moduli space.

Following \cite{CANDELAS199121}, one can solve
the Picard-Fuchs equations around the Landau-Ginzburg point where an independent set of solutions is given by\footnote{Note that a different basis of the periods can be obtained from the fundamental period \eqref{eq:fundamental_period_quintic} by acting with the $\mathbb{Z}_5$ symmetry of this model, see \cite{CANDELAS199121}.}

\eq{
\omega_k(\psi)=(5\psi)^{k} \sum_{n=0}^\infty  \left({\Gamma(k/5+n)\over
    \Gamma(k/5)}\right)^5
         {\Gamma(k)\over \Gamma(k+5n)} \, (5\psi)^{5n}
}
for $k=1,2,3,4$. These infinite sums converge for $|\psi|<1$ and have
the fundamental domain $0\le \text{Arg}(\psi)<2\pi/5$. In the \kahler{}
moduli space of the quintic, this corresponds to the non-geometric
phase. They can be rewritten in terms of hypergeometric functions,
whose  analytic continuation to the region $|\psi|>1$, corresponding to the large volume region, is straightforward.
The transformation to a symplectic basis is performed by applying
a  matrix $\mathrm{m}$, so that $\Pi=\mathrm{m}\, \omega$, whose inverse
was explicitly stated in \cite{Bizet:2016paj}.

Thus, the symplectic basis of the periods can be expanded around the Landau-Ginzburg point as
\eq{
F^0&=2.937558i\, \psi-4.289394 i \,\psi^{2}+1.462601 i\,
\psi^{3} +O(\psi^{4})\\[0.1cm]
F^1&=(7.314220-11.691003 i)\,\psi-(0.963029-6.520577
i) \,\psi^{2}\\
&\phantom{=}-(0.328374+2.223391 i) \,\psi^{3} +O(\psi^{4})\\[0.1cm]
X_0&=(2.021600-1.468779 i) \,\psi-(0.696854-2.144697 i)\,
\psi^{2}\\
&\phantom{=}-(0.237613+0.731300i) \,\psi^{3}+O(\psi^{4})\\[0.1cm]
X_1&=2.125637i\, \psi-1.185559 i\, \psi^{2}+0.404253 i\, \psi^{3}+O(\psi^{4})
}
where we have actually expanded them up to order $100$.
Writing $\psi=|\psi|\exp(i\theta)$, the resulting K\"ahler potential takes the simple form
\eq{
           K&=-\log\Big[ -i(X^\Lambda \ov F_\Lambda-\ov X^\Lambda  F_\Lambda)\Big]\\[0.1cm]
 &=-\log\Big(    19.217617\,|\psi|^{2}
           -3.694710\,|\psi|^{4} + 0.429576\,|\psi|^{6} \\
       &\phantom{=-\log\Big(}+ 0.320294\,|\psi|^{7} \cos(5\theta) +O(|\psi|^{8})  \Big)\;.
}
We notice that the first three terms feature a continuous
phase shift symmetry $\theta\to \theta+\Delta\theta$ that is broken by the
fourth and higher order terms to a discrete shift symmetry $\theta\to
\theta+2\pi n/5$. This behavior is very similar to the large complex
structure point where also the tree-level K\"ahler potential features
a continuous shift symmetry that is broken to a discrete one by
non-perturbative instanton corrections\footnote{Note that here we have
  used the wording from the mirror dual K\"ahler moduli perspective.}.

\subsubsection*{Swampland Distance Conjecture}

The Swampland Distance Conjecture says that any quantum gravity
effective action has only a finite range of validity in field space. In its
refined version this critical distance is of the order of the
Planck-length. Recall that the intuition for this conjecture came from reasoning
in the volume regime. There, the proper field
distance $\Theta$  was related to the distance $t$ appearing in the masses of KK
modes 
as $\Theta\sim \lambda^{-1} \log t$ so that for $\Theta>\Theta_\lambda=\lambda^{-1}$
infinitely
many states became exponentially light, thus  spoiling the validity of the effective field theory.

The question we can now approach is, how this picture generalizes to the 
stringy quantum regime of small volumes/radii. Here, where the asymptotic logarithmic scaling of the proper distance is not present, we may potentially accumulate a contribution $\Theta_0$ to the proper distance without a significant change in mass scales. If this was already above the Planck scale, it would be in contradiction with the RSDC.

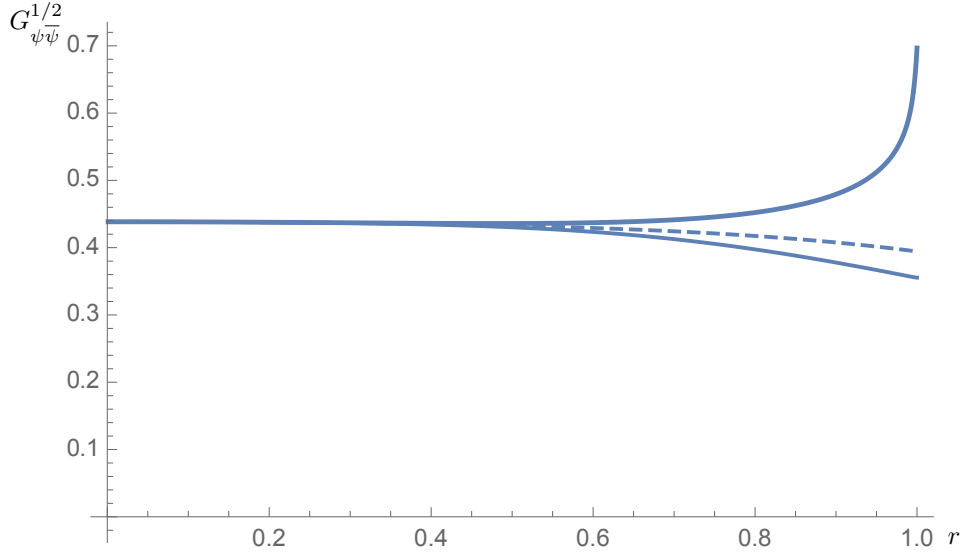
\begin{figure}[ht]
  \centering
  \includegraphics[width=0.8\textwidth]{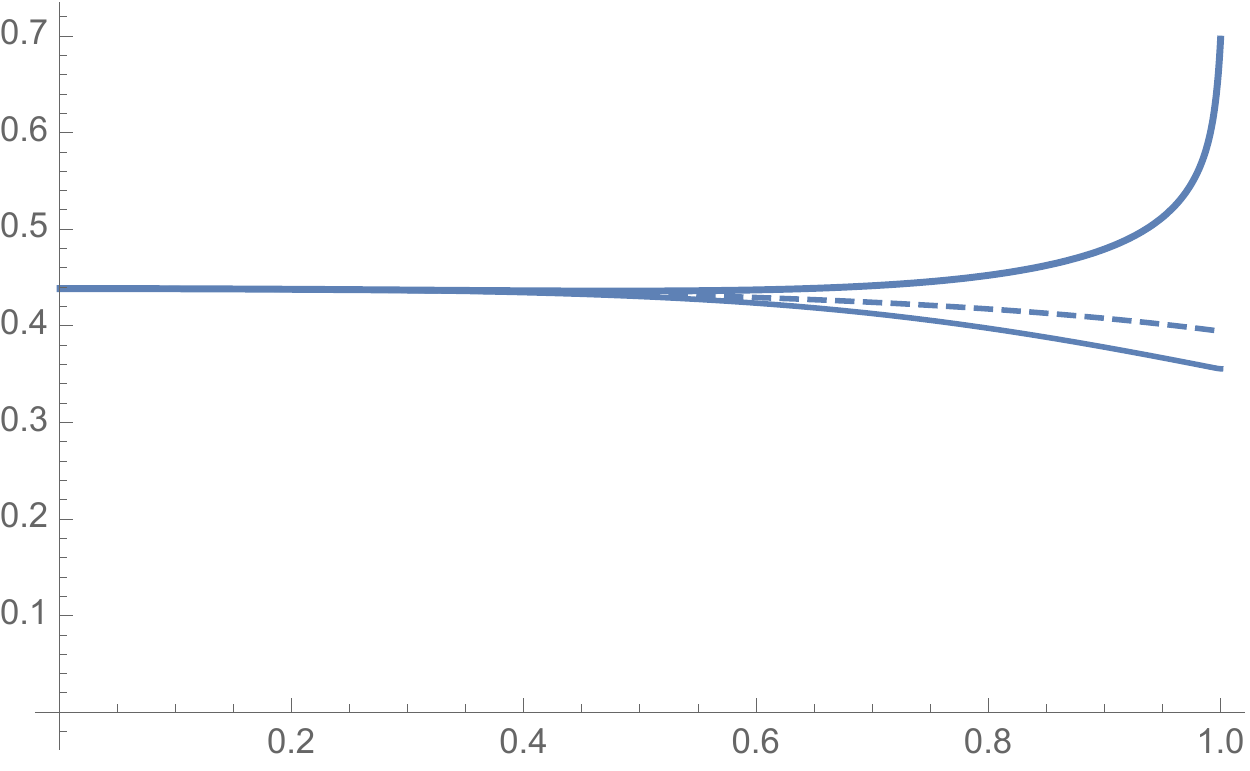}
  \begin{picture}(0,0)
  \put(0,5){\footnotesize$r$}
  \put(-355,200){\footnotesize$G^{1/ 2}_{\psi\ov \psi}$}
\end{picture}
  \caption{Plot of ${G^{1/ 2}_{\psi\ov \psi}(r,\theta)}$. Dashed line: up to order
      $r^6$. Solid upper line: up to order $r^{100}$ for
      $\theta=0$. Solid lower line:  up to order $r^{100}$ for
      $\theta={2\pi\over 10}$.}
  \label{fig:sqrtk}
\end{figure}

First, we can compute the proper field distance between the Landau-Ginzburg
point $\psi=0$ and a point $\psi=r e^{i\theta}$ inside the radius of
convergence $0\le r<1$.
As we will see shortly, for $r<0.5$ we can approximate the K\"ahler potential
by its first three shift symmetric terms. In this case the geodesic 
is just a ray of constant $\theta$ so that the proper field distance is
\eq{
            \Delta \Theta \sim\int_0^{1}  dr\, \sqrt{G_{\psi\ov \psi}(r)}
}
with $r=|\psi|$. At lowest order in $r$ the K\"ahler metric is just a constant
\eq{
                     G_{\psi\ov \psi}(r)\sim   {\beta\over \alpha}\,
}
with $\alpha=19.217617$ and $\beta=3.694710$ so that
\eq{
            \Delta \Theta =\int_0^{1}  dr\, \sqrt{G_{\psi\ov \psi}(r)}\sim
            \sqrt{\beta\over\alpha} \sim 0.44\,.
}
Including also the term of order $|\psi|^{6}$ in the K\"ahler
potential, only changes this result slightly to $0.43$. 
With the periods computed up to order $O(100)$ we have numerically
evaluated the integral. In figure \ref{fig:sqrtk} we show the behavior
of  the integrand $\sqrt{G_{\psi\ov \psi}(r,\theta)}$ along the two rays
$\theta=0$ and $\theta={2\pi\over 10}$.
In the figure we also included the contribution from the first three
shift symmetric terms. It shows that up to $r=1/2$ this is a good
approximation and the angle dependence sets in once one reaches
the boundary. The divergence for the line $\theta=0$ at $r=1$
reflects the fact that here one  hits the conifold singularity, where
the metric diverges. For the  mid point $\theta={2\pi\over 10}$, the metric
even becomes smaller close to the radius of convergence. 
However, for the proper field distance we find the numerical results
\eq{
          \Delta \Theta(\theta=0)=0.45\,, \qquad\Delta \Theta(\theta={\textstyle
            {2\pi\over 10}})=0.42\,.
}
\noindent
Let us make two  remarks
\begin{itemize}
\item{Unlike the large volume point, the Landau-Ginzburg point is
    at finite proper field distance. Therefore, one does not get the
    $\log$-behavior for large field distances, as for the  large
    complex structure  point. As a consequence the points at large
    distance are still those in the large volume region.
}
\item{In order to reach those points, one has to cross the chart $0\le
    |\psi|<1$. Therefore, if this region had already  a trans-Planckian
    radius, the RSDC would be violated. 
The Refined Swampland Distance Conjecture in such a case means that the
  radius of convergence   should not be larger
    than $M_{\rm pl}$. This is {\it consistent} with the $\Delta \Theta$ we
  found. 
}
\end{itemize}

\subsubsection*{The conifold point}

As we have seen,
even though the metric diverges at the conifold point $\psi^5=1$,  the distance to it converges. The conifold point is not
as interesting as the Landau-Ginzburg point with regards to violating
the RSDC because it sits right at the boundary between the two
phases. It is at maximum distance to the Landau-Ginzburg point, but
this distance is still finite and below the Planck scale. The distance
to the large volume phase is infinitesimal. This means that
independent of the direction in which we displace, there will be no
tension with the RSDC. In the following discussion we will therefore
focus on geodesics which start at the point deepest inside the
non-geometric phase, i.e. the Landau-Ginzburg point,
and continue into  the large volume phase.

\subsubsection*{Trajectories traversing multiple patches}

After locally checking some necessary conditions for the Refined
Swampland Distance Conjecture to hold, we will now consider geodesics
which traverse multiple coordinate patches. 
For that purpose, we use the following general procedure
\begin{itemize}
  \item Determine the metric on the moduli space and compute the corresponding geodesics $x^\mu(\Theta)$, parametrized by proper distance $\Theta$.
  \item Identify a tower of states whose mass should decrease along the geodesics.
  \item Find the mass $M_\text{KK}(\psi)$ of this tower as a function of the position in moduli space.
  \item Express the mass $M_\text{KK}(\Theta)$ in terms of the proper distance along the geodesics.
\end{itemize}

\begin{figure}[ht]
  \centering
  \includegraphics[width=\textwidth]{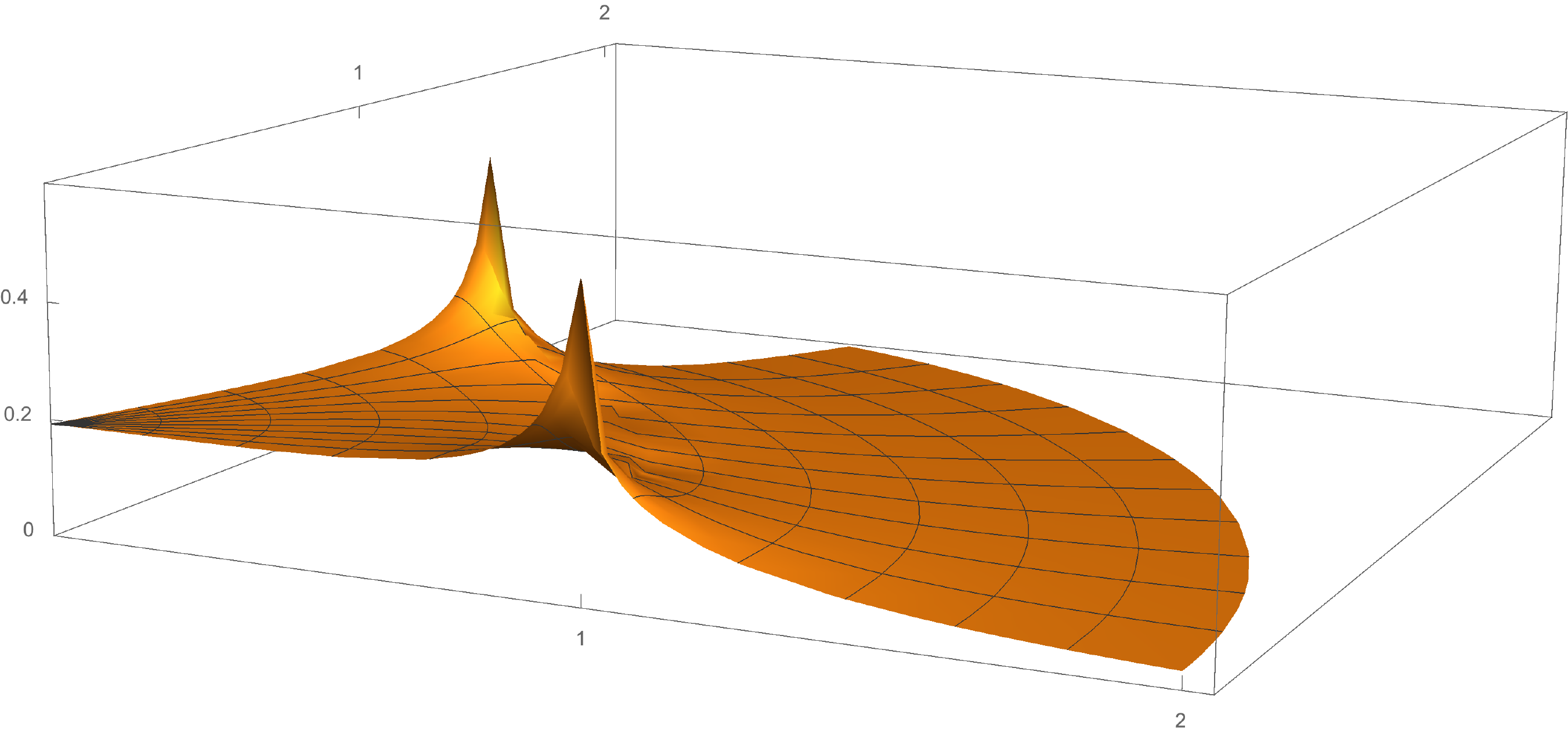}
  \begin{picture}(0,0)
  \put(130,20){\footnotesize$\text{Re}(\psi)$}
  \put(-30,205){\footnotesize$\text{Im}(\psi)$}
  \put(-200,175){\footnotesize$G_{\psi\overline{\psi}}(\psi)$}
\end{picture}
  \caption{The metric on the \kahler{} moduli space of the quintic.}
  \label{fig:metric}
\end{figure}
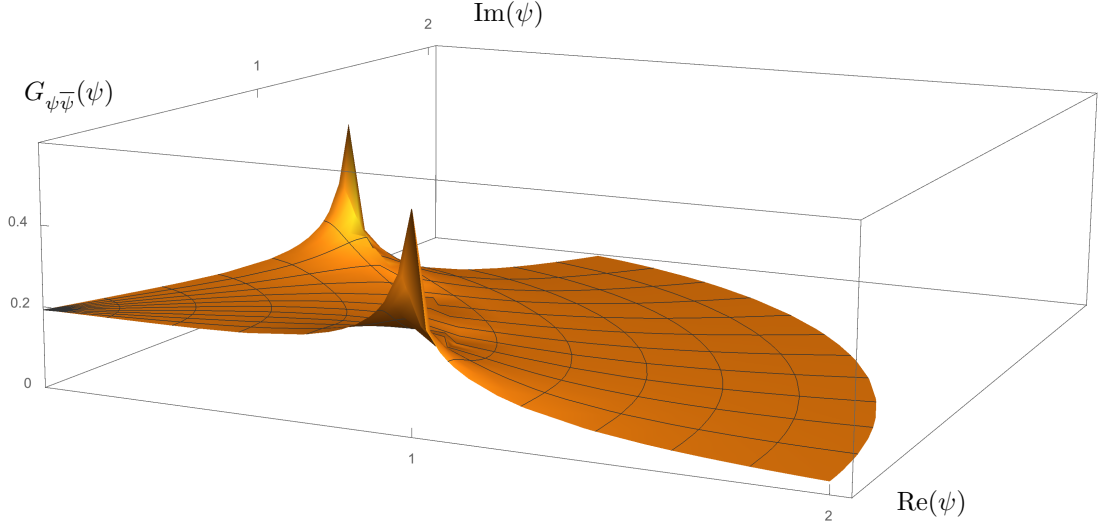

\vspace{0.2cm}
The moduli space metric of the quintic is obtained patchwise from the periods in the Landau-Ginzburg and large volume regions respectively up to order $\psi^{50}$, or alternatively by the GLSM construction of section \ref{sec_KahlerMetric}. The resulting metric is illustrated in figure \ref{fig:metric}.

\noindent
The geodesics can be obtained numerically by solving the geodesic equation
\begin{equation}
\label{geodesicEq}
  \frac{d^2x^\mu}{d\tau^2}+\Gamma^\mu_{\alpha\beta} \frac{dx^\alpha}{d\tau} \frac{dx^\beta}{d\tau}=0\;,
\end{equation}
where $\tau=a \Theta+b$ is an affine parameter and in our case $x=(r,\theta)$. We set $b=0$ without loss of generality. The parameter $a$ can then be determined from the initial conditions by computing the square root of the pullback of the metric onto the geodesic

\begin{equation}
  a=\frac{d\Theta}{d\tau}=\left.\sqrt{G_{\alpha\beta}(x(\tau))\frac{dx^\alpha}{d\tau}\frac{dx^\beta}{d\tau}}\, \right|_{\tau=0}\;.
\end{equation}

We first want to investigate the fate of geodesics going radially
outward from the Landau-Ginzburg point $\psi=0$. While the coordinate
$\psi=r\exp(i \theta)$ is periodic, $\theta\equiv \theta+ 2 \pi/5$,
the metric additionally enjoys an enhanced $\mathbb{Z}_2$ reflection
symmetry along the ray $\theta=\pi/5$. This allows us to restrict to
geodesics in the angular region $\theta\in(0,\pi/5)$. We can also
directly infer that the rays $\theta=0,\pi/5$ are exact geodesics. 

The
behavior of the radially outgoing geodesics can be qualitatively
deduced as follows. The geodesic equation implies
\begin{equation}\label{eq:geodesiceqn_phi}
  \ddot{\theta}=-\Gamma^\theta_{rr}\,\dot{r}^2-2\Gamma^{\theta}_{r\theta}\,\dot{r}\,\dot{\theta}=\frac12 G^{\theta\theta}G_{rr,\theta}\,\dot{r}^2-G^{\theta\theta}G_{\theta\theta,r}\,\dot{r}\,\dot{\theta}\;.
\end{equation}
For small $r$ the metric is approximately constant so that, with the
initial condition $\dot{\theta}(0)=0$, $\theta$ stays approximately
constant while $r$ increases. When $r\simeq 1$, the $\theta$ gradient
of the metric becomes important and equation
(\ref{eq:geodesiceqn_phi}) implies that the initially straight line is
attracted towards the region of increasing $G_{rr}$, i.e. the conifold
point. Once the geodesic passes into the region $r\gg 1$, the metric
becomes approximately flat in the $\theta$ direction but the $r$
gradient becomes important. Since now
$\dot{\theta}<0,\dot{r}>0,G_{\theta\theta,r}<0$, the minus sign in the
second term of (\ref{eq:geodesiceqn_phi}) implies that the geodesic
continues towards decreasing $\theta$ until it hits the
$\text{Re}(\psi)=0$ axis. It then re-enters the moduli space from the
$\theta=2 \pi/5$ ray.

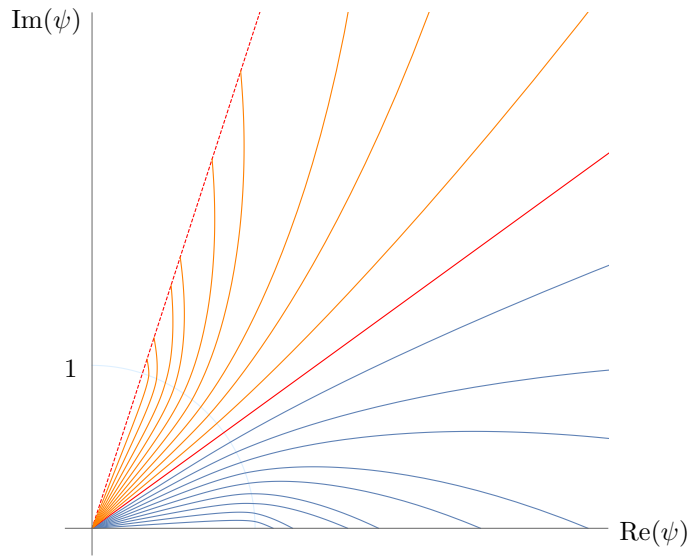
\begin{figure}[ht!]
\centering
\vspace{2ex}
\includegraphics[width=0.5\textwidth]{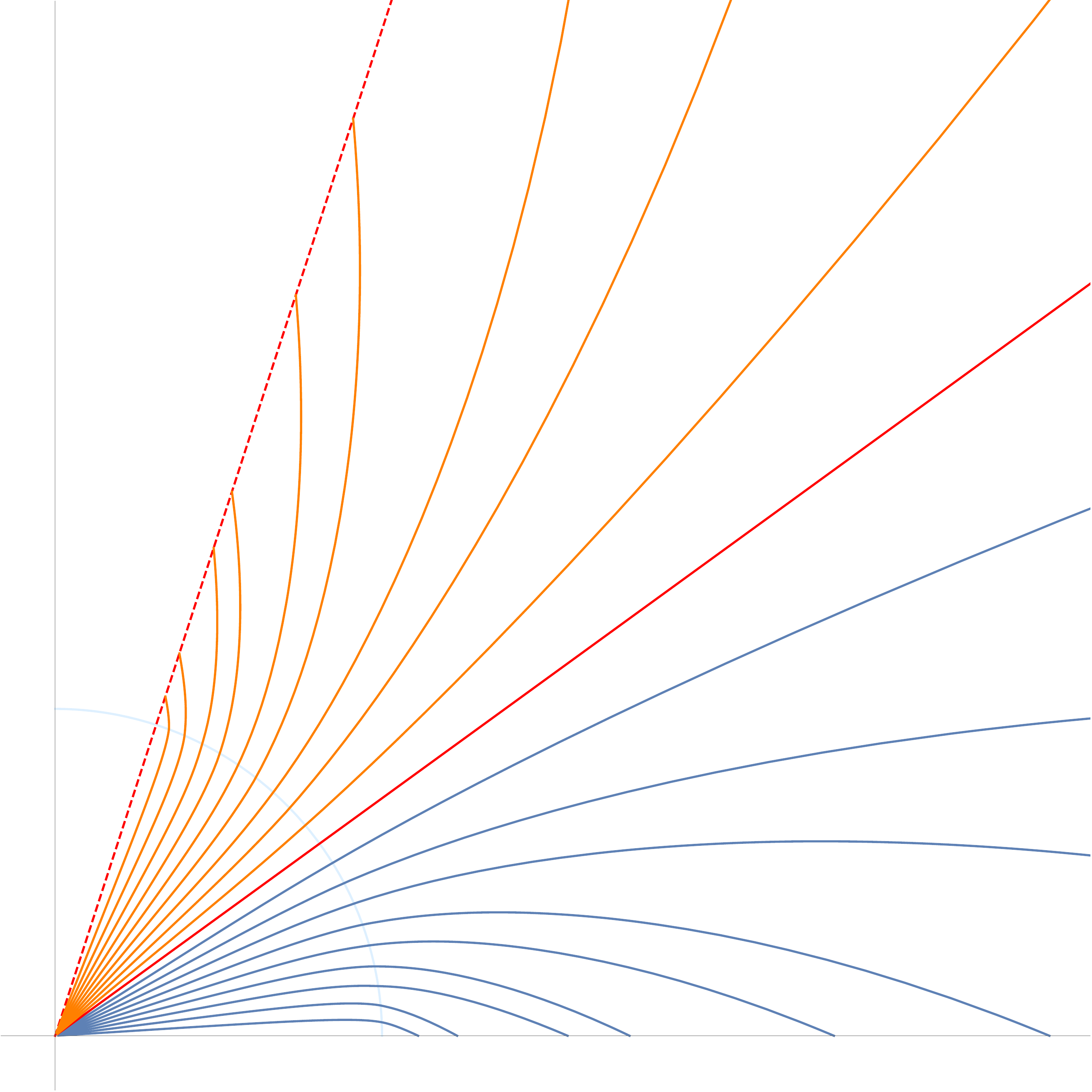}
\begin{picture}(0,0)
  \put(0,6){\footnotesize$\text{Re}(\psi)$}
  \put(-230,200){\footnotesize$\text{Im}(\psi)$}
  \put(-210,68){\footnotesize$1$}
\end{picture}
\caption{Geodesics for the initial data $(r,\dot{r},\theta,\dot{\theta})=(0,1,i\cdot\pi/50,0)$, for $i=1,\dots,10$. The orange geodesics are the $\mathbb{Z}_2$ images.}
\label{fig:geodesics_rep}
\end{figure}

Since we are interested in the geodesic distance and hence shortest geodesics between points, we can stop integrating the geodesic when it hits the axis. This is because of the symmetry properties of the metric we will always find a shorter geodesic connecting the relevant points in the ``upper'' half-cone of the moduli space. Figure \ref{fig:geodesics_rep} shows a few representative geodesics. 

We can see directly from the discussion of the radius of the
Landau-Ginzburg region that geodesics passing too close to the
conifold point will not be interesting, since they hit the
$\text{Re}(\psi)=0$ axis shortly after crossing the phase
boundary. This means that they will not have total length much bigger
than the distance between the Landau-Ginzburg and conifold point
$\Delta \Theta\simeq0.45$. In fact for the geodesic with
$\theta=\pi/50$ we find the numerical result $\Delta \Theta\approx
0.5$. 


In order to test the RSDC in this moduli space we will consider another set of geodesics with a slightly finer scanning of the angle, $\theta=\pi/5-i\pi/60$, for $i=0,\dots,11$. Disregarding the geodesic $\theta=\pi/5$, which continues straight to the large volume point, the longest geodesic in this family has $\theta=\tfrac{11}{60}\pi$, which hits the axis at $\text{Re}(\psi)\approx 110$, after traveling for a total proper distance of $\Delta \Theta\approx 1.53$.

By performing an asymptotic expansion of the metric in the large
volume phase, we realize that it has the asymptotic form
\begin{equation}
  g_{rr}(r)\approx\frac{3}{4\left(r\log r\right)^2}\;
\end{equation}
with $\lambda=2/\sqrt{3}$.
From this one can see that the geodesic distance from the Landau-Ginzburg
point asymptotically grows as the double 
logarithm
\begin{equation}
\label{eq:doublelog}
    \Theta(r)\simeq\frac{1}{\lambda}\log\big(\log (r)\big)\;.
\end{equation}

After identifying a family of relevant geodesics, the next step is to
identify a tower of states, whose mass we expect to display the
exponential behavior predicted by the Refined Swampland Distance
Conjecture. Through mirror symmetry, the complex structure moduli
space of the mirror quintic is mapped to the \kahler{} moduli space of
the quintic. The single complex structure modulus $\psi$ is mapped to
the overall volume modulus $t=\int B+i\int J$ of the quintic. Working
in the \kahler{} moduli space of the quintic, we have as a natural
candidate the Kaluza-Klein tower associated to the overall volume. 
As we have computed in section 2.2, the associated mass scale is 
then
\begin{equation}
  M_\text{KK}(t)\sim\frac{1}{(\text{Im}(t))^2}\;.
\end{equation}
In order to express this in terms of the proper field distance, one
needs the mirror map $t=t(\psi)$.

\begin{figure}[ht!]
  \centering
  \includegraphics[width=0.3\textwidth]{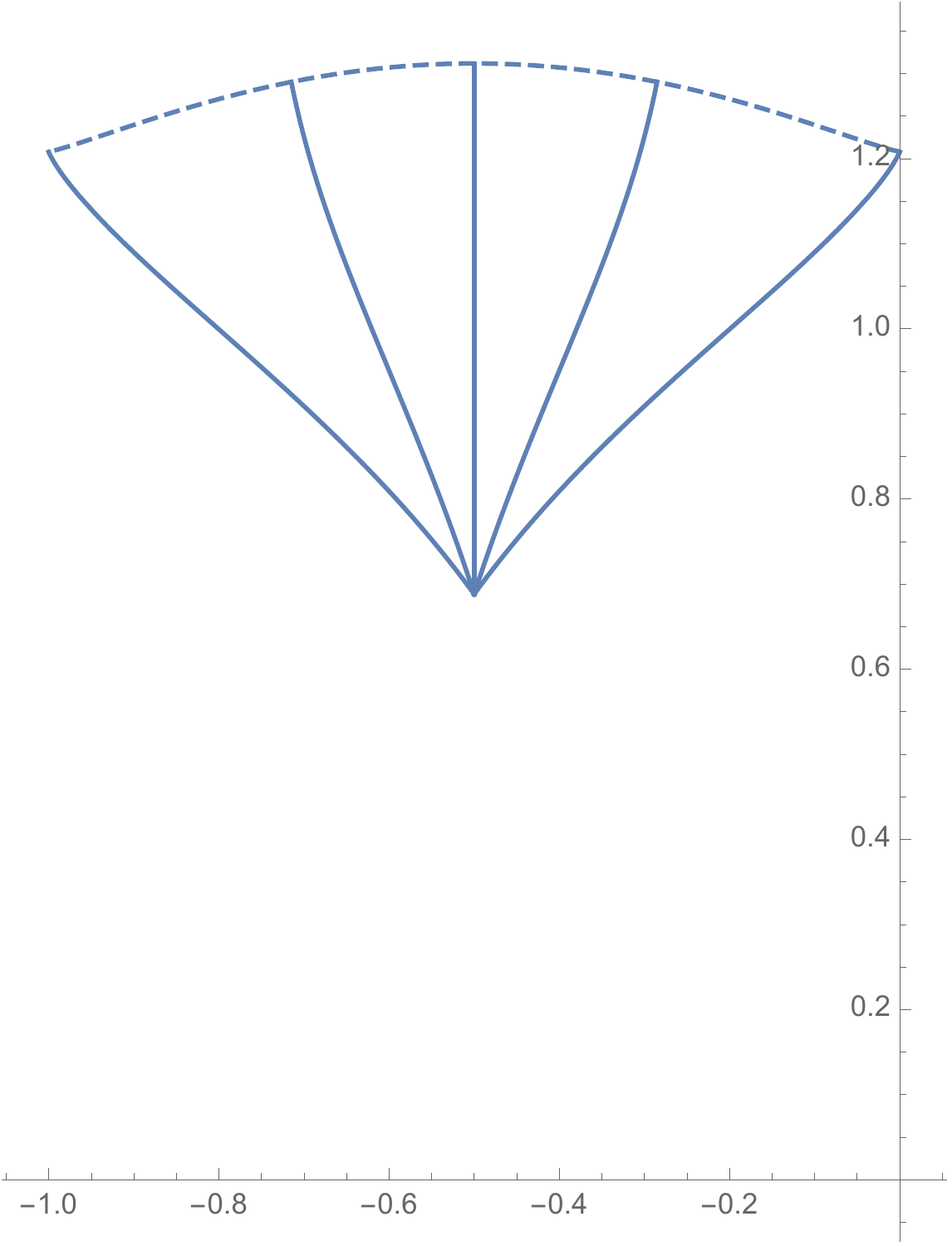}
  \begin{picture}(0,0)
  \put(-155,5){\footnotesize$\text{Re}(t)$}
  \put(-7,154){\footnotesize$\text{Im}(t)$}
\end{picture}
  \caption{The fundamental domain of the Landau-Ginzburg region in the
    coordinate $t_{\rm LG}= X_1/X_0$ for various ${\rm Arg} (\psi)$. }
  \label{fig:wgctest2}
\end{figure}
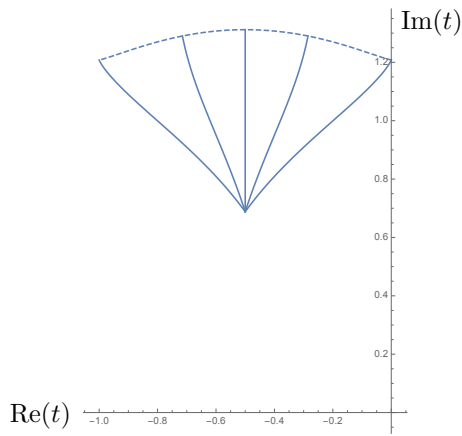

Let us now be more precise about the computation of the mirror map,
which  is defined by
\eq{
  t(\psi) = \int B + i \int J = \frac{X^1(\psi)}{X^0(\psi)} \, .
  }
Using the explicit form of the periods $X^0$ and $X^1$, we find for the mirror map near the Landau-Ginzburg point
\eq{
  t_\text{LG}(\psi) = - \frac{1}{2} + 0.688 \, i + \left( 0.279 + 0.384 \, i \right) \, \psi + \dots
  \qquad {\rm for} \qquad
  |\psi| < 1 \, .
  }
Figure \ref{fig:wgctest2} shows $t_{\rm LG}$ for different ${\rm Arg} (\psi)$, which agrees with the more qualitative plot in figure \ref{fig:toymodulispace}.

\noindent
Similarly, one can compute the mirror map near the conifold
\eq{
  t_C(\psi) = {\textstyle \frac{0.540 \, i + 0.754 \, i \, \psi}{1.071 - (0.124 + 0.890 \, i) \, (\psi - 1) + 0.283 \, (\psi - 1) \, \log \left( \frac{1}{\psi^5} - 1 \right)} \, +\, \dots}
  \quad {\rm for} \quad
  |\psi| \approx 1 \,
  }
and in the large volume regime
\eq{
\label{eq:mirrormap_quintic_LCS}
  t_M(\psi) = \frac{5}{2 \pi} \, \log (5) + \frac{5}{2 \pi} i \, \log \psi + \dots
  \qquad {\rm for} \qquad
  |\psi| > 1 \, ,
  }
where we used results from section \ref{mirror}.

At this point we can already see the exponential relation between
$M_\text{KK}$ and $\Theta$. Combining the doubly logarithmic behavior
of $\Theta(r)$ \eqref{eq:doublelog} with the logarithmic one of
$t_M(\psi)$ \eqref{eq:mirrormap_quintic_LCS}, we have\footnote{The
  factor of two in the exponential should not be taken too seriously,
  as  we just gave a rough estimate for  the KK mass scale, $M_{\rm
    KK}\sim {M_s/{\rm Vol}^{1\over 6}}$. We note that in \cite{Grimm:2018ohb}
    another proposal for the infinite tower of exponentially light states has been
    given. There, these were BPS wrapped D-branes,
    i.e. non-perturbative states.} 
\begin{equation}
    \Theta\simeq\frac{1}{\lambda}\log\left(\text{Im}(t_M)\right)\qquad\Rightarrow\qquad M_\text{KK}\simeq\frac{1}{(\text{Im}(t_M))^2}\simeq e^{-2\lambda\,\Theta}\;.
\end{equation}
This is precisely the behavior predicted by the RSDC.

Since we know  both the proper distance $\Theta$ and  the value of the
complexified K\"ahler modulus $t=\int B+i\int J$ along the geodesics,
we can plot the logarithm of $\text{Im}(t)$ against $\Theta$.

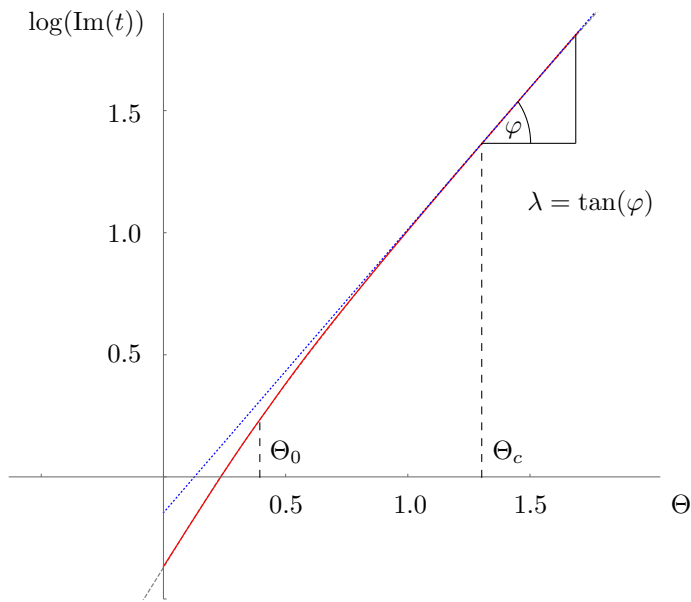
\begin{figure}[ht!]
  \centering
  \vspace{2ex}
  \begin{tikzpicture}(0,0)
  \node[inner sep=0pt] at (0,0)
{\includegraphics[width=0.6\textwidth]{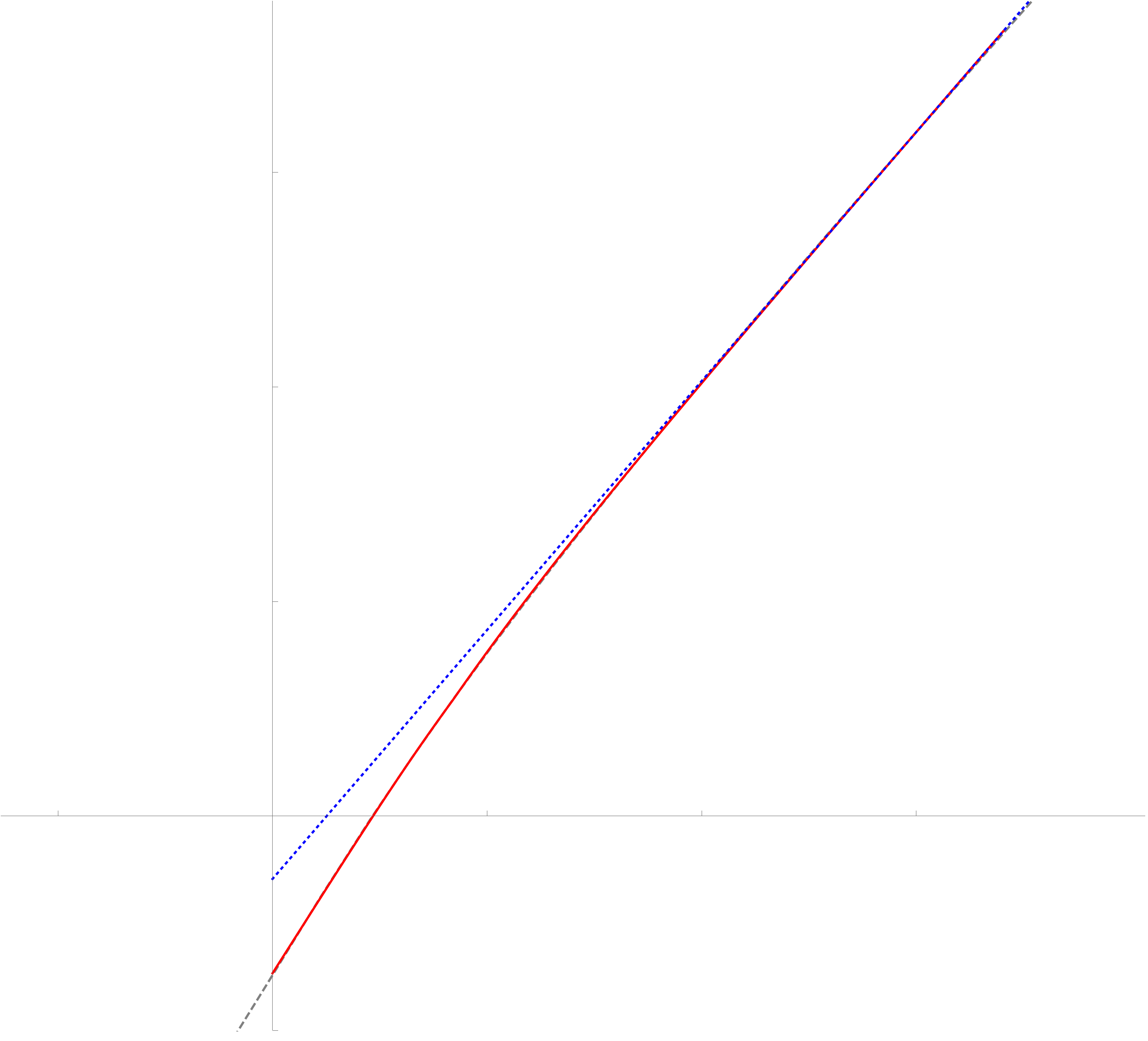}};
    \node at (4.6,-2.6) {\footnotesize$\Theta$};
    \node at (-0.65,-2.6) {\footnotesize$0.5$};
    \node at (1,-2.6) {\footnotesize$1.0$};
    \node at (2.6,-2.6) {\footnotesize$1.5$};

    \node at (-3.3,3.8) {\footnotesize$\log(\text{Im}(t))$};
    \node at (-2.8,-0.6) {\footnotesize$0.5$};
    \node at (-2.8,1) {\footnotesize$1.0$};
    \node at (-2.8,2.6) {\footnotesize$1.5$};

    \draw[dashed] (-1,-2.25) -> (-1,-1.5);
    \node at (-0.65,-1.9) {\footnotesize$\Theta_0$};

    \draw[dashed] (1.95,-2.25) -> (1.95,2.2);
    \node at (2.3,-1.9) {\footnotesize$\Theta_c$};

    \draw (1.95,2.2) -> (3.2,2.2);
    \draw (3.2,2.2) -> (3.2,3.65);

    \draw (2.6,2.2) arc (0:34:1);
    \node at (2.37,2.38) {\footnotesize$\varphi$};

    \node at (3.4,1.4) {\footnotesize$\lambda=\tan(\varphi)$};
  \end{tikzpicture}
\caption{The logarithm of $\text{Im}(t)$ against $\Theta$. }
\label{fig:SC_test_1}
\end{figure}

The Refined Swampland Distance Conjecture predicts a linear behavior after
some critical distance $\Theta_0\lesssim 1$. Figure \ref{fig:SC_test_1} shows that this is precisely the case. The expected linear behavior is reached for
  $\Theta=\Theta_0\lesssim \mathcal{O}(1)$. The depicted red graph
  corresponds to the central geodesic with initial angle
  $\theta=\pi/5$. We find that this is the geodesic for which
  $\Theta_0$ is the largest. The dotted blue line shows the fit to the
  asymptotic linear behavior, while the dashed grey fit also captures corrections to $\Theta\simeq\frac{1}{\lambda}\log(\text{Im}(t))$ up to order $1/\text{Im}(t)^3$.

The parameters $\lambda$ and $\Theta_0$ are determined as follows. The
value of $\Theta_0$ is defined to be the value of the proper distance
along the geodesics from the Landau-Ginzburg point to the phase
boundary at $|\psi|=1$. We also define $t_0$ to be the value of the \kahler{} modulus at the phase boundary, $t_0(\theta)\equiv t(r=1,\theta)$. To determine $\lambda$, we perform a fit of
the asymptotic behavior of the proper distance as a function of the
\kahler{} modulus according to the leading order terms
\begin{equation}
  \label{eq:fitted_model}
  \Theta(t)\simeq\frac{1}{\lambda}\log(t)+\alpha_0+\frac{\alpha_1}{t^3}\;.
\end{equation}
The angular distribution of the fit-parameters as well as $\Theta_\lambda, \Theta_0$ and $\Theta_c$ is shown in Table \ref{tab:critdistslope}. 
\begin{table}[ht!]
  \centering
  \begin{tabular}{c c c c c c}
  $\theta_\text{init} \cdot 60/\pi$ & $\alpha_0$ & $\alpha_1$ & $\lambda^{-1}$ & $\Theta_0$ & $\Theta_c$\\
  \hline\hline
3 & 0.1315 & 0.2043 & 0.9605 & 0.4262 &
1.3866 \\ 4 & 0.1127 & 0.2099 & 0.9865 & 0.4261 & 1.4125 \\ 5 &
0.0998 & 0.2213 & 0.9780 & 0.4260 & 1.4040 \\ 6 & 0.0955 & 0.2294 &
0.9567 & 0.4259 & 1.3827 \\ 7 & 0.0818 & 0.2475 & 0.9611 & 0.4259 &
1.3869 \\ 8 & 0.0877 & 0.2592 & 0.9275 & 0.4258 & 1.3533 \\ 9 &
0.0808 & 0.2825 & 0.9253 & 0.4257 & 1.3510 \\ 10 & 0.0929 & 0.3093 &
0.8969 & 0.4257 & 1.3226 \\ 11 & 0.0998 & 0.3497 & 0.8845 & 0.4257 &
1.3102 \\ 12 & 0.1234 & 0.1662 & 0.8657 & 0.4256 & 1.2914\\
\end{tabular}
  \caption{Values of the fit-parameters $\alpha_0,\alpha_1,\lambda^{-1}$, critical distance $\Theta_0$ and combined critical distance $\Theta_c$ for the family of geodesics with initial angles $\theta_\text{init}=\pi/5-i\pi/60$, for $i=2,\dots,11$. We see that $\Theta_0$ is approximately constant for the quintic. The total critical distance varies mostly because of the angular dependence of $\lambda$.}
  \label{tab:critdistslope}
\end{table}

\noindent
We have excluded the geodesics with $i=10,11$ from the analysis because the fact that they hit the $\text{Re}(\psi)$ axis almost immediately after the phase transition did not allow for a good fit.
As a result we find that the critical distance is always of order one and satisfies the bound
\begin{equation}
  \Theta_c\leq1.387\;.
\end{equation}
The amount of variation of $\Theta_0$ is minimal, while $\lambda$ deviates noticably between the different geodesics. 

As a cross-check of our method, for the central geodesic with
$\theta=\pi/5$ we can compare the result for $\lambda$ and $\alpha_1$
with the analytic result (\ref{eq:Theta_expansion}). The value for
$\Theta_\lambda\simeq 0.866$ agrees perfectly with the expected value
of $\Theta_\lambda=\sqrt{3/4}$. For $\alpha_1$, we insert the Euler
characteristic of the quintic ($\chi=-200$) into the formula
(\ref{eq:Theta_expansion}) and obtain $\alpha_1\simeq0.168$, which
deviates from the fit value only by one percent. This can be explained
by the fact that we neglected the higher order $1/t$ corrections in
our fit.

We observe that those geodesics passing closer to the conifold and
thus deviate the most from being straight lines in the $\psi$-plane
have the largest $\Theta_\lambda$. This is because of the fact that
while both the real and imaginary part of the complexified \kahler{}
modulus contribute to the proper distance, only the imaginary part
controls a mass scale. The imaginary part of the \kahler{} modulus is
(asymptotically) mapped to the absolute value $|\psi|$ through the
mirror map, while the real part is mapped to
$\text{Arg}(\psi)$. Curving into the ``axionic'' direction in moduli
space thus decreases the rate of the exponential mass fall-off. The
fact that we still find $\Theta_\lambda<M_\text{pl}$ for all geodesics
is a non-trivial test of the RSDC. It seems to be not unrelated to the
statement that periodic directions of the moduli space should have a
sub-Planckian  periodicity.

For the average over all sampled geodesics of $t_0$, the value of the \kahler{} modulus at the phase transition, we obtain the value $\text{Im}(t_0)\simeq 1.31$.
The average values of the characteristic proper distances of the geodesics turn out to be 
\begin{equation}
    \Theta_0 \simeq 0.4259 \, , \qquad 
  \Theta_{\lambda} \simeq 0.9343 \, , \qquad
  \text{and} \qquad
  \Theta_c \simeq 1.3601 \, ,
\end{equation}
in perfect agreement with the RSDC.

\subsection{Hypersurfaces: \texorpdfstring{$\mathbb P^4_{11112} [6]$}{P11112}, \texorpdfstring{$\mathbb P^4_{11114} [8]$}{P11114} and \texorpdfstring{$\mathbb P^4_{11125} [10]$}{P11125}}

As pointed out in section \ref{sec_1paraquinticmodels} there are only four Calabi-Yau manifolds with $h^{11} = 1$ defined by a single polynomial constraint in a weighted projective space, 
namely $\mathbb P^4_{11111} [5]$, $\mathbb P^4_{11112} [6]$, $\mathbb P^4_{11114} [8]$ and $\mathbb P^4_{11125} [10]$.
After having discussed the quintic rather detailed in the last section, we are now turning towards the other three manifolds.
These are per construction very similar to the quintic so that the results we are going to compute will be qualitatively equivalent and only differ by slightly altered numerical values.

In particular the structure of the moduli space agrees exactly with the one of the quintic.
That is, there exists a conifold singularity at $\psi = 1$ with $\psi$ being the coordinate of the moduli space.
The regime $|\psi| > 1$ is covered by the large volume chart, whereas the region $|\psi| <1$ corresponds to the (orbifolded) Landau-Ginzburg phase.
There is again a residual $\mathbb Z_d$ symmetry depending on the degree $d$ of the analyzed projective space.

The behavior of the metric for $\mathbb P^4_{11112} [6]$ is
qualitatively the same as for the quintic. For these two CY threefolds the metric
is approximately flat around the origin $G_{\psi\overline\psi}\simeq
\text{const}$, whereas  the asymptotic behavior of the metric is like
$G_{\psi\overline\psi}\simeq 1/(|\psi|^2\log|\psi|^2)$. For the  other two
threefolds $\mathbb
P^4_{11114} [8]$ and $\mathbb P^4_{11125} [10]$ the metrics differ slightly around
the origin, in that they have $G_{\psi\overline\psi}\simeq
\text{const}\cdot |\psi|^2$, but show the same asymptotic 
behavior.

In the following we shall determine average values for $\lambda$, $\Theta_0$ and $\Theta_c$ based on characteristic geodesic trajectories for each of the three moduli spaces.
Analogously to figure \ref{fig:geodesics_rep} we will investigate geodesics starting close to $\psi = r \exp (i \theta)= 0$ and moving outwards in radial direction.
All of them will transit from the Landau-Ginzburg phase into the large volume regime.
More precisely, we analyze 12 geodesics $\gamma_j$, $j = 0, \dots, 11$ with start points $(r_i, \theta_i) = \left( r_i ,\frac{\pi}{d} \, \left(1 - \frac{j}{12} \right) \right)$ and choose an initial velocity $(r'_i, \theta'_i) = \left( 1 ,0 \right)$.
Note that $r_i$ has to be adjusted model by model since numerical fluctuations disturb the metric near the origin.
Apparently $\gamma_0$ corresponds to the angle bisector and $\gamma_{11}$ comes close to the conifold singularity.

Before presenting the results for each model, let us point out that we
determined the K\"ahler metric from the partition function of the corresponding GLSM, as described in section \ref{sec_KahlerMetric}.
A formula to calculate the mirror map of one-parameter models was given in section \ref{mirror}.
The evaluation of the geodesics follows the procedure described for the quintic.
Hence, in order to determine the slope parameter $\lambda$, we will again fit the ansatz
\begin{equation}
\label{log_ansatz}
  \Theta(t)\simeq\frac{1}{\lambda}\log(t)+\alpha_0 + \frac{\alpha_1}{t^3}\;.
\end{equation}
Now, let us present our results case by case.
\paragraph{$\mathbf{\mathbb P^4_{11112}[6]}$}

We take the initial value $r_i = 0.01$ and add to every geodesic an initial length of 
\eq{
	\Theta_i = \int_0^{r_i} dr \, \sqrt{G_{\psi \ov \psi}} \simeq 0.0039\, .
	}
Computing the proper distance for each geodesic and fitting the ansatz \eqref{log_ansatz}, we obtain the values listed in table \ref{tab_P6}.
Notice that our fits led to an average threshold ${\rm Im}\, t_0
\simeq 1.49$ calculated as the transition point to the large volume
phase.

The average proper distance collected in the non-geometric phase is $\Theta_0\simeq 0.40$. For the distance between the Landau-Ginzburg and conifold points we find the value $\Delta\Theta\simeq 0.41$.
\begin{table}[ht!]
  \centering
  \begin{tabular}{c c c c c c}
  $\theta_\text{init} \cdot 72/\pi$ & $\alpha_0$ & $\alpha_1$ & $\lambda^{-1}$ & $\Theta_0$ & $\Theta_c$\\
  \hline\hline
1 & -0.081 & 0.414 & 0.957 & 0.405 & 1.362 \\ 2 & -0.058 & 0.347 &
0.934 & 0.404 & 1.338 \\ 3 & -0.057 & 0.329 & 0.929 & 0.402 & 1.331
\\ 4 & -0.052 & 0.319 & 0.911 & 0.400 & 1.311 \\ 5 & -0.056 & 0.327 &
0.906 & 0.399 & 1.305 \\ 6 & -0.067 & 0.347 & 0.914 & 0.398 & 1.312
\\ 7 & -0.074 & 0.368 & 0.914 & 0.397 & 1.311 \\ 8 & -0.068 & 0.389 &
0.896 & 0.396 & 1.292 \\ 9 & -0.060 & 0.423 & 0.882 & 0.396 & 1.278 \\
10 & -0.060 & 0.459 & 0.881 & 0.395 & 1.276 \\ 11 & -0.060 & 0.502 &
0.879 & 0.395 & 1.274 \\ 12 & -0.042 & 0.283 & 0.866 & 0.395 & 1.260\\
\end{tabular}
  \caption{Fitting the ansatz \eqref{log_ansatz} to a plot of the proper length of geodesics $\gamma_j$ depending on the mirror map coordinate $t$.
  The table lists all fitting parameters including the critical distance $\Theta_c = \Theta_0 + \lambda^{-1}$ for the model $\mathbb{P}^4_{11112}[6]$.}
  \label{tab_P6}
\end{table}
These results are not in contradiction with the RSDC  as all values are $O (1)$
\eq{
	\Theta_0 \simeq 0.3984 \, , \qquad 
	\Theta_{\lambda} \simeq 0.9056 \, , \qquad
	\text{and} \qquad
	\Theta_c \simeq 1.3041 \, .
	}

\newpage
\paragraph{$\mathbf{\mathbb P^4_{11114}[8]}$}
For this moduli space we have computed the geodesics starting from $r_i = 0.1$ and as a consequence had to add $\Theta_i \simeq 0.0023$ to the proper lengths of the geodesics.
The critical value of the K\"ahler modulus at the phase transition is on average
${\rm Im}\, t_0 \simeq 1.77$.

The distance between the Landau-Ginzburg and conifold points is $\Delta\Theta\simeq0.23$.
All results can be found in table \ref{tab_P8}.
\begin{table}[ht!]
  \centering
  \begin{tabular}{c c c c c c}
  $\theta_\text{init} \cdot 96/\pi$ & $\alpha_0$ & $\alpha_1$ & $\lambda^{-1}$ & $\Theta_0$ & $\Theta_c$\\
  \hline\hline
1 & -0.426 & 0.747 & 0.933 & 0.225 & 1.158 \\ 2 & -0.409 & 0.673 &
0.920 & 0.224 & 1.144 \\ 3 & -0.399 & 0.629 & 0.908 & 0.223 & 1.130 \\
4 & -0.393 & 0.619 & 0.893 & 0.221 & 1.115 \\ 5 & -0.398 & 0.640 &
0.892 & 0.221 & 1.113 \\ 6 & -0.409 & 0.668 & 0.900 & 0.220 & 1.120 \\ 7
& -0.417 & 0.700 & 0.904 & 0.219 & 1.123 \\ 8 & -0.414 & 0.737 & 0.893
& 0.218 & 1.112 \\ 9 & -0.409 & 0.782 & 0.883 & 0.218 & 1.101 \\ 10 &
-0.410 & 0.827 & 0.882 & 0.218 & 1.100 \\ 11 & -0.408 & 0.885 & 0.878 &
0.218 & 1.096 \\ 12 & -0.388 & 0.613 & 0.865 & 0.217 & 1.082\\
\end{tabular}
  \caption{Fitting the ansatz \eqref{log_ansatz} to a plot of the proper length of geodesics $\gamma_j$ depending on the mirror map coordinate $t$.
  The table lists all fitting parameters including the critical distance $\Theta_c = \Theta_0 + \lambda^{-1}$ for the model $\mathbb P^4_{11114}[8]$.}
  \label{tab_P8}
\end{table}

\noindent
The RSDC is thus in agreement with the analyzed  geodesics in the moduli space of $\mathbb P^4_{11114}[8]$. In average we end up with the following values
\eq{
	\Theta_0 \simeq 0.2201 \, , \qquad
	\Theta_{\lambda} \simeq 0.8961 \, , \qquad
	\text{and} \qquad
	\Theta_c \simeq 1.1162 \, .
	}
\paragraph{$\mathbf{\mathbb P^4_{11125}[10]}$}
Here we have assumed $r_i = 0.14$, leading to $\Theta_i \simeq 0.0040$ which needs to be added to the proper lengths. For this model the proper distance between the Landau-Ginzburg and conifold points has the smallest value, $\Delta\Theta\simeq 0.21$.
Fitting the ansatz \eqref{log_ansatz} gives the values summarized in table \ref{tab_P10}.
Note that the critical K\"ahler modulus is ${\rm Im}\, t_0 \simeq 2.17$.
\begin{table}[ht!]
  \centering
  \begin{tabular}{c c c c c c}
  $\theta_\text{init} \cdot 120/\pi$ & $\alpha_0$ & $\alpha_1$ & $\lambda^{-1}$ & $\Theta_0$ & $\Theta_c$\\
  \hline\hline
1 & -0.655 & 1.482 & 0.949 & 0.213 & 1.162 \\ 2 & -0.616 & 1.289 &
0.919 & 0.212 & 1.131 \\ 3 & -0.593 & 1.179 & 0.899 & 0.210 & 1.109 \\
4 & -0.583 & 1.151 & 0.885 & 0.210 & 1.094 \\ 5 & -0.587 & 1.182 &
0.885 & 0.209 & 1.094 \\ 6 & -0.597 & 1.219 & 0.891 & 0.208 & 1.099
\\ 7 & -0.602 & 1.263 & 0.892 & 0.208 & 1.100 \\ 8 & -0.598 & 1.322 &
0.884 & 0.207 & 1.091 \\ 9 & -0.593 & 1.392 & 0.876 & 0.207 & 1.083
\\ 10 & -0.594 & 1.449 & 0.875 & 0.207 & 1.082 \\ 11 & -0.592 & 1.522
& 0.873 & 0.207 & 1.080 \\ 12 & -0.578 & 1.195 & 0.865 & 0.206 & 1.071
\\
\end{tabular}
  \caption{Fitting the ansatz \eqref{log_ansatz} to a plot of the proper length of geodesics $\gamma_j$ depending on the mirror map coordinate $t$.
  The table lists all fitting parameters including the critical distance $\Theta_c = \Theta_0 + \lambda^{-1}$ for the model $\mathbb P^4_{11125}[10]$.}
  \label{tab_P10}
\end{table}
Also in this model the average values agree with the RSDC
\eq{
	\Theta_0 \simeq 0.2086 \, , \qquad
	\Theta_{\lambda} \simeq 0.8911 \, , \qquad
	\text{and} \qquad
	\Theta_c \simeq 1.0997 \, .
	}


\section{RSDC for CY manifolds with $h^{11}=2$}
\label{sec:two_parameter}

In this section we will extend our analysis to CY threefolds with two
K\"ahler moduli. 
As a consequence,  in addition to the LG and large volume phase (LV) we obtain also two \emph{hybrid} regimes.
As a new feature, these hybrid phases have solely one complex parameter $\psi$ or $\phi$
bounded, whereas the other one is able to reach infinite
distances. 
Our goals are to determine the precise proper length of finite
directions in the different phases of the moduli space, extract the logarithmic behavior of proper distances along infinite directions and last but not least compute the critical distance $\Theta_\lambda$ for certain accessible regimes.

More precisely, if we consider an infinite direction, we will always encounter a $\log$-structure in the mirror map.
Thus the corresponding K\"ahler modulus may grow only logarithmically past some critical distance, which is expected to cause a Kaluza-Klein state becoming exponentially light.
Therefore, the appearance of the logarithm is in agreement with the Swampland Distance Conjecture \cite{Ooguri:2006in}, where the state in question is given by this Kaluza-Klein mode.

Recall that the RSDC makes a stronger statement by predicting the invalidity of the effective theory after having traversed at most $O (1) \, M_{\rm pl}$ in moduli space.
Hence, the $\log$-term behavior in the proper distance has to occur
roughly at proper distance one and finite directions in the moduli
space have to have proper length  less than one.
In this section, we will explicitly confirm the latter and
asymptotically approach the $\log$ scaling.

Similar to the one-parameter section we shall discuss one example,
that is $\mathbb P^4_{11222} [8]$, quite extensively and briefly list
the results for other two parameter CY threefolds.

\subsection{An illustrative example: \texorpdfstring{$\mathbb
    P^4_{11222} [8]$}{P11222}}

Let us at first focus on the weighted projective space $\mathbb P^4_{11222} [8]$ which was studied in great detail in the literature, see for instance \cite{Candelas:1993dm, Aspinwall:1994ay,Berglund:1993ax}.
Recall the construction of the K\"ahler metric on the mirror dual  in section
\ref{sec_KahlerMetric}, where we had already introduced basic facts
about the geometry of $\mathbb P^4_{11222} [8]$.

In terms of the homogeneous coordinates $[x_1, x_2, x_3, x_4, x_5]$, 
consider the Calabi-Yau hypersurface
\eq{
\label{hyperconstraint11222}
  P = x_1^8 + x_2^8 + x_3^4 + x_4^4 +x_5^4
  - 8 \psi\, x_1 x_2 x_3 x_4 x_5 - 2 \phi\, x_1^4 x_2^4 \, ,
  }
with two complex parameters $(\psi,\phi)$ corresponding to complex structure moduli.
The smooth family of threefolds given by the quotient $\{ P=0 \}/(\mathbb Z_4)^3$ identifies the mirror of $\mathbb P^4_{11222} [8]$.
Following \cite{Candelas:1993dm} it is convenient to mod the hypersurfaces $\{ P =0\}$ by an even larger group, which requires to mod the parameter space $\big\{ (\psi,\phi) \big\}$ by a $\mathbb Z_8$.
Its generators act according to
\eq{
  (\psi,\phi) \mapsto (\alpha \psi, - \phi)
  }
with $\alpha$ being the 8th root of unity.

The special points of the moduli space appear when the hypersurface constraint \eqref{hyperconstraint11222} becomes singular, that is, for nontrivial solutions to $P=0$ and $\partial P/ \partial x_i =0$ for all $x_i$.
One finds a conifold singularity at 
\eq{
  (\phi + 8 \psi^4)^2 \, = \, 1
}
as well as another singularity for\footnote{The singular threefolds at
  the locus $\phi =1$ are birationally equivalent to the mirror of the
  one-parameter space $\mathbb P^5_{11111} [2 \ 4]$ (see \cite{Berglund:1993ax}).}
\eq{
\phi^2 \, = \, 1 \, .
}
The two singularities above split the moduli space into four phases: a smooth Calabi-Yau, a Landau-Ginzburg (orbifold) and two hybrid regimes, which we call hybrid $\mathbb P^1$ and hybrid orbifold.

\vspace{0.4cm}
\noindent
{\bf 5.1.1. The phase structure of the K\"ahler moduli space}
\vspace{0.5cm}

\noindent
Let us explain the origin and connection of these four-phases.
We start in the LG phase.
In this phase both complex structures are bounded and 
all homogeneous coordinates $x_i$ are classically vanishing, such that the target space is simply a point.
However, there are massless quantum fluctuations around their vacuum expectation values or in other words, there exists a residual $\mathbb Z_8$ symmetry on the coordinates $x_i$.
Hence effectively we have a Landau-Ginzburg (orbifold) theory living on  $\mathbb C^5 / \mathbb Z_8$.

The singularity of $\mathbb C^5 / \mathbb Z_8$ can be ``blown-up'' by replacing the singularity with an exceptional divisor. Here, it turns out that this divisor has two irreducible components \cite{Aspinwall:1994ay}: $\mathbb C^3 \times \mathbb P^1$ and $\mathbb P^4$.
The four different phases are obtained by separately blowing up to these components of the divisor.
For instance, consider blowing-up to the component $\mathbb C^3 \times \mathbb P^1$.
This leads to a one dimensional target space given by a Landau-Ginzburg (orbifold) bundle over a $\mathbb P^1$ space.
That is why we denote this regime as hybrid $\mathbb P^1$ phase.
Again, some of the $x_i$ were only fixed classically, such that one
still faces a residual $\mathbb Z_4$ symmetry at every point of
$\mathbb P^1$.

In a second step, also blowing-up along the second component $\mathbb P^4$ of the exceptional divisor resolves the $\mathbb Z_4$ singularities in each fiber.
Satisfying in addition a hypersurface constraint 
gives  a $K3$ surface fibered over the $\mathbb P^1$ base.
Thus, one  arrives at a smooth Calabi-Yau manifold.
The full procedure can be summarized as follows

\vspace{0.2cm}
\eq{
\renewcommand{\arraystretch}{1.3}
\begin{tabular}{@{}c|ccccc@{}}
  \toprule
  \multirow{2}{*}{phase} 
  & LG theory
  &\multirow{2}{*}{$\xrightarrow[\mathbb C^3 \times \mathbb P^1]{\text{resolved by}}$}
  & hybrid theory 
  &\multirow{2}{*}{$\xrightarrow[ \mathbb P^4]{\text{resolved by}}$} 
  & smooth\\
  & on $\mathbb C^5 / \mathbb Z_8$ & & on $\mathbb C^4 / \mathbb Z_4$ & & Calabi-Yau\\
  & & & & & \\
  target & \multirow{2}{*}{point} & & \multirow{2}{*}{$\mathbb P^1$} & & $K3$ fibration\\
  space & & & & & over $\mathbb P^1$ base\\
\bottomrule
\end{tabular} 
\nonumber
}

\vspace{0.2cm}
\noindent
Alternatively, one may first blow-up the $\mathbb P^4$ component. 
In this case one  ends up in a  hybrid orbifold phase. This regime is
an  orbifold because its target space is still equipped with $\mathbb Z_2$ quotient singularities.

\begin{figure}[ht!]
\centering
\begin{tikzpicture}[xscale=0.7,yscale=0.7]
\node[inner sep=0pt] at (0,0)
{\includegraphics[scale=0.95]{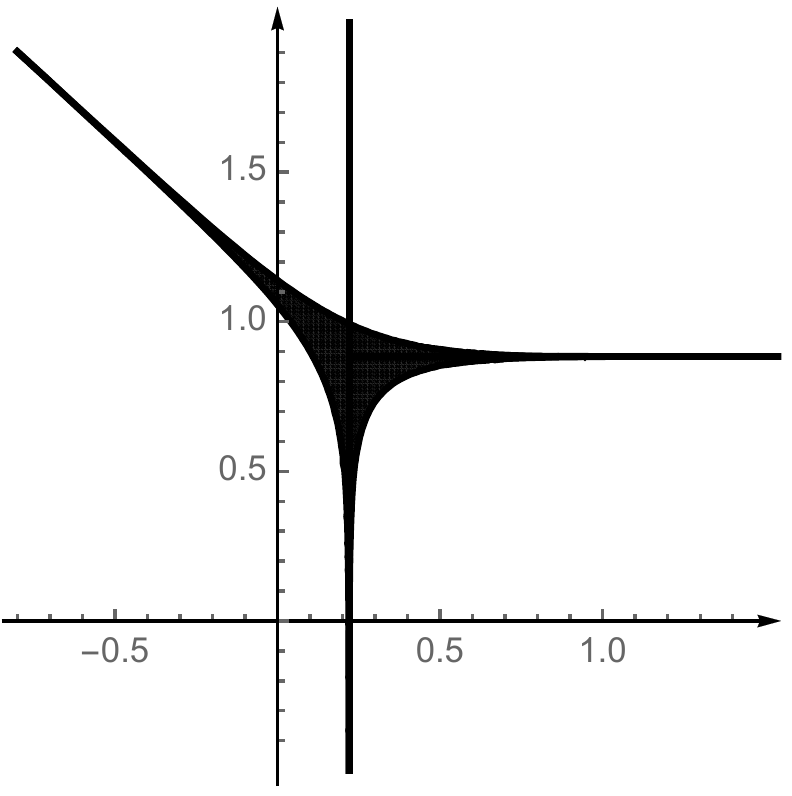}};
\node at (5,-3.6) {$\rho_1$};
\node at (-1.1,5) {$\rho_2$};
\node at (0.3,4.2) {\scriptsize $\phi^2=1$};
\node at (4,0) {\scriptsize $(\phi + 8 \psi^4)^2=1$};
\node[align=center,below] at (3.5,3.5) {Large volume\\  phase};
\node[align=center,below] at (3.5,-1.3) {$\mathbb P^1$ phase};
\node[align=center,below] at (-5,0) {Landau-Ginzburg\\ phase};
\node[align=center,below] at (-3.2,5.4) {orbifold\\ phase};
\end{tikzpicture}
\caption{\small The singular loci divide the K\"ahler moduli space of $\mathbb P^4_{11222} [8]$ in four regimes \cite{Aspinwall:1994ay}.
Note that one is in principle able to transit between the phases if the moduli are equipped with a non-zero imaginary part.}
\label{fig_Aspinwall11222}
\end{figure}
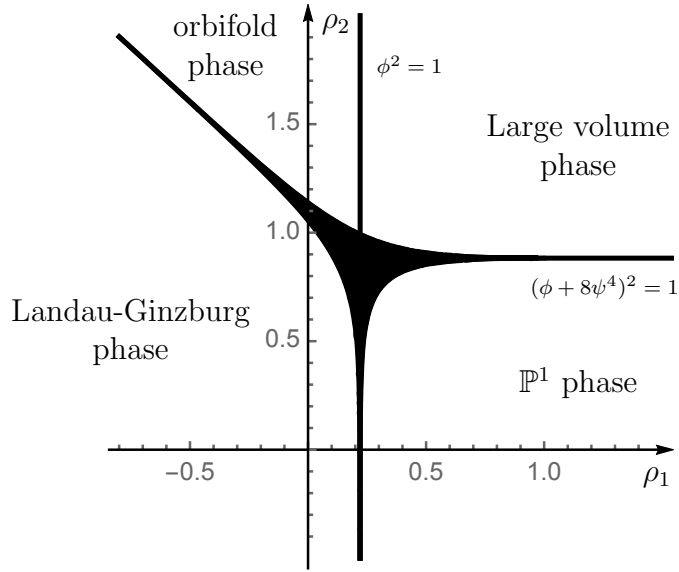

Following \cite{Aspinwall:1994ay} let us introduce the coordinates
\eq{
  \rho_1 \, = \, \frac{1}{2 \pi} \, \log |4 \phi^2| \, , \qquad
  \rho_2 \, = \, \frac{1}{2 \pi} \, \log \left| \frac{2^{11}\, \psi^4}{\phi} \right| \, .
  }
The separation of the four phases of the complex structure  moduli space can be nicely depicted in these coordinates, see figure \ref{fig_Aspinwall11222}.
Moreover, the $\rho_{1,2}$ coordinates make the spatial extension of the conifold singularity $(\phi + 8 \psi^4)^2 =1$ apparent, which is marked by the shaded area in the plot.

\vspace{0.1cm}
\vspace{0.1cm}
An obvious question is whether one can circumvent the singular area in plot \ref{fig_Aspinwall11222} or in other words, transit between any of the four phase of the moduli space.
In principle this is indeed possible by computing the periods in charts covering the  whole moduli space.
However, our periods derived in section \ref{app_metric} do not converge for the entire moduli space.
Instead they cannot be trusted arbitrarily close to the conifold singularity.
Let us explain this point in more detail for the case of $\mathbb P^4_{11222} [8]$.

The singular loci are fixed by $(\phi + 8 \psi^4)^2 =1$ as well as $\phi^2 =1$, hence one can clearly circumvent the singularities simply by giving $\psi$ or $\phi$ a non-zero imaginary part.
So, there are in fact trajectories starting and ending in different phases.
However, in practice the period computation of section \ref{app_metric} converges for $|8 \, \psi^4| \lessgtr |\phi \pm 1|$, which agrees with the conifold constraint only for real moduli.
Obviously, rotating $\psi$ by some phase shift does not affect the
convergence relation,  but adding a phase to $\phi$ does.
\vspace{0.1cm}
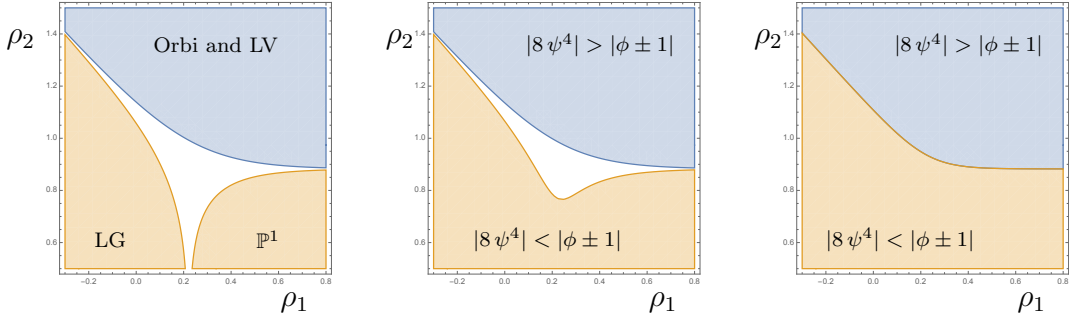
\begin{figure}[ht!]
\centering
\begin{tikzpicture}[xscale=0.7,yscale=0.7]
\node[inner sep=0pt] at (-7,0)
{\includegraphics[scale=0.3]{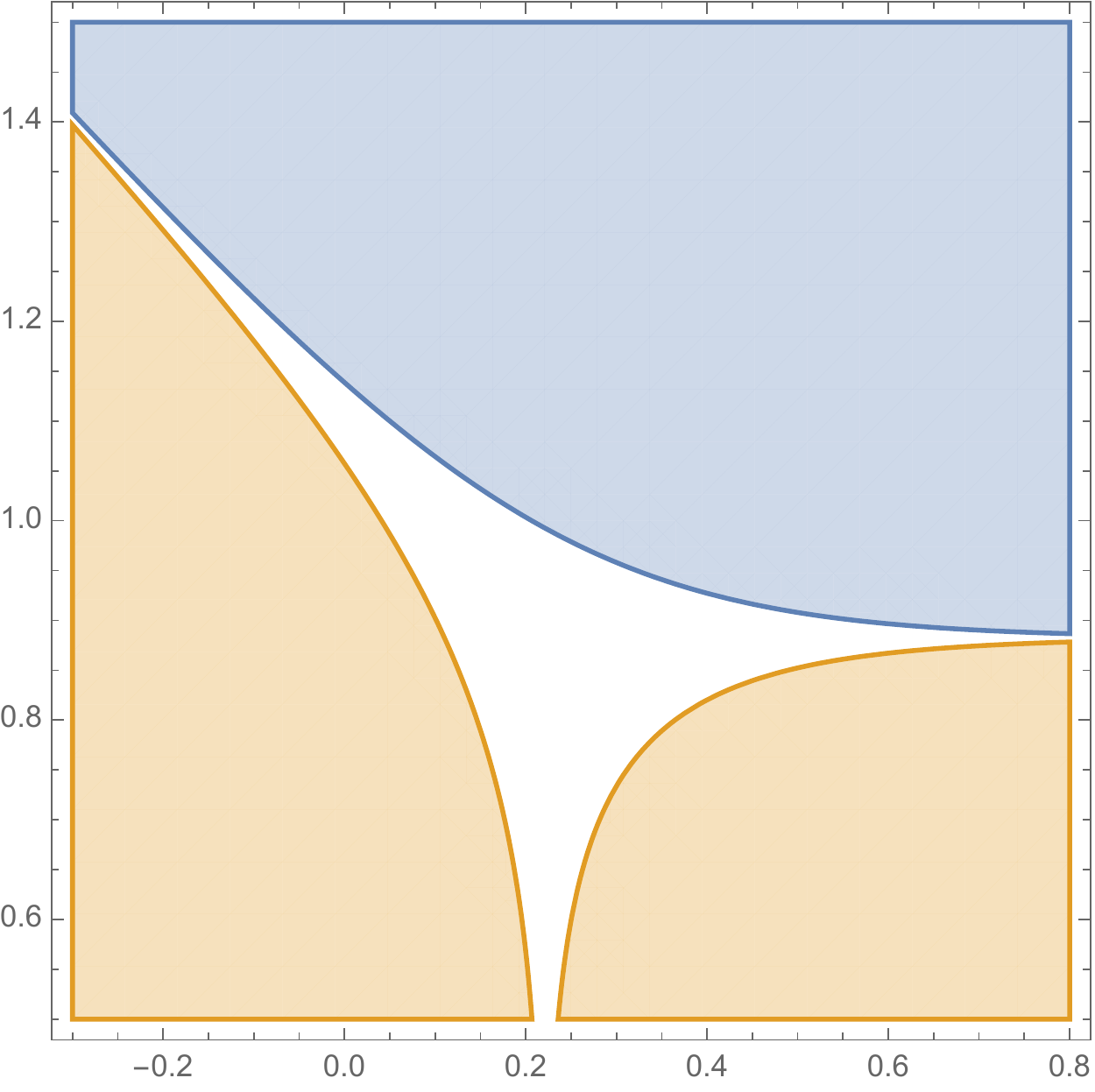}};
\node[inner sep=0pt] at (0,0)
{\includegraphics[scale=0.3]{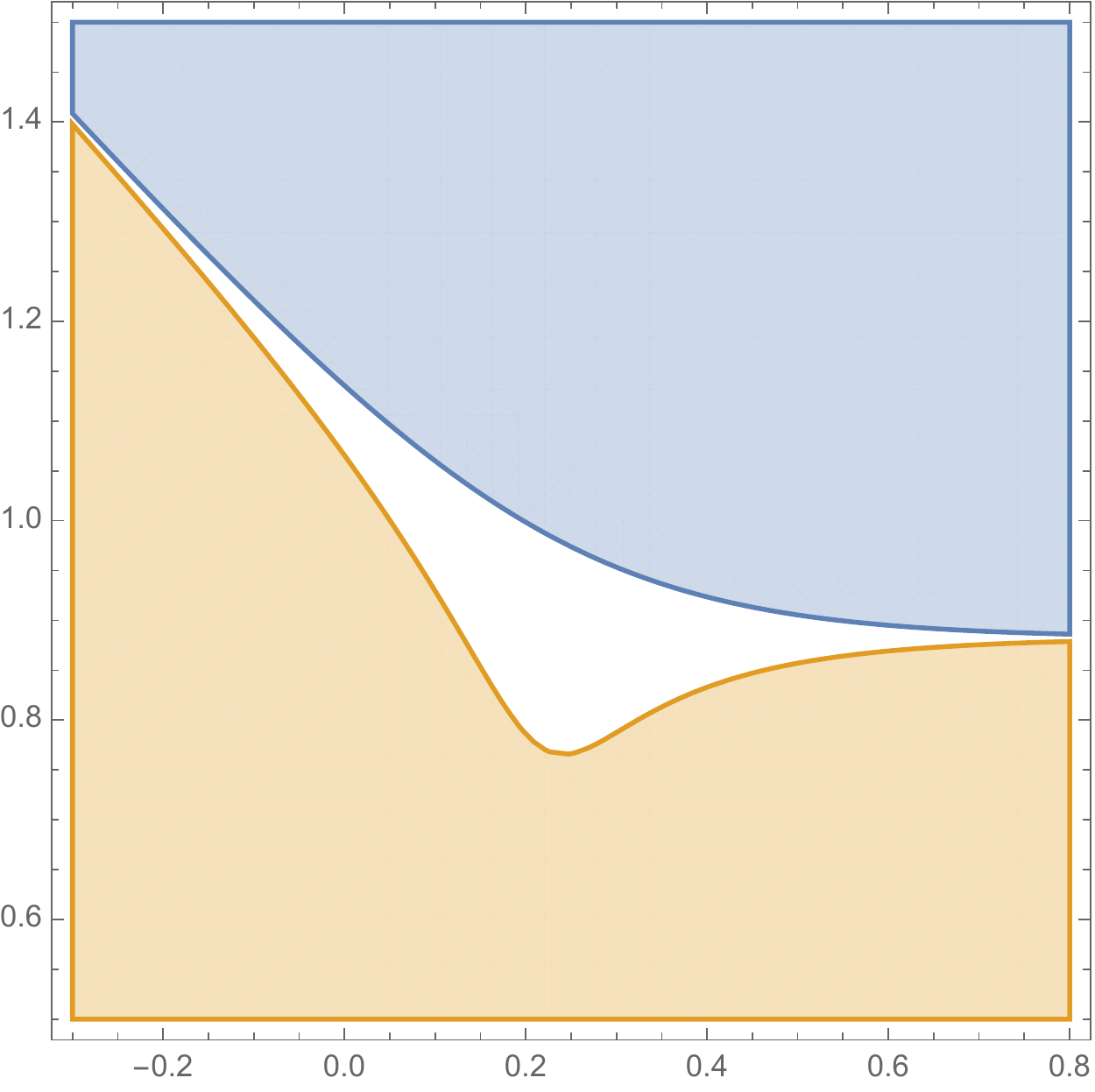}};
\node[inner sep=0pt] at (7,0)
{\includegraphics[scale=0.3]{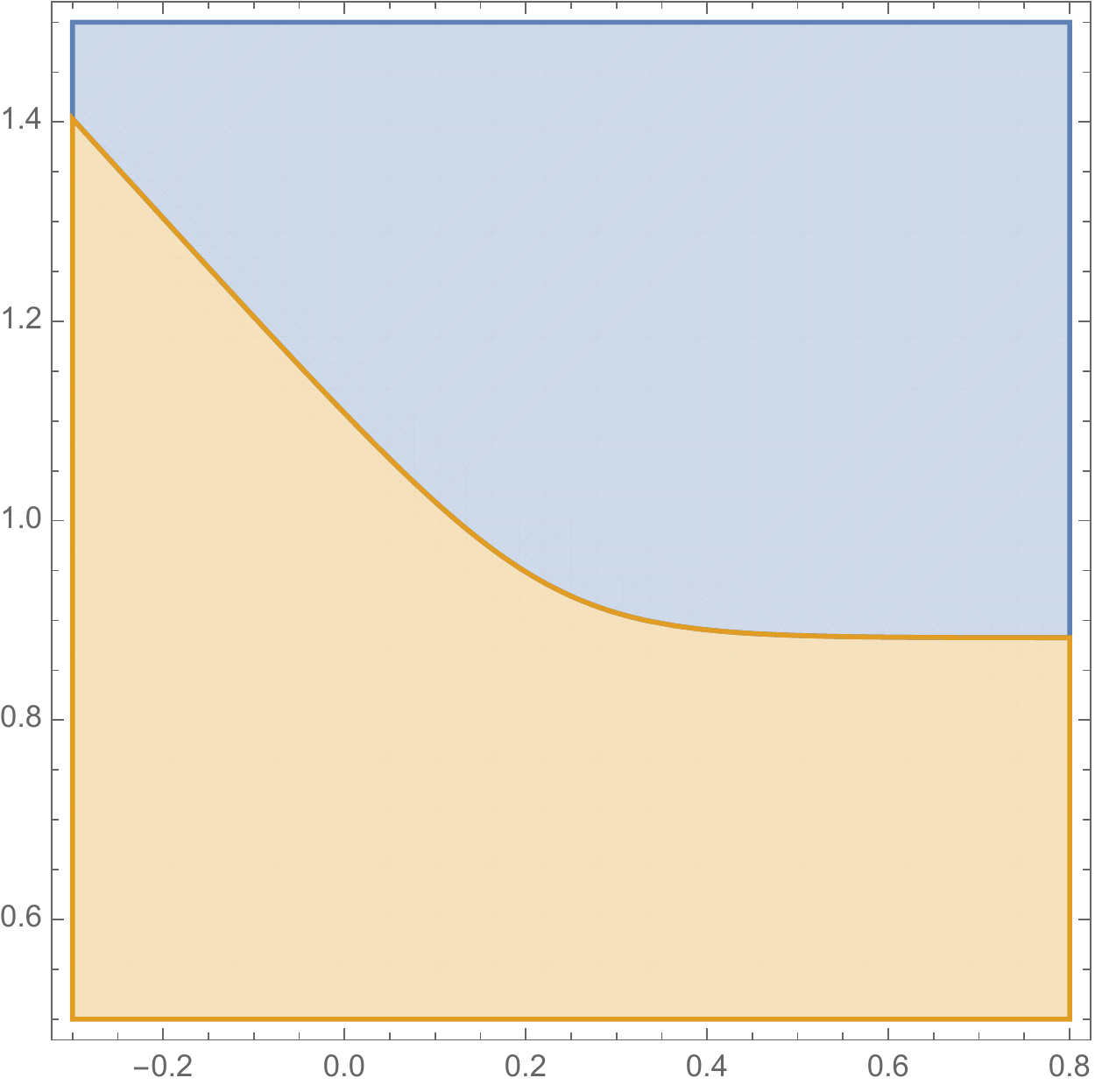}};
\node at (9,-3) {$\rho_1$};
\node at (-3,2) {$\rho_2$};
\node at (2,-3) {$\rho_1$};
\node at (4,2) {$\rho_2$};
\node at (-5,-3) {$\rho_1$};
\node at (-10.2,2) {$\rho_2$};
\node at (-8.5,-1.8) {\scriptsize LG};
\node at (-6.5,1.9) {\scriptsize Orbi and LV};
\node at (-5.5,-1.8) {\scriptsize $\mathbb P^1$};
\node at (7.8,1.9) {\scriptsize $|8 \, \psi^4| > |\phi \pm 1|$};
\node at (6.5,-1.8) {\scriptsize $|8 \, \psi^4| < |\phi \pm 1|$};
\node at (0.8,1.9) {\scriptsize $|8 \, \psi^4| > |\phi \pm 1|$};
\node at (-0.2,-1.8) {\scriptsize $|8 \, \psi^4| < |\phi \pm 1|$};
\end{tikzpicture}
\caption{\small These plots depict the convergence regions of the periods for $\mathbb P^4_{11222} [8]$, where the blue regions covers $|8 \, \psi^4| > |\phi \pm 1|$ and yellow $|8 \, \psi^4| < |\phi \pm 1|$.
Shown are three different phases of $\phi$.
In the left graph we have chosen $\phi = |\phi|$, hence the convergence area agrees with the phase picture from the conifold constraint $(\phi + 8 \psi^4)^2=1$.
In the middle graph we have $\phi = |\phi| \, e^{\frac12 \, i}$, i.e. the periods derived in section \ref{app_metric} cannot describe transitions between blue and yellow phases. 
As shown in the right graph, the non-convergence area vanishes precisely for choosing $\phi = |\phi| \, e^{\frac{\pi}{2} \, i}$.
In all plots $\psi$ is real.}
\label{fig_ConvReg}
\end{figure}
Figure \ref{fig_ConvReg} shows three plots of the convergence regions of the periods differing by the phase of $\phi$.
As one can see, a small phase for $\phi$ opens the
border between Landau-Ginzburg and hybrid $\mathbb P^1$ phase.
Only if the phase is precisely ${\rm Arg} \, (\phi) = \frac{\pi}{2}$, the convergence relations reduce to a single condition $|8 \, \psi^4| \lessgtr |\phi|^2 + 1$, such that 
one is able to traverse between any of the four phases.
This is only possible due to the $\mathbb Z_2$ symmetry of $\phi$ determining the convergence relation.

The same statement holds true for $\mathbb P^4_{11226} [12]$, but not for $\mathbb P^4_{11169} [18]$ as one can observe later.
In the case of $\mathbb P^4_{11169} [18]$, $\phi$ has a $\mathbb Z_3$ symmetry and thus the convergence relations cannot reduce to a single one (see equation \ref{convergence11169} for details). As a consequence, it is impossible to cover the entire moduli space with the charts derived in section \ref{app_metric}.

\vspace{0.5cm}
\noindent
{\bf 5.1.2. Tests of the RSDC: computing $\Theta_0$ and $\log$-behavior}
\vspace{0.4cm}

\begin{figure}[ht!]
\centering
\begin{tikzpicture}[xscale=0.8,yscale=0.8]
\node at (3,-0.3) {$\phi$};
\node at (-3.3,6) {$\psi$};
\node[align=left,below] at (-3,0) {$0$};
\node[align=center,below] at (0,0) {\scriptsize $\phi=1$};
\node[align=center,above] at (2.5,1.5) {\scriptsize $|8 \psi^4| \lessgtr |\phi \pm 1|$};

\draw[-] (-3,0) -> (-3,4.5);
\draw[-] (-3,4.5) -> (-3,5.5);
\draw[->] (-3,5.5) -> (-3,6);

\draw (-3,0) -> (1.5,0);
\draw[-] (1.5,0) -> (2.5,0);
\draw[->] (2.5,0) -> (3,0);

\draw[dashed,blue] (0,0) -> (0,6);
\coordinate (A) at (-3,3);
\coordinate (B) at (3,1.5);
\draw[dashed, blue]    (A) to[out=-35,in=180] (B);

\node[align=center,below] at (1.7,5) {large volume};
\node[align=center,below] at (1.5,1.2) {$\mathbb P^1$};
\node[align=center,below] at (-1.5,1.7) {Landau\\ Ginzburg};
\node[align=center,below] at (-1.5,5) {orbifold};
\end{tikzpicture}
\caption{\small A schematic plot of the four phases of $\mathbb P^4_{11222} [8]$ in real $(\phi,\psi)$ coordinates.
The singular locus $|8 \psi^4| \sim |\phi \pm 1|$ is actually a two dimensional surface due to the $\pm$ indicating a logical ``and".
Directions bounded in field space do not necessarily have to be bounded in their proper distance.}
\label{fig_Plot11222}
\end{figure}
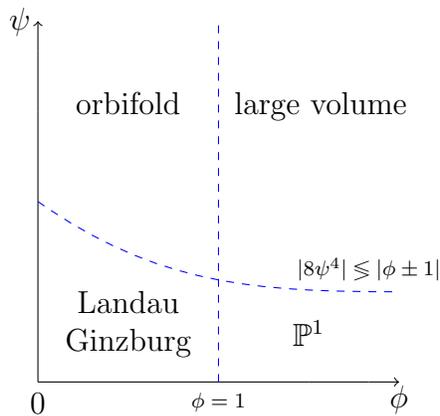

\noindent
As shown in figure \ref{fig_Plot11222},
schematically the different phases of the moduli space can also be
depicted in the coordinates $(\psi, \phi)$.
Here we are assuming to have real fields, i.e. setting ${\rm Im} (\psi) = {\rm Im} (\phi) = 0$ for simplicity.

Before starting the discussion about the lengths of curves in the moduli space, let us stress that we are distinguishing three different types of curves:
we call any arbitrary path a \emph{trajectory}. If the trajectory is additionally satisfying the geodesic equation it is called \emph{geodesic}. 
The length of the globally shortest geodesic is denoted as the \emph{distance} or \emph{proper length}.
The reason for making such a clear separation between different types of curves will become apparent in the following.

Figure  \ref{fig_Plot11222} might be misleading in the sense that finite sized
directions in the fields $\psi$ and  $\phi$ do not necessarily have to be finite in their proper length.
In the case of $\mathbb P^4_{11222} [8]$, it will eventually turn out that all these seemingly finite directions in $\psi$, $\phi$ are in fact bounded in their proper lengths.
To compare with the prediction of RSDC, in the following we shall
calculate the proper length of such finite directions.

Secondly, in this section we will be concerned with trajectories along
infinite directions. Both in the large volume and in the hybrid phases the
Swampland Distance Conjecture implies that they show a $\log$ behavior in their distances.
One has to be cautious here, as actually, this has to hold for the globally shortest geodesics between two arbitrary points in the moduli space.
Hence, one would have to solve the geodesic differential equations for certain start and end points and minimize the set of solutions regarding their proper lengths.
Due to the huge variety of possible trajectories in the four real
dimensional moduli space of $\mathbb P^4_{11222} [8]$, this is
obviously very elaborate and exceeds our computational capabilities.
Therefore, we proceed by doing the analysis in asymptotic regimes of the moduli space where we are able to determine the shortest geodesics.

To see the exponentially light KK-modes it is important to express the
final result in terms for the K\"ahler coordinates $t_i$ that can be
determined via the mirror map.
As discussed, the mirror map can be found by analyzing the monodromy
properties of the periods. 
For the case at hand this has been done by \cite{Candelas:1993dm} and the mirror map in all phases is given by
\begin{equation}
\label{mirrormap11222}
  \begin{aligned}
    t_1&=-\frac14 +\frac{2\omega_2+\omega_4}{4\omega_0}\;,\\
    t_2&=\frac14 -\frac{2\omega_2+\omega_4}{4\omega_0}+\frac{3\omega_1+2\omega_3+\omega_5}{4\omega_0}\;,
  \end{aligned}
\end{equation}
where the periods $\omega_i$ have been computed in section \ref{app_metric}.
Later, we will find it useful to state the asymptotic behavior of the
K\"ahler coordinates  at special points in moduli space.
Let us emphasize that the K\"ahler metrics that we are going to employ, have all been computed with the period method presented in section \ref{app_metric}.
The GLSM method introduced in section \ref{sec_KahlerMetric} was
primarily used to cross check results.

\subsubsection*{Landau-Ginzburg phase}

The Landau-Ginzburg regime is similar to the example of the  quintic \ref{sec_quintic_pheno} in the sense that the algebraic parameters $|\phi|$ and $|\psi|$ are both bounded
\eq{
  0 \, \le \, \phi \, < \, 1 \, , \qquad
  0 \, \le \, \psi \, < \psi_{c} \, .
  }
The upper bound $\psi_c$ is determined as a solution of the convergence condition $|8 \psi^4| < |\phi \pm 1|$ for fixed $\phi$.
The reader is referred to section \ref{app_metric} for details.

\begin{figure}[ht!]
\centering
\begin{tikzpicture}[xscale=0.8,yscale=0.8]
\node at (3,-0.3) {$\phi$};
\node at (-3.3,6) {$\psi$};
\node[align=left,below] at (-3,0) {$0$};
\node[align=center,below] at (0,0) {\scriptsize $\phi=1$};
\node[align=center,above] at (2.5,1.5) {\scriptsize $|8 \psi^4| \lessgtr |\phi \pm 1|$};

\draw[-] (-3,0) -> (-3,4.5);
\draw[dashed] (-3,4.5) -> (-3,5.5);
\draw[->] (-3,5.5) -> (-3,6);

\draw (-3,0) -> (1.5,0);
\draw[dashed] (1.5,0) -> (2.5,0);
\draw[->] (2.5,0) -> (3,0);

\draw[dashed,blue] (0,0) -> (0,6);
\coordinate (A) at (-3,3);
\coordinate (B) at (3,1.5);
\draw[dashed, blue]    (A) to[out=-35,in=180] (B);

\draw[<->] (-2.7,0.3) to (-0.3,0.3);
\draw[<->] (-2.7,0.4) to (-2.7,2.7);
\draw[<-] (-2.6,0.4) to (-2.3,0.544);
\draw[->] (-1.3,1) to (-0.1,1.6);

\node at (-0.4,0.6) {\scriptsize $\gamma_2$};
\node at (-2.4,2) {\scriptsize $\gamma_1$};
\node at (-0.3,1.2) {\scriptsize $\gamma_3$};

\node[align=center,below] at (1.5,5) {\scriptsize LV};
\node[align=center,below] at (1.5,1.2) {$\mathbb P^1$};
\node[align=center,below,red] at (-1.5,1.7) {\scriptsize Landau- \\ \scriptsize{Ginzburg}};
\node[align=center,below] at (-1.5,5) {\scriptsize orbifold};
\end{tikzpicture}
\caption{\small The Landau-Ginzburg phase has finite length in every direction.
To show this we compute the lengths of the following three paths in $(\phi,\psi)$:
$\gamma_1 : (0,0) \rightarrow (0,0.59)$, $\gamma_2 : (0,0) \rightarrow (1,0)$, $\gamma_3 : (0,0) \rightarrow (0.5,0.5)$.}
\label{fig_11222LG}
\end{figure}
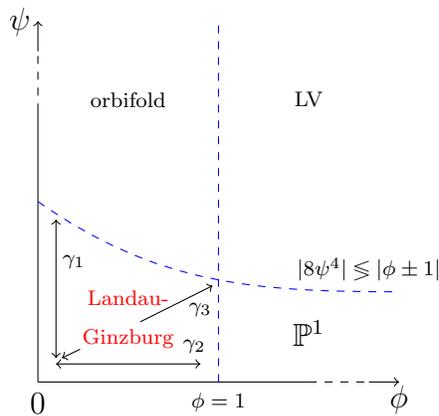

As already pointed out,
whether all geodesics in the LG phases stay finite is
a priori not obvious. If they are finite, it is important to compute
their proper lengths.
For that purpose let us investigate three trajectories as depicted in figure \ref{fig_11222LG}.
All of these trajectories start close to the smallest possible values of the algebraic moduli $\psi = \phi =0$.
Then, we consider one trajectory only in direction of $\psi$ or $\phi$, respectively.
Additionally, there will be one moving directly towards the conifold singularity.
The labeling of the curves follows figure  \ref{fig_11222LG}.
Using the metric computed in section \ref{sec_KahlerMetric}, we obtain the following lengths for these trajectories
\eq{
\label{dists_P11222_LG}
  \Delta \Theta_1 \, &= \, \int_{\gamma_1} d \psi \, \sqrt{G_{\psi \ov\psi} (\psi)} \, 
  = \, 0.40 \, ,\\
  \Delta \Theta_2 \, &= \, \int_{\gamma_2} d \phi \, \sqrt{G_{\phi \ov\phi} (\phi)} \, 
  = \, 0.24 \, ,\\
  \Delta \Theta_3 \, &= \, \int_{\gamma_3} d \tau \, 
    \sqrt{G_{\mu \ov \nu} \left( \psi (\tau),\phi (\tau) \right) \, 
    \frac{d x^\mu}{d \tau} \frac{d \ov{x}^{\ov{\nu}}}{d \tau}} \, 
  = \, 0.36 \, ,
  }
where we have denoted $x^\mu = \{ \psi, \phi \}$.
All these directions are finite and smaller than one, as expected from the RSDC.

For the sake of completeness, let us point out that the mirror map coordinates in the Landau-Ginzburg regime approach the finite values $t_1 = \frac{1}{2} (-1 + i)$ and $t_2 \simeq 0.5 + 0.21 \, i$ for $\phi , \psi \rightarrow 0$.

\subsubsection*{Hybrid phase - \texorpdfstring{$\mathbb P^1$}{P1}}

The conceptually new regimes of this moduli space are clearly the hybrid phases $\mathbb P^1$ and orbifold.
We begin with the hybrid $\mathbb P^1$ phase, where the two complex structure parameters are limited by the regime
\eq{
  1 \, < \, \phi \, < \, \infty \, , \qquad
  0 \, \le \, \psi \, < \psi_{c} \, .
  }
Again, the upper bound $\psi_c$ is determined as solution of the convergence condition $|8 \psi^4| < |\phi \pm 1|$ for fixed $\phi$.

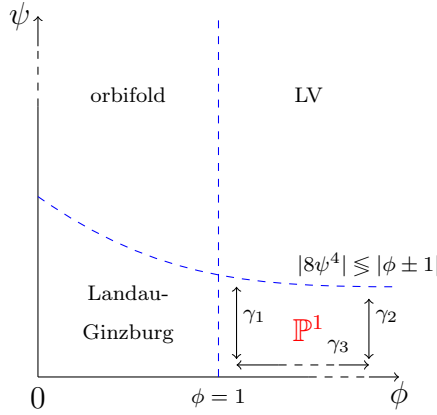
\begin{figure}[ht!]
\centering
\begin{tikzpicture}[xscale=0.8,yscale=0.8]
\node at (3,-0.3) {$\phi$};
\node at (-3.3,6) {$\psi$};
\node[align=left,below] at (-3,0) {$0$};
\node[align=center,below] at (0,0) {\scriptsize $\phi=1$};
\node[align=center,above] at (2.5,1.5) {\scriptsize $|8 \psi^4| \lessgtr |\phi \pm 1|$};

\draw[-] (-3,0) -> (-3,4.5);
\draw[dashed] (-3,4.5) -> (-3,5.5);
\draw[->] (-3,5.5) -> (-3,6);

\draw (-3,0) -> (1.5,0);
\draw[dashed] (1.5,0) -> (2.5,0);
\draw[->] (2.5,0) -> (3,0);

\draw[dashed,blue] (0,0) -> (0,6);
\coordinate (A) at (-3,3);
\coordinate (B) at (3,1.5);
\draw[dashed, blue]    (A) to[out=-35,in=180] (B);

\draw[<->] (0.3,0.3) to (0.3,1.5);
\draw[<->] (2.5,0.3) to (2.5,1.3);
\draw[<-] (0.3,0.2) to (1,0.2);
\draw[dashed] (1,0.2) to (2,0.2);
\draw[->] (2,0.2) to (2.5,0.2);

\node at (0.6,1) {\scriptsize $\gamma_1$};
\node at (2.8,1) {\scriptsize $\gamma_2$};
\node at (2,0.5) {\scriptsize $\gamma_3$};

\node[align=center,below] at (1.5,5) {\scriptsize LV};
\node[align=center,below,red] at (1.5,1.2) {$\mathbb P^1$};
\node[align=center,below] at (-1.5,1.7) {\scriptsize Landau- \\ \scriptsize{Ginzburg}};
\node[align=center,below] at (-1.5,5) {\scriptsize orbifold};
\end{tikzpicture}
\caption{\small The hybrid $\mathbb P^1$ fibration regime has one finite as well as one infinite direction.
The dashed arrow and coordinate axes symbolize their extension to infinity.
For the calculation we used $\gamma_1 : (1.1,0) \rightarrow (1.1,0.33)$.}
\label{fig_11222P1}
\end{figure}

The phase diagram of figure \ref{fig_Aspinwall11222} leads naturally to the following questions:
First, does really only one K\"ahler modulus $t_i$
exhibit a logarithmic behavior for large distances in field space? That is to be expected for trajectories parallel to the $\phi$ axis, but not for the ones parallel to the Landau-Ginzburg regime.
Second, does there exist a geodesic moving towards $|\rho_1| \rightarrow \infty$ or is the attractive effect of the conifold singularity strong enough to bend any geodesic into the singularity?
Finally, will the RSDC  hold true even for geodesic passing the Landau-Ginzburg and a hybrid phase before entering the large volume regime?
It seems to be challenging for arbitrary geodesics to collect less than $O (1) \, M_{\rm pl}$ in the proper distance after crossing two regimes.
In order to answer the latter two questions one has to consider actual geodesics. Thus we will comment on these questions later and focus now on the first one by analyzing the trajectories shown in figure \ref{fig_11222P1}.
Note that we assume again for simplicity $\psi$, $\phi$ to be real-valued.
\vspace{0.1cm}

The curve $\gamma_1$ starts near $\psi=0$, $\phi=1$ and moves only in $\psi$ direction, hence along the Landau-Ginzburg regime towards the singularity.
Via integration we find its length
\eq{
  \Delta \Theta_1 \, &= \, \int_{\gamma_1} d \psi \, \sqrt{G_{\psi \ov\psi} (\psi)} \, 
  = \, 0.24 \, .
  }
This result is consistent  with the RSDC  and with our expectations,
as it is close to to the Landau-Ginzburg phase.
Besides, we can estimate the asymptotic distance for large $\phi$ as depicted by curve $\gamma_2$ in figure \ref{fig_11222P1}.
According to section \ref{app_metric}, in the $\mathbb{P}^1$ phase the asymptotic behavior of the metric for $(\phi \rightarrow 0,\psi \rightarrow \infty)$ is
\begin{equation}
\label{metricP11222_P1}
  G^{\rm asymp}_{\mathbb{P}^1}\simeq
  \begin{pmatrix}
    G^{\rm asymp}_{\phi \ov\phi} & 0\\
    0 & G^{\rm asymp}_{\psi \ov\psi}
  \end{pmatrix}
  \simeq\begin{pmatrix}
    \frac{0.25}{|\phi|^2 \, (\log|\phi|)^2} & 0\\
    0 & \frac{0.5905}{\sqrt{|\phi|}}
  \end{pmatrix}\; .
\end{equation}
Integrating this asymptotic metric for large $\phi$ leads to a finite and non-zero value
\eq{
  \Delta \Theta_2 \, = \, \int_{\gamma_2} d \psi \, 
  \sqrt{G^{\rm asymp}_{\mathbb{P}^1, \, \psi \ov\psi} (\phi)} \,
  \simeq \, \sqrt{\frac{0.5905}{\sqrt{|\phi|}}} \cdot 
  \sqrt[4]{\frac{|\phi|}{8}} \,
  = \, 0.46 \, .
  }
Finally, we want to point out that the trajectory $\gamma_3$ ``parallel''
to  the large volume  phase has infinite length.
To see that, we would have to integrate the K\"ahler metric of
$\mathbb P^1$ in the $\phi$ direction towards $\infty$.
But asymptotically this integral is given by
\eq{
  \Theta\sim \int d \phi \, \sqrt{G^{\rm asymp}_{\mathbb{P}^1, \, \phi \ov\phi} (\phi)} \sim \log \left( \log \phi \right) \, .
  }
As a consequence, the direction $\phi$ in $\mathbb P^1$ is infinite, in contrast to the $\psi$ direction.

As we have seen, it is the K\"ahler coordinates ${\rm Im}(t_i)$ (following from the
mirror map) that control the exponential drop-off of the KK-modes.
As explained in section \ref{app_metric} and using eq. \eqref{mirrormap11222}, deep inside the $\mathbb{P}^1$ phase ($\psi\to 0,\phi\to\infty$), we find
\begin{equation}
\label{mm11222P1}
  \begin{aligned}
    t_1&\simeq\frac12 \, (-1 + i)+\dots\;,\\
    t_2&\simeq \left( 1- \frac{i}{2} \right) 
    + \frac{8 i \, \log (2)}{2 \pi} + \frac{i}{\pi}\log(\phi)+\dots\;.
  \end{aligned}
\end{equation}
This is just right, as the logarithmic scaling behavior with respect
to $t_2$ becomes
\eq{
                     \Theta \sim     \log \left( \log \phi \right)
                     \sim \log ({\rm Im}\, t_2)\,.
}

\subsubsection*{Hybrid phase - orbifold}

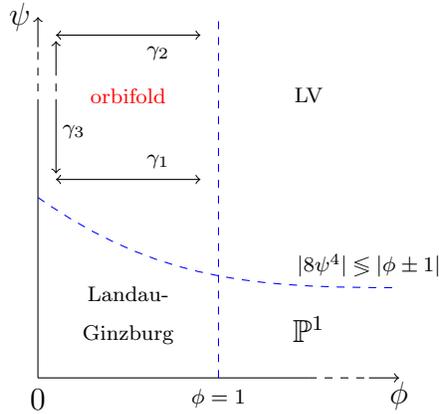
\begin{figure}[ht!]
\centering
\begin{tikzpicture}[xscale=0.8,yscale=0.8]
\node at (3,-0.3) {$\phi$};
\node at (-3.3,6) {$\psi$};
\node[align=left,below] at (-3,0) {$0$};
\node[align=center,below] at (0,0) {\scriptsize $\phi=1$};
\node[align=center,above] at (2.5,1.5) {\scriptsize $|8 \psi^4| \lessgtr |\phi \pm 1|$};

\draw[-] (-3,0) -> (-3,4.5);
\draw[dashed] (-3,4.5) -> (-3,5.5);
\draw[->] (-3,5.5) -> (-3,6);

\draw (-3,0) -> (1.5,0);
\draw[dashed] (1.5,0) -> (2.5,0);
\draw[->] (2.5,0) -> (3,0);

\draw[dashed,blue] (0,0) -> (0,6);
\coordinate (A) at (-3,3);
\coordinate (B) at (3,1.5);
\draw[dashed, blue]    (A) to[out=-35,in=180] (B);

\draw[<->] (-2.7,3.3) to (-0.3,3.3);
\draw[<->] (-2.7,5.7) to (-0.3,5.7);
\draw[<-] (-2.7,3.4) to (-2.7,4.5);
\draw[dashed] (-2.7,4.5) to (-2.7,5.1);
\draw[->] (-2.7,5.1) to (-2.7,5.6);

\node at (-1,3.6) {\scriptsize $\gamma_1$};
\node at (-1,5.4) {\scriptsize $\gamma_2$};
\node at (-2.4,4.1) {\scriptsize $\gamma_3$};

\node[align=center,below] at (1.5,5) {\scriptsize LV};
\node[align=center,below] at (1.5,1.2) {$\mathbb P^1$};
\node[align=center,below] at (-1.5,1.7) {\scriptsize Landau- \\ \scriptsize{Ginzburg}};
\node[align=center,below,red] at (-1.5,5) {\scriptsize orbifold};
\end{tikzpicture}
\caption{\small In the hybrid orbifold phase we find again an infinite
  direction $\gamma_3$ plus a finite one $\gamma_1 : (0,0.59)
  \rightarrow (1,0.59)$. The distance in the $\phi$ direction asymptots (indicated by dashed arrow) to a finite $\gamma_2$, more precisely the distance will be zero.}
\label{fig_11222Orbi}
\end{figure}

As one might guess from figure \ref{fig_Plot11222}, the hybrid orbifold phase is qualitatively quite similar to the $\mathbb P^1$ hybrid phase.
Now the algebraic coordinates may vary in the interval
\eq{
  0 \, \le \, \phi \, < \, 1 \, , \qquad
  \psi_c \, < \, \psi \, < \infty \, ,
  }
with $\psi_c$ as in the other phases.
Again, we focus on three trajectories: a finite one $\gamma_1$ along
the Landau-Ginzburg regime, its asymptotic equivalent $\gamma_2$ for
large $\psi$ and an infinite direction $\gamma_3$ along the large
volume regime.
Figure \ref{fig_11222Orbi} summarizes these curves schematically.

\vspace{0.1cm}
\vspace{0.1cm}

\noindent
We begin by computing the distance from $\phi=0$ to the conifold singularity.
Keeping $\psi$ fixed, the integral leads to the length
\eq{
  \Delta \Theta_1 \, &= \, \int_{\gamma_1} d \phi \, \sqrt{G_{\phi \ov\phi} (\phi)} \, 
  = \, 0.21 \, ,
  }
which is very close to our result \eqref{dists_P11222_LG} in the Landau-Ginzburg phase.

Let us now check whether this asymptots to a finite distance $\gamma_2$ as in the $\mathbb P^1$ fibration regime.
The asymptotic metric for real moduli near $\phi \simeq 0$ (see section \ref{app_metric} for details) reads
\begin{equation}
\label{metricP11222_orbi}
  G^{\rm asymp}_\text{orbi}\simeq
  \begin{pmatrix}
    \frac{0.09}{|\log(\psi)|^2} & 0\\
    0 & \frac{0.75}{|\psi|^2 \, (\log |\psi|)^2}
  \end{pmatrix}\;.
\end{equation}
Consequently, the asymptotic distance $\gamma_2$ is indeed finite
\eq{
  \Delta \Theta_2 \, = \, \int_{\gamma_2} d \phi \, 
  \sqrt{G^{\rm asymp}_{\text{orbi}, \, \phi \ov\phi} (\psi)} \,
  \sim \, \frac{\sqrt{0.09}}{|\log \psi|} \int_0^1 d\phi \,
  \xrightarrow{\psi \rightarrow \infty} \, 0 \, .
  }
This is interesting since the distance to the large volume  phase vanishes. In other words, for large $\psi$ we end up at the $\phi=1$ singularity and hence on the one-parameter subspace $\mathbb P^5_{11111} [2 \ 4]$.

From the metric one can see that the pure $\psi$ direction $\gamma_3$ of the hybrid orbifold phase is infinite.
The metric in the $\psi$ direction leads again to a $\Theta\sim\log (\log  \psi)$ growth of the distance.
The asymptotic form of the mirror map confirms exactly our observation.
For $(\psi\to\infty,\phi\to 0)$, we find
\begin{equation}
\label{mm11222Orbi}
  \begin{aligned}
      t_1&\simeq \frac{i}{2 \pi} \, \log(8 \psi)^4+\dots\;,\\
      t_2&\simeq \frac{1}{2}+\dots\;.
  \end{aligned}
\end{equation}
Thus, the distance is in agreement with the Swampland Distance Conjecture as now $\Theta \sim \log \left( \log \psi \right) \sim \log ({\rm Im}\, t_1)$.

Only $t_1$ grows logarithmically for large values of $\psi$, whereas $t_2$ approaches a constant.
Note that this was exchanged in the $\mathbb P^1$ regime.
In addition, the imaginary part ${\rm Im} (t_2) = \int_{\Sigma_2}J$ asymptotes to
zero\footnote{The (at a first glance surprising) possibility for zero minimum distances in string target spaces was already observed in \cite{Aspinwall:1994zu}.
In fact, there it was argued that for ``sufficiently complicated'' topologies a vanishing exceptional divisor demands some other part of the target space to become infinitely large.
This is in line with the $\log (\psi)$ term in $t_1$.}. This means that one has a vanishing two-cycle $\Sigma_2$, validating the interpretation of being located at the one-parameter subspace $\mathbb P^5_{11111} [2 \ 4]$.

\subsubsection*{Large volume  phase}

In the remaining phase of the moduli space under investigation, the algebraic parameters may take values in
\eq{
  1 \, < \, \phi \, < \, \infty \, , \qquad
  \psi_c \, < \, \psi \, < \infty \, .
  }
and $\psi_c$ again as in the other phases.
By definition it is clear that both directions, $\phi$ and $\psi$, are infinite.
Therefore we expect the $\Theta\sim \log\log (-)$-behavior, demanded by the swampland
distance conjecture, in any of these directions.

The mirror map confirms this expectation.
Around the large volume point, according to \cite{Candelas:1993dm} they are given by
\begin{equation}
  \begin{aligned}
    t_1&\simeq\frac12 +\frac{i}{2\pi}\log\left(\frac{(8\psi)^4}{2\phi}\right)+\dots\;,\\
    t_2&\simeq  \frac{i}{\pi}\log\left(2\phi\right)+\dots\;,
  \end{aligned}
\end{equation}
where the dots denote polynomial corrections.
So, whenever one of the algebraic coordinates becomes large, the
corresponding K\"ahler coordinate will grow logarithmically.

\vspace{0.5cm}
\noindent
{\bf 5.1.3. Tests of the RSDC: computing $\Theta_\lambda$}
\vspace{0.4cm}

\noindent
So far, we have only motivated the logarithmic behavior of proper distances by analyzing the mirror maps of each phase in the moduli space.
Let us now turn towards the critical distance $\Theta_\lambda$ where
the logarithm is significant, i.e. challenge the  RSDC.
For that purpose, two different approaches are presented: calculating K\"ahler potentials and geodesics in asymptotic regions of the moduli space.
Both methods will turn out to give the same results for $\Theta_\lambda$.

\subsubsection*{Asymptotic K\"ahler potentials}

Before commenting on geodesics in two-parameter models, recall that
the value for $\Theta_\lambda$ deep inside the large volume regime
(LV) can be derived from the generic form \eqref{Kpotlcs} 
of the asymptotic K\"ahler potential.
It is interesting to compare this K\"ahler potential in the large
volume region to the asymptotic ones in the two hybrid phases.

For the mirror of $\mathbb P^4_{11222} [8]$ the prepotential depending on the K\"ahler coordinates $t_i$ is asymptotically given by \cite{Candelas:1993dm}
\eq{
  \mathcal F = - \frac43 \, (t_1)^3 + 2 \, (t_1)^2 t_2 + \dots \, ,
  }
where the dots denote subleading corrections. 
Employing the standard formula for the periods $\Pi (t_1, t_2) =
(1,t_1, t_2, 2 \mathcal F - \partial_{t_1} \mathcal F -\partial_{t_2} \mathcal F, \partial_{t_1}
\mathcal F, \partial_{t_2} \mathcal F)$ 
(in inhomogeneous coordinates), we obtain the  K\"ahler potential 
\eq{
\label{Kpotasy}
  K^{\rm asymp}_{\rm LV} \simeq - \, \log \left[ \frac{4 \, i}{3} \, (t_1 - \ov t_1)^3 
  +2 \, i \, (t_1 - \ov t_1)^2 \, (t_2 - \ov t_2) \right] \, .
  }
In the hybrid phases we are able to compute an asymptotic expression for the K\"ahler potential as well:

\paragraph*{Hybrid $\mathbb P^1$ phase}
Expanding the K\"ahler potential $K_{\mathbb P^1}$ of this regime around $\psi \approx 0$ and $\phi \rightarrow \infty$, we find asymptotically
\eq{
  \exp \left( -K^{\rm asymp}_{\mathbb P^1} \right) &\simeq 
   6.99 \,i \, \frac{\psi \overline{\psi}}{(\phi \overline{\phi})^\frac14} \,
   \bigg[5.54 - 2.46 \frac{\psi \overline{\psi}}{(\phi \overline{\phi})^\frac14}\\
&\phantom{aaaaaaaaaaaaaaa}+\Big(1 - 0.59 \frac{\psi \overline{\psi}}{(\phi \overline{\phi})^\frac14} \Big) \, \log (\phi \, \overline{\phi})
\bigg]
}
which is consistent  with the metric \eqref{metricP11222_P1}.
Taking the mirror map \eqref{mm11222P1} into account,  in
leading-order leads to the simple result
\eq{
  K^{\rm asymp}_{\mathbb P^1} \simeq - \, \log (t_2 - \ov t_2) \, .
  }
Obviously, the large volume K\"ahler potential \eqref{Kpotasy} reduces to the $\mathbb P^1$ expression for small $t_1$ and very large $t_2$.
It also confirms the target space geometry since the K\"ahler modulus $t_2$ measures the size of the $\mathbb P^1$ space.

Moreover, even without calculating geodesics one can derive a value for $\Theta_\lambda$ in this asymptotic region.
That is, using the induced metric
\eq{
  G_{t_2\overline{t}_2} = \frac{1}{4 \,({\rm Im} \, t_2)^2}
  }
and carrying out an analogous calculation as in section \ref{sec:RSC}, we obtain a critical proper distance $\Theta_\lambda = \sqrt{0.25} \simeq 0.5 $.
As $\Theta_\lambda < O(1)$, this is  in agreement with the RSDC.

\paragraph*{Hybrid orbifold phase}
Similarly, we approximate the K\"ahler potential $K_{\rm orbi}$
in the hybrid orbifold phase around $\phi \approx 0$ and $\psi
\rightarrow \infty$ by
\eq{
  \exp \left( -K^{\rm asymp}_{\rm orbi} \right) \, &\simeq \,
  26.32  i - 0.45  i\, \phi \ov \phi
  + 0.13 i \, \phi \ov \phi \, \log (\psi \ov \psi)\\
  &\,+ 17.85 i  \log (\psi \ov \psi)
  + 4.29 i \left[ \log (\psi \ov \psi) \right]^2
  + 0.34i  \left[ \log (\psi \ov \psi) \right]^3 \, 
  }
which is again in line with the asymptotic metric \eqref{metricP11222_orbi}.
By plugging in the mirror map \eqref{mm11222Orbi}, we get the leading-order result
\eq{
  K^{\rm asymp}_{\rm orbi} \simeq - 3\, \log (t_1 - \ov t_1) \, ,
  }
which is as expected from the large volume expression \eqref{Kpotasy} for $t_1$ much larger than $t_2$.
From the target space point of view, $t_1$ corresponds to the overall volume of the orbifold.
The critical proper distance is the same as already computed in
section \ref{sec:RSC}, $\Theta_\lambda = \sqrt{0.75} \simeq 0.87$.

Next, we challenge the refined conjecture from the perspective of geodesics.

\vspace{2cm}

\vspace{0.5cm}
\noindent
{\bf 5.1.4. Tests of the RSDC: asymptotic geodesics}
\vspace{0.4cm}
\label{2param_geodesic}

\noindent
In order to investigate the refined version of the Swampland Distance
Conjecture one has to check whether the proper lengths of
\emph{geodesics} grow at best logarithmically after having traversed $\Theta_\lambda \sim O (1) M_{\rm pl}$ distances.
This behavior is expected to occur for trajectories moving sufficiently far along one of the infinite directions pointed out above.
Obviously, in the two-parameter model the most interesting geodesics cross various phases.
It is crucial to note that not every geodesic is
automatically the globally shortest path between two points.

Basically the reason boils down to the attractive effect of the singularities in the moduli space at hand.
However, the RSDC  holds only for the shortest geodesic, which leads to technical difficulties.
In fact, it is quite delicate to find the solution to the geodesic equation with minimal proper length for given start and end points.
In particular so, as we are now working in real four dimensional space.
One could only determine rough upper bounds for $\Theta_\lambda$ from several typical trajectories.

Here we will instead pursue a different approach, where the goal is to compute geodesics in asymptotic regions of the moduli space, where we have already derived a simple analytic expression for the K\"ahler metric on the moduli space.

\paragraph*{Hybrid $\mathbb P^1$ phase}
At first consider the hybrid $\mathbb P^1$ phase and in particular the regime $\psi \approx 0$ and $\phi \rightarrow \infty$. 
In this regime a trajectory purely in the $\phi$ direction will be a geodesic and moreover it will be the shortest one.
We show in the following that such a trajectory solves the geodesic equation and compute the critical distance $\Theta_\lambda$.
Note that if it is a geodesic, it is automatically the shortest one due to the symmetric influence of the singularities.

Taking into account that we want to keep $\psi \approx 0$, we assume the initial values $\psi_i = d \psi_i / d \tau = 0$ with an affine proper time parameter $\tau$.
As a consequence the two interesting geodesic equations simplify to
\begin{align}
\frac{d^2 \phi}{d \tau^2} + \Gamma^\phi_{\phi \phi} \left( \frac{d \phi}{d \tau}\right)^2 &= 0 \label{GEphi} \\
\frac{d^2 \psi}{d \tau^2} + \Gamma^\psi_{\phi \phi} \left( \frac{d \phi}{d \tau}\right)^2 &= 0 \label{GEpsi}  \, .
\end{align}
Recall that the metric in this regime is asymptotically given by eq. \eqref{metricP11222_P1}
\begin{equation}
  G^{\rm asymp}_{\mathbb{P}^1}\simeq
  \begin{pmatrix}
    \frac{0.25}{|\phi|^2 \, (\log |\phi|)^2} & 0\\
    0 & \frac{0.5905}{\sqrt{|\phi|}}
  \end{pmatrix}\; .
\end{equation}
Due to the simple structure of the metric, the Christoffel symbols are straightforward to compute
\eq{
  \Gamma^\phi_{\phi \phi} = - \frac{1}{\phi} \, \left( 1 + \frac{1}{\log \phi}\right) \, ,
  \qquad\quad
  \Gamma^\psi_{\phi \phi} = 0 \, .
  }
Hence the geodesic equation \eqref{GEpsi} for $\psi$ shows that there
is no movement in the $\psi$ direction, i.e. we stay at $\psi \approx 0$.
In other words, a trajectory $\gamma$ purely in $\phi$ direction is indeed a geodesic.
The geodesic equation \eqref{GEphi} for $\phi$ can be solved analytically with two integration constants $C_1$, $C_2$.
The solution turns out to be a double exponential of the form
\eq{
\label{GEphiSol}
  \phi (\tau) = \exp \left[ \exp \left( C_1 \tau + C_1 C_2 \right)\right] \, .
  }
If we then evaluate the proper distance $\Theta$ for the geodesic $\gamma$, we find a direct proportionality to $\tau$. This is clear since $\Theta$ and $\tau$ are affine parameters.
The crucial factor is the $\sqrt{0.25}$ from the numerator of $G^{\rm asymp}_{\mathbb{P}^1, \, \phi \ov\phi}$.
More precisely we find the following proper distance
\eq{
  \Theta = \int_\gamma d \tau \, 
  \sqrt{G^{\rm asymp}_{\mathbb{P}^1, \, \phi\ov\phi} \, \left( \frac{d \phi}{d \tau} \right)^2}
  =  \frac{1}{2} \, C_1 \, \tau \, .
  }
The SDC predicts an exponential growth of
the K\"ahler coordinate depending on the proper distance.
Here, one can observe this directly for the geodesic \eqref{GEphiSol} and the mirror map $t_2$ from eq. \eqref{mm11222P1}
\eq{
  {\rm Im} (t_2) = -\frac12 + \frac{8 \log 2}{2 \pi} + \frac{1}{\pi} \, \exp \left(2\, \Theta + C_1 C_2 \right) \, .
  }
The exponential factor becomes relevant beyond  a critical distance $\Theta_\lambda = 0.5$ according to our definition in section \ref{sec:RSC}.
Since we satisfy $\Theta_\lambda < 1$, we are clearly in agreement
with the  RSDC.

\paragraph*{Hybrid orbifold phase}
One can perform the same analysis for the hybrid orbifold phase with similar results.
There, we consider a trajectory at $\phi \approx 0$ in the limit $\psi
\rightarrow \infty$, i.e. moving far from the origin purely in the $\psi$ direction.
The asymptotic metric \eqref{metricP11222_orbi} was again diagonal and
included the component $G^{\rm asymp}_{{\rm orbi}, \, \psi \ov\psi} = 0.75/|\psi \log \psi|^2$.
Assuming now $d \psi / d \tau \approx 0$, the geodesic equations and its solutions are equivalent to the hybrid $\mathbb P^1$ phase (with two constants $C_1$, $C_2$).
The proper distance is now given by $\Theta = \sqrt{0.75} \, C_1 \tau$, such that the mirror map \eqref{mm11222Orbi} obeys the following relation 
\eq{
  {\rm Im} (t_1) = \frac{6 \log 2}{\pi} + \frac{2}{\pi} \, 
   \exp \left(\frac{\Theta}{\sqrt{0.75}} + C_1 C_2 \right) \,.
   }
The critical distance is now found to be $\Theta_\lambda = \sqrt{0.75}
\sim 0.87 < 1$ and hence satisfies the RSDC, as well.
Note that all results from investigating geodesics agree precisely with the one derived from asymptotic K\"ahler potentials.

To consolidate our results, we   consider other two-parameter moduli
spaces and  perform a similar analysis.

\subsection{\texorpdfstring{$\mathbb P^4_{11226} [12]$}{P11226}}
In this section we analyse, by analogy with $\mathbb P^4_{11222} [8]$, the moduli space of $\mathbb P^4_{11226} [12]$, which was also investigated in \cite{Candelas:1993dm}.
The defining hypersurface polynomial is now given by
\eq{
  P = x_1^{12} + x_2^{12} +x_3^6 + x_4^6+x_5^6 - 12 \psi \, x_1 x_1 x_3 x_4 x_5 - 2 \phi \, x_1^6 x_2^6 \, .
  }
The mirror of $\mathbb P^4_{11226} [12]$ can again be identified with the Calabi-Yau threefold satisfying the constraint $\{ P =0 \}/G$, where in this case the group $G$ may be enlarged to include a $\mathbb Z_{12}$ symmetry.
Its action on the algebraic parameters reads
\eq{
  (\psi, \phi) \mapsto (\alpha \psi, - \phi)
  }
with $\alpha = \exp \left( 2 \pi i /12 \right)$.
As explained in \cite{Candelas:1993dm}, the structure of the moduli space is very similar to the one of $\mathbb P^4_{11222} [8]$.
In particular, the four different phases appear again:
Landau-Ginzburg, large volume, hybrid $\mathbb P^1$ and hybrid orbifold.
The singular loci in case of $\mathbb P^4_{11226} [12]$ are $\phi^2=1$ and the conifold surface $(\phi + 864 \psi^6)^2 =1$.
The singular threefolds at $\phi^2=1$ are again birationally equivalent to the mirror of a one-parameter model.
In this case they correspond to the CICY $\mathbb P^5_{111113} [2 \ 6]$.

The periods may be calculated with the techniques described in section  \ref{app_metric}.
The series expansion of the fundamental period converges for
\eq{
  |864 \psi^6| \, \lessgtr \, |\phi \pm 1| \,,
  }
depending on the regime we want to investigate.
Note that the $\pm$ stands again for a logical ``and''.

Let us now briefly investigate various distances in the moduli space of the mirror $\mathbb P^4_{11226} [12]$ in analogy with the schematic drawings in figures \ref{fig_11222LG}, \ref{fig_11222P1}, \ref{fig_11222Orbi}.
In particular we want to determine infinite directions and estimate characteristic finite ones.
Let us point out that we assume again real moduli $\psi$, $\phi$ for simplicity.

Furthermore, we shall calculate the mirror map and analyze its asymptotic behavior for each phase.
Eventually, we will encounter the same structure as for the $\mathbb P^4_{11222} [8]$ including the logarithms.
The derivation of the formula for the mirror map follows \cite{Candelas:1993dm} and is summarized in section \ref{app_mirror11226}.
The result turns out to be
\begin{equation}
\label{mirrormap11226}
  \begin{aligned}
    t_1&=- \frac12 +\frac{\omega_2+\omega_4}{2 \omega_0}\;,\\
    t_2&= \frac12 -\frac{\omega_2+\omega_4}{2 \omega_0}+\frac{\omega_1 + \omega_3+\omega_5}{2 \omega_0}\;,
  \end{aligned}
\end{equation}
where the periods $\omega_i$ have been computed in section \ref{app_metric}.

\paragraph{Landau-Ginzburg phase}
The boundaries of this phase are governed by the constraints $0 \le \phi < 1$ and $0 \le \psi < \psi_c$, where the maximal value $\psi_c$ is the real solution to $|864 \psi_c^6| < |\phi \pm 1|$.
Then, we start at $\psi = \phi = 0$ deep in the Landau-Ginzburg regime and consider two trajectories:
one purely in direction $\phi$ keeping $\psi$ fixed and one vice versa.
By integration we find the maximal length in these directions
\eq{
  \Delta \Theta_1 \, &= \, \int_0^{\psi_c} d \psi \, \sqrt{G_{\psi \ov\psi} (\psi)} \, 
  = \, 0.21 \, ,\\
  \Delta \Theta_2 \, &= \, \int_0^1 d \phi \, \sqrt{G_{\phi \ov\phi} (\phi)} \, = \, 0.32 \, .
  }
Hence, we conclude that the Landau-Ginzburg phase is again of finite
size in any direction with distances shorter than $O (1) \, M_{\rm pl}$.
The mirror map \eqref{mirrormap11226} asymptotes to finite values as well.
For $\phi, \psi \rightarrow 0$ we end up with
$t_1 \simeq  -\frac12 +  0.866 \, i $ and $t_2 \simeq \frac12 + 0.134 \, i$.

\paragraph{$\mathbb P^1$ hybrid phase}
The algebraic parameters of $\mathbb P^1$ are constrained by the intervals $1 < \phi < \infty$ and $0 \le \psi < \psi_c$ with $\psi_c$ as above.
We begin with a trajectory along the boundary to the Landau-Ginzburg phase.
That is, we start at $\psi = 0$, $\phi = 1.1$ and integrate in $\psi$ direction up to $\psi_c$ without altering $\phi$.
The length turns out to be finite and quite small
\eq{
  \Delta \Theta_1 \, &= \, \int_0^{\psi_c} d \psi \, \sqrt{G_{\psi \ov\psi} (\psi)} \, 
  = \, 0.1 \, .
  }
According to example $\mathbb P^4_{11222} [8]$, we expect this length to asymptote to a finite value.
Therefore, we approximate the asymptotic metric for $(\psi, \phi) \rightarrow (0, \infty)$. At leading order we find
\begin{equation}
\label{metric11226P1}
  G^{\rm asymp}_{\mathbb{P}^1}\simeq
  \begin{pmatrix}
    \frac{0.25}{|\phi|^2 \, (\log |\phi|)^2} & 0\\
    0 & \frac{27.23 \, |\psi|^2}{|\phi|^{\frac{2}{3}}}
  \end{pmatrix}\; .
\end{equation}
Indeed we arrive at a finite result by integration and using  $|\psi_c|^2 \sim |\phi|^{\frac{1}{3}} / (864^{\frac13})$ 
\eq{
  \Delta \Theta_2 \, = \, \int_0^{\psi_c} d \psi \, 
  \sqrt{G^{\rm asymp}_{\mathbb{P}^1, \, \psi \ov\psi} (\psi, \phi)} \,
  \simeq \, \frac{\sqrt{27.23}}{|\phi|^{\frac{1}{3}}} \cdot 
  \int_0^{\psi_c} d \psi \, \psi^2 \,
  = \, 0.27 \, .
  }
Instead of this approximate analytic approach, one may also evaluate the integral over the full metric numerically, which confirms our result for $\Delta \Theta_2$. 
In contrast, distances of the hybrid $\mathbb P^1$ phase in $\phi$ may become infinite.
One way to see this, is that the integral
\eq{
 \Theta \sim \int d \phi \, \sqrt{G^{\rm asymp}_{\mathbb{P}^1, \, \phi \ov\phi} (\phi)} \sim \log \left( \log \phi \right)
  }
does obviously not converge to a finite value.

Next, we evaluate the formulae \eqref{mirrormap11226} for the mirror map.
For $\phi \rightarrow \infty$ and $\psi \simeq 0$, we obtain asymptotically
\begin{equation}
  \begin{aligned}
    t_1& \simeq -\frac12 +  0.866 \, i +\dots\;,\\
    t_2&\simeq 1 + 0.10 \, i  + \frac{i}{ \pi} \log(2 \phi)+\dots\;.
  \end{aligned}
\end{equation}
On the one hand, $t_1$ is finite as in the Landau-Ginzburg phase.
On the other hand, $t_2$ growths logarithmically in $\phi$ and hence the proper distance is $\Theta \sim \log t_2$ in agreement with the SDC.

\paragraph{Orbifold hybrid phase}
In this regime the algebraic parameter may vary within $0 \le \phi < 1$ and $\psi_c < \psi < \infty$, employing the constraint $|864 \, \psi_c^6| \ge |\phi \pm 1|$.
At first, we compute the trajectory from $\phi =0$ to $\phi =1$ keeping $\psi$ constant at a minimal value $\psi \sim 0.32$.
Basically we integrate along the border to the Landau-Ginzburg phase.
A simple calculation shows:
\eq{
  \Delta \Theta_1 \, &= \, \int_0^1 d \phi \, \sqrt{G_{\phi \ov\phi} (\phi)} \, 
  = \, 0.16 \, .
  }
Recall that asymptotically for large $\psi$ this length was decreasing to zero in the example $\mathbb P^4_{11222} [8]$.
In fact, we will observe the same behavior here.
The metric near $\phi \simeq 0$ and $\psi \rightarrow \infty$ leads to the expression
\begin{equation}
\label{metric11226Orbi}
  G^{\rm asymp}_\text{orbi}\simeq
  \begin{pmatrix}
    \frac{0.04}{|\log(\psi)|^2} & 0\\
    0 & \frac{0.75}{|\psi|^2 \, (\log |\psi|)^2}
  \end{pmatrix}\;.
\end{equation}
Thus, the asymptotic proper distance in $\phi$ direction is again vanishing due to its scaling $\int_0^1 d \phi \,\sqrt{G^{\rm asymp}_{\text{orbi}, \, \phi \ov\phi} (\psi)} \, \sim \, 1/ |\log \psi|$.
One then ends up at the $\phi=1$ locus which is birational equivalent to the complete intersection CY $\mathbb P^5_{111113} [2 \ 6]$.
Note that one may traverse infinite distances in $\psi$ direction as expected.
This is easy to see from the metric and in direct analogy with $\mathbb P^4_{11222} [8]$.

This time, for $\psi \rightarrow \infty$ and $\phi \simeq 0$  the mirror map \eqref{mirrormap11226} approaches the values
\begin{equation}
  \begin{aligned}
    t_1&\simeq - \frac34   + \frac{i}{2 \pi} \, \log (12 \psi )^6 +\dots\;,\\
    t_2&\simeq \frac12 +\dots\; ,
  \end{aligned}
\end{equation}
indicating a logarithmic growth of the proper distance depending on the K\"ahler modulus $t_1$.
Analogously to $\mathbb P^4_{11222} [8]$, the imaginary part of $t_2$ vanishes asymptotically, i.e. the two-cycle $\Sigma_2$ shrinks to zero \cite{Aspinwall:1994zu}.
Again, we arrive then on the $\phi =1$ locus which corresponds to the one-parameter CICY $\mathbb P^5_{111113} [2 \ 6]$.

\paragraph{Large volume phase}
Finally, the large volume regime $1 < \phi < \infty$ and
$\psi_c < \psi < \infty$ ($\psi_c$ as above) is infinite in both
directions $\phi, \psi$.
As expected, we find a logarithmic growth of the proper distances in
any direction, after employing the mirror map \cite{Candelas:1993dm}
\begin{equation}
  \begin{aligned}
    t_1&\simeq\frac12 +\frac{i}{2\pi}\log\left(\frac{(12\psi)^6}{2\phi}\right)+\dots\;,\\
    t_2&\simeq  \frac{i}{\pi}\log\left(2\phi\right)+\dots\;.
  \end{aligned}
\end{equation}
The dots indicate polynomial corrections that are sub-leading at infinity.

\paragraph{Summary}

The different regimes of the moduli space of the mirror $\mathbb P^4_{11226} [12]$ with its finite lengths are schematically depicted in figure \ref{fig_11226}.
The results are qualitatively very similar to $\mathbb P^4_{11222} [8]$.
 In accordance with the  RSDC, all finite characteristic lengths are smaller than $O
(1) \, M_{\rm pl}$.
Infinite directions show a $\log$-behavior in their corresponding K\"ahler modulus. This confirms again the predictions from the SDC.

In addition, we were able to determine the critical distance $\Theta_\lambda$, where the logarithmic behavior becomes essential.
The computation is indeed equivalent to the one of section (5.1.4).
By using the metrics \eqref{metric11226P1} and \eqref{metric11226Orbi}, we find therefore in the hybrid $\mathbb P^1$ phase $\Theta_\lambda = 1/2$ and in the hybrid orbifold phase $\Theta_\lambda = \sqrt{0.75}$.
\vspace{0.1cm}
\begin{figure}[ht!]
\centering
\begin{tikzpicture}[xscale=1.2,yscale=1.2]
\node at (3,-0.3) {$\phi$};
\node at (-3.3,6) {$\psi$};
\node[align=left,below] at (-3,0) {$0$};
\node[align=center,below] at (0,0) {\scriptsize $\phi=1$};
\node[align=center,above] at (2.5,1.5) {\scriptsize $|864 \psi^6| \lessgtr |\phi \pm 1|$};

\draw[-] (-3,0) -> (-3,4.5);
\draw[dashed] (-3,4.5) -> (-3,5.5);
\draw[->] (-3,5.5) -> (-3,6);

\draw (-3,0) -> (1.5,0);
\draw[dashed] (1.5,0) -> (2.5,0);
\draw[->] (2.5,0) -> (3,0);

\draw[dashed,blue] (0,0) -> (0,6);
\coordinate (A) at (-3,3);
\coordinate (B) at (3,1.5);
\draw[dashed, blue]    (A) to[out=-35,in=180] (B);

\draw[<->] (-2.7,0.3) to (-0.3,0.3);
\draw[<->] (-2.7,0.4) to (-2.7,2.7);
\node at (-1.4,0.6) {\footnotesize 0.21};
\node at (-2.3,2) {\footnotesize 0.32};

\draw[<->] (0.3,0.3) to (0.3,1.5);
\draw[<->] (2.5,0.3) to (2.5,1.3);
\draw[<-] (0.3,0.2) to (1,0.2);
\draw[dashed] (1,0.2) to (2,0.2);
\draw[->] (2,0.2) to (2.5,0.2);
\node at (0.7,1) {\footnotesize 0.10};
\node at (2.9,1) {\footnotesize 0.27};
\node at (1.5,0.4) {\footnotesize $\infty$};

\draw[<->] (-2.7,3.3) to (-0.3,3.3);
\draw[<->] (-2.7,5.7) to (-0.3,5.7);
\draw[<-] (-2.7,3.4) to (-2.7,4.5);
\draw[dashed] (-2.7,4.5) to (-2.7,5.1);
\draw[->] (-2.7,5.1) to (-2.7,5.6);
\node at (-1.5,3.6) {\footnotesize 0.16};
\node at (-1.5,5.4) {\footnotesize $0$};
\node at (-2.4,4.5) {\footnotesize $\infty$};

\node[align=center,below] at (1.5,4.5) {\footnotesize LV};
\node[align=center,below] at (1.5,1.3) {$\mathbb P^1$};
\node[align=center,below] at (-1.5,1.8) {\footnotesize Landau- \\ \footnotesize{Ginzburg}};
\node[align=center,below] at (-1.5,4.5) {\footnotesize orbifold};
\end{tikzpicture}
\caption{\small Schematic plot of the moduli space of the mirror $\mathbb P^4_{11226} [12]$ with finite as well as infinite directions as calculated above.}
\label{fig_11226}
\end{figure}
\vspace{0.1cm}

\subsection{\texorpdfstring{$\mathbb P^4_{11169} [18]$}{P11169}}
The final example in this section is the moduli space of  $\mathbb P^4_{11169} [18]$.
An extensive analysis of this model can be found in \cite{Candelas:1994hw}.
In the literature it was shown that the most general possible polynomial of degree 18 employed to describe this moduli space may be reduced to
\eq{
  P = x_1^{18} + x_2^{18} +x_3^{18} + x_4^3+x_5^2 - 18 \psi \, x_1 x_1 x_3 x_4 x_5 - 3 \phi \, x_1^6 x_2^6 x_3^6 \, .
  }
The moduli space is again defined by the hypersurface constraint $\{ P =0 \}/G$.
There exists a scaling symmetry preserving the form of $P$, which is identified with an enlarged group $\hat G$.
This group induces a $\mathbb Z_{18}$ action on the algebraic parameters
\eq{
  (\psi, \phi) \mapsto (\alpha \psi, \alpha^6 \phi)
  }
with the nontrivial $18^{\rm th}$ root of unity $\alpha = \exp \left( 2 \pi i /18 \right)$.
In fact, there are additional transformations leaving the hypersurface constraint invariant. For these and further details we simply refer to the literature.

The qualitative structure of the moduli space is similar to the one of $\mathbb P^4_{11222} [8]$, in the sense that there are four extended phases:
Landau-Ginzburg, orbifold, large volume and an additional hybrid phase.
However, now we do not have a hybrid $\mathbb P^1$ phase, but instead the exceptional divisor has a $\mathbb P^2$ component \cite{Hosono:1993qy} and thus for large $\phi$, small $\psi$ we face a $\mathbb P^2$ target space.
Moreover, the large volume phase corresponds to a different geometry, that is not a $K3 \times \mathbb P^1$ fibration that we ended up with in $\mathbb P^4_{11222} [8]$.
The singular loci in case of $\mathbb P^4_{11169} [18]$ are slightly different as well.
Both singularities correspond to conifolds and are located at $\phi^3=1$ and
 $(\phi + 2^2 3^8 \psi^6)^3 =1$.
 
For the calculation of the periods we refer to section \ref{sec_KahlerMetric} and in particular to section \ref{app_metric}.
We will restrict our analysis again to the computation of finite distance in the moduli space according to the schematic curves plotted in figures \ref{fig_11222LG}, \ref{fig_11222P1} and \ref{fig_11222Orbi}.
One can show that the series expansion of the fundamental period converges for
\eq{
\label{convergence11169}
  |2^2 3^8 \, \psi^6| \, \lessgtr \, |\phi - \alpha^{-6\tau}| \,,
  }
depending on the regime we want to investigate.
Now we have three constraints as $\tau$ may vary between 0,1,2 with $\alpha= \exp \left( 2 \pi i/ 18\right)$.
Let us point out that we assume again real moduli $\psi$, $\phi$ for simplicity.

Infinite directions will show a logarithmic behavior in their proper distances depending on the K\"ahler moduli.
The mirror map for the $\mathbb P^4_{11169} [18]$ model is given by \cite{Candelas:1994hw}
\begin{equation}
\label{mirrormap11169}
  \begin{aligned}
    t_1&= \frac{\omega_3 - \omega_0}{\omega_0}\;,\\
    t_2&= \frac{\omega_4+ \omega_5 - 2 \omega_3 + 2 \omega_0}{\omega_0}\;.
  \end{aligned}
\end{equation}

\paragraph{Landau-Ginzburg phase}
Here, the moduli take values within the intervals $0 \le \phi < 1$ and $0 \le \psi < \psi_c$, with $\psi_c$ being the maximal real solution to condition \eqref{convergence11169}.
We compute the following two trajectories:
one purely in direction $\phi$ keeping $\psi$ fixed and one vice versa, starting at $\psi = \phi = 0$.
Their lengths are given by
\eq{
  \Delta \Theta_1 \, &= \, \int_0^{\psi_c} d \psi \, \sqrt{G_{\psi \ov\psi} (\psi)} \, 
  = \, 0.074 \, ,\\
  \Delta \Theta_2 \, &= \, \int_0^1 d \phi \, \sqrt{G_{\phi \ov\phi} (\phi)} \, = \, 0.087 \, .
  }
Again, both directions are finite and less than $O (1) \, M_{\rm pl}$.
This result is confirmed by the mirror map because they converge to finite values as well.
More precisely, for $\phi, \psi \rightarrow 0$ we obtain the asymptotic mirror map \eqref{mirrormap11169} coordinates
$t_1 \simeq \omega^{-2} = -\frac12 +  0.866 \, i $ and $t_2 \simeq 1 + 0.238 \, i$.

\paragraph{$\mathbb P^2$ hybrid phase}
Considering the parameter space
$1 < \phi < \infty$ and $0 \le \psi < \psi_c$, we compute at first the proper distance of a trajectory along the Landau-Ginzburg phase.
Taking $\psi = 0$, $\phi = 1.1$ as initial values, we integrate along the
$\psi$ direction up to $\psi_c$ without altering $\phi$
\eq{
  \Delta \Theta_1 \, &= \, \int_0^{\psi_c} d \psi \, \sqrt{G_{\psi \ov\psi} (\psi)} \, 
  = \, 0.015 \, .
  }
One can approximate
the K\"ahler metric for $(\psi, \phi) \rightarrow (0, \infty)$
\begin{equation}
\label{metric11169p1}
  G^{\rm asymp}_{\mathbb{P}^2}\simeq
  \begin{pmatrix}
    \frac{0.5}{|\phi|^2 \, (\log |\phi|)^2} & 0\\
    0 & \frac{185837.8 \, |\psi|^6}{|\phi|^{\frac{4}{3}}}
  \end{pmatrix}\; 
\end{equation}
and compute the maximal distance in the $\psi$ direction  asymptotically 
\eq{
  \Delta \Theta_2 \, = \, \int_0^{\psi_c} d \psi \, 
  \sqrt{G^{\rm asymp}_{\mathbb{P}^2, \, \psi \ov\psi} (\psi, \phi)} \,
  \simeq \, \frac{\sqrt{185837.8}}{|\phi|^{\frac{2}{3}}} \cdot 
  \int_0^{\psi_c} d \psi \, \psi^3 \,
  = \, 0.12 \, ,
  }
with $|\psi_c|^4 \sim  |\phi|^{\frac{2}{3}} / (2^2 3^8)^{\frac23}$.
A numerical evaluation of the integral with the full (not asymptotic)
metric gives the same  result.
Due to the other component of the metric above, distances 
in the $\phi$ direction may become infinite.

Computing the mirror maps reveals a logarithmic growth of $t_2$ in $\phi$ and hence a logarithmic growth of the proper distance $\Theta$ in $t_2$ as expected due to the Swampland Distance Conjecture.
On the other hand, $t_1$ approaches the finite value which we also
found in the Landau-Ginzburg phase.
Indeed, in the limit $\phi \rightarrow \infty$ and keeping $\psi
\simeq 0$, inserting the periods computed in section \ref{app_metric}
into  \eqref{mirrormap11169} leads to (recall $\omega = \exp (2 \pi i/3)$)
\begin{equation}
  \begin{aligned}
    t_1&\simeq \omega^{-2} + \dots \simeq -\frac12 +  0.866 \, i +\dots\;,\\
    t_2&\simeq \frac32  + \frac{3 \, i}{2 \pi} \log(3 \phi)+\dots\;.
  \end{aligned}
\end{equation}

\paragraph{Orbifold hybrid phase}
In this regime the algebraic parameters are fixed by $0 \le \phi < 1$ and $\psi_c < \psi < \infty$.
As before, we begin with computing a trajectory from $\phi =0$ to $\phi =1$ keeping $\psi$ constant at a minimal value $\psi \sim 0.32$.
\eq{
  \Delta \Theta_1 \, &= \, \int_0^1 d \phi \, \sqrt{G_{\phi \ov\phi} (\phi)} \, 
  = \, 0.065 \, .
  }
In the former examples  $\mathbb P^4_{11222} [8]$ and  $\mathbb P^4_{11226} [12]$ this distance was asymptotically vanishing for large $\psi$.
Thus, we approximate the K\"ahler metric near $\phi \simeq 0$ and $\psi \rightarrow \infty$
\begin{equation}
\label{metric11169orbi}
  G^{\rm asymp}_\text{orbi}\simeq
  \begin{pmatrix}
    \frac{1}{|\log(\psi)|^6} & 0\\
    0 & \frac{0.75}{|\psi|^2 \, (\log |\psi|)^2}
  \end{pmatrix}\;.
\end{equation}
Thus, also here the distance in $\phi$ direction is asymptotically zero.
However, in $\psi$ direction it is possible to travel infinite distances.

The mirror map coordinates \eqref{mirrormap11169} for $\psi \rightarrow \infty$ and $\phi \simeq 0$ are approximately given by
\begin{equation}
  \begin{aligned}
    t_1&\simeq - \frac23   + \frac{i}{2 \pi} \, \log (18 \psi )^6 +\dots\;,\\
    t_2&\simeq 1 +\dots\;.
  \end{aligned}
\end{equation}
Since the proper distance scales like $\Theta \sim \log \log \psi$, we end up once again with a logarithmic growth of $\Theta$ in $t_1$.

As opposed to the $\mathbb P^2$  hybrid phase,
now it is $t_1$ which may become infinitely large.
Again, the imaginary part of $t_2$ vanishes asymptotically and hence the two-cycle $\Sigma_2$ shrinks to zero \cite{Aspinwall:1994zu}.
In this limit one finally hits the conifold singularity at $\phi = 1$ as the proper distance to the singularity approaches zero. 
Recall that in the case of $\mathbb P^4_{11222} [8]$ we were running into a non-singular one-parameter subspace instead.

\paragraph{Large volume phase}
Eventually, the large volume regime $1 < \phi < \infty$ and
$\psi_c < \psi < \infty$ ($\psi_c$ as above) is infinite in both
directions $\phi, \psi$ with the mirror map given by \cite{Candelas:1994hw}
\begin{equation}
  \begin{aligned}
    t_1&\simeq - \frac12 +\frac{i}{2\pi}\log\left(\frac{(18\psi)^6}{3\phi}\right)+\dots\;,\\
    t_2&\simeq \frac{3}{2} + \frac{3 \, i}{2 \pi}\log\left(3\phi\right)+\dots\;.
  \end{aligned}
\end{equation}

\begin{figure}[ht!]
\centering
\begin{tikzpicture}[xscale=1.2,yscale=1.2]
\node at (3,-0.3) {$\phi$};
\node at (-3.3,6) {$\psi$};
\node[align=left,below] at (-3,0) {$0$};
\node[align=center,below] at (0,0) {\scriptsize $\phi=1$};
\node[align=center,above] at (2.5,1.5) {\scriptsize $|2^2 3^8 \, \psi^6| \, \lessgtr \, |\phi - \alpha^{-6\tau}|$};

\draw[-] (-3,0) -> (-3,4.5);
\draw[dashed] (-3,4.5) -> (-3,5.5);
\draw[->] (-3,5.5) -> (-3,6);

\draw (-3,0) -> (1.5,0);
\draw[dashed] (1.5,0) -> (2.5,0);
\draw[->] (2.5,0) -> (3,0);

\draw[dashed,blue] (0,0) -> (0,6);
\coordinate (A) at (-3,3);
\coordinate (B) at (3,1.5);
\draw[dashed, blue]    (A) to[out=-35,in=180] (B);

\draw[<->] (-2.7,0.3) to (-0.3,0.3);
\draw[<->] (-2.7,0.4) to (-2.7,2.7);
\node at (-1.4,0.6) {\footnotesize 0.087};
\node at (-2.3,2) {\footnotesize 0.074};

\draw[<->] (0.3,0.3) to (0.3,1.5);
\draw[<->] (2.5,0.3) to (2.5,1.3);
\draw[<-] (0.3,0.2) to (1,0.2);
\draw[dashed] (1,0.2) to (2,0.2);
\draw[->] (2,0.2) to (2.5,0.2);
\node at (0.7,1) {\footnotesize 0.015};
\node at (2.9,1) {\footnotesize 0.12};
\node at (1.5,0.4) {\footnotesize $\infty$};

\draw[<->] (-2.7,3.3) to (-0.3,3.3);
\draw[<->] (-2.7,5.7) to (-0.3,5.7);
\draw[<-] (-2.7,3.4) to (-2.7,4.5);
\draw[dashed] (-2.7,4.5) to (-2.7,5.1);
\draw[->] (-2.7,5.1) to (-2.7,5.6);
\node at (-1.5,3.6) {\footnotesize 0.065};
\node at (-1.5,5.4) {\footnotesize $0$};
\node at (-2.4,4.5) {\footnotesize $\infty$};

\node[align=center,below] at (1.5,4.5) {\footnotesize LV};
\node[align=center,below] at (1.5,1.3) {$\mathbb P^1$};
\node[align=center,below] at (-1.5,1.8) {\footnotesize Landau- \\ \footnotesize{Ginzburg}};
\node[align=center,below] at (-1.5,4.5) {\footnotesize orbifold};
\end{tikzpicture}
\caption{\small Schematic plot of the moduli space of the mirror $\mathbb P^4_{11169} [18]$ with finite as well as infinite directions as calculated above.
The constraint for one conifold uses $\tau = 0, 1, 2$ and $\alpha= \exp \left( 2 \pi i/ 18\right)$.}
\label{fig_11169}
\end{figure}
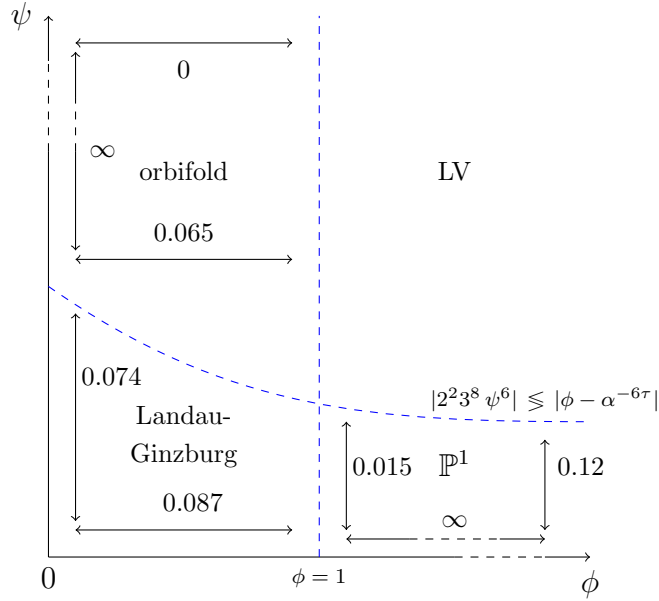

\paragraph{Summary}
Overall, distances of the moduli space of $\mathbb P^4_{11169} [18]$ behave similar to the ones of both other two-parameter models, which we discussed above.
Figure \ref{fig_11169} summarized all distances of $\mathbb P^4_{11169} [18]$.
The logarithmic dependence of the different mirror maps underline the Swampland Distance Conjecture and the small finite lengths its refined version.
Furthermore, we can compute the critical distance $\Theta_\lambda$ in asymptotic regions of the moduli space.
Following section \ref{2param_geodesic} the metrics
\eqref{metric11169p1} and \eqref{metric11169orbi} lead to
$\Theta_\lambda = \sqrt{0.5} \approx 0.71$ for the hybrid $\mathbb
P^2$ phase and $\Theta_\lambda = \sqrt{0.75}$ for the hybrid orbifold phase.
In accordance with the RSDC  both distances are less than $M_{\rm pl}$.
\vspace{0.1cm}
\vspace{0.1cm}

\section{RSDC for a CY manifold with $h^{11}=101$}

After a thorough analysis of the geometry of the moduli space of one-
and two-parameter CYs, we now turn to the question of the description of
higher dimensional moduli spaces. While the methods used so far are in
principle applicable to models with many moduli, the computational
effort increases drastically, rendering this approach 
unfeasible.

A new method, recently developed by Aleshkin and Belavin in
\cite{Aleshkin:2017fuz} and \cite{Aleshkin:2017oak}, allows the
computation of the periods of the complete 101-dimensional \kahler{} moduli
space of the mirror quintic in the Landau-Ginzburg phase. We will apply this method with
slight modifications to calculate the radii of the Landau-Ginzburg
phase of the \kahler{} sector of the 
mirror quintic.

The defining polynomial for the quintic can be written as

\begin{equation}
P=x_1^5+x_2^5+x_3^5+x_4^5+x_5^5+\sum_{s=0}^{100}\Phi_{s}e^{s}
\end{equation}
with
$e^{s}=x_1^{{s}_1}{x}_2^{{s}_2}x_3^{{s}_3}x_4^{{s}_4}x_5^{{s}_5}$. The
$s$ are 101 5-dimensional vectors labeling the deformations, with each
entry representing the exponent of one coordinate. 
In the following formulas,  the index $\mu$  labels the
$2h^{21}+2=204$ periods and has
to be understood in the same way. The deformations of the polynomial
can be grouped into all possible permutations of 5 $s$-vectors,
$(1,1,1,1,1)$, $(2,1,1,1,0)$, $(2,2,1,0,0)$, $(3,1,1,0,0)$ and
$(3,2,0,0,0)$. The number of permutations of these are
$(1,20,30,30,20)$. Note that $\Phi_0$ corresponds to the fundamental
${s}=0=(1,1,1,1,1)$ deformation and is the same as the $\psi$ parameter of
the quintic up to a  rescaling $-5\Phi_0={\psi}$.

The periods are given by \cite{Aleshkin:2017fuz}
\begin{equation}
\label{sigma1}
\sigma_\mu(\phi)=\sum_{n_1\geq 0}\dotsc\sum_{n_5\geq 0}\left[\prod_{i=1}^5\left({\mu_i+1\over 5}\right)_{n_i}\right]\sum_{m\in \Sigma_n}\prod_{s=0}^{100}{\Phi_s^{m_s}\over m_s!}\;,
\end{equation}
where $(x)_a={\Gamma[x+a]/ \Gamma[x]}$ denotes the Pochhammer symbol, $n$ is a 5-dimensional vector with integer entries, $m$ is a 101-dimensional vector with scalar entries $m_s$ and
\begin{equation}
\label{constraint}
\Sigma_n=\left\{m\, \left|\, \sum_s m_ss_i=5n_i+\mu_i\quad\forall i\in 1..5\right.\right\}\;.
\end{equation}
The \kahler{} potential is
 
\begin{equation}
e^{-K}=\sum_{\mu=0}^{203}(-1)^{\text{deg}(\mu)/5}\left[\prod_{i=1}^5{\Gamma\left({\mu_i+1\over 5}\right)\over \Gamma\left({4-\mu_i\over 5}\right)}\right]|\sigma_\mu(\phi)|^2\;,
\end{equation}
where $\text{deg}(\mu)=\sum_{i=1}^5\mu_i$ is the degree of the corresponding deformation.
All moduli corresponding to the permutations of the same t-vector have the same periods and therefore the \kahler{} potential is symmetric. For computational feasibility we restrict to this five-dimensional subspace, i.e. from now on $s=0,\dots,4$ and $\mu=0,\dots,11$. Including the $(0,0,0,0,0)$ deformation corresponding to  $H^{3,0}$ and the conjugated cycles, we obtain twelve periods. The \kahler{} potential becomes

\begin{equation}
\label{Kahlerfinal}
e^{-K}=\sum_{\mu=0}^{11}\pi_{\mu}(-1)^{\text{deg}(\mu)/5}\left[\prod_{i=1}^5{\Gamma\left({\mu_i+1\over 5}\right)\over \Gamma\left({4-\mu_i\over 5}\right)}\right]|\sigma_\mu(\phi)|^2\;,
\end{equation}
where $\pi_{\mu}$ denote the numbers of permutations of the deformations, i.e. $\pi_{\mu}=\{1,1,20,30,30,20,1,1,20,30,30,20\}$. In addition a combinatorial coefficient $c(m_s)$ has to be included in the periods: 
\begin{equation}
\label{sigma2}
\sigma_\mu(\phi)=\sum_{n_1\geq 0}\dotsc\sum_{n_5\geq 0}\left[\prod_{i=1}^5\left({\mu_i+1\over 5}\right)_{n_j}\right]\sum_{m\in \Sigma_n}\prod_{s=0}^4{\Phi_s^{m_s}\cdot c(m_s)}\;.
\end{equation}
This coefficient can be calculated as follows. In principle one has to
find 101-dimensional vectors $m$, which solve \eqref{constraint}. If
one uses the symmetry of the problem to reduce to 5 dimensions, one
deformation represents $\pi_s$ different deformations. 

The calculation is most easily understood in an example. We look at a deformation which has 20 permutations, e.g. the (2,1,1,1,0). If this deformation contributes for example $m_s$=3, there are 3 different ways to obtain this. Three deformations could contribute 1, one deformation contributes 1 and another one 2 or one single deformation contributes 3. Then one has to choose these deformations out of the 20 available, resulting in 
${20\choose 3}+{20\choose 1}{19\choose 1}+{20\choose 1}$ different
combinations. But these deformations are not of the same order in the
fields, therefore one has to include the $1\over m_s!$ term of
\eqref{sigma1}. The final coefficient then 
is 
\eq{
{{20\choose 3}\over 1!1!1!}+{{20\choose 1}{19\choose 1}\over
    1!2!}+{{20\choose 1}\over 3!}={4000\over 3}\,.
}
The $c(m_s)$ in \eqref{sigma2} calculates exactly this number. 

For the general case, 
first we define the set of all possible integer partitions with less then $\pi_s$ integers of $s$ as $P(s)$. Then we count how often each integer appears in each permutation of this set. The resulting set is denoted $A$. We then define a set $L_k$ for each $m_s$ with elements $L_{k,i}$ as
\begin{equation}
L_{k,i}=\pi_s-\sum_{j=1}^{i-1} A_{k,j}
\end{equation}
In the example above, 
\eq{
A&=\{\{3,0,0\},\{1,1,0\},\{0,0,1\}\}\qquad {\rm and} \\  
L&=\{\{20,17,17\},\{20,19,18\},\{20,20,20\}\}\,.
}
With these definitions the coefficient can be written as
\begin{equation}
c(m_s)=\sum_{k=1}^{|P(s)|}\prod_{j=1}^{m_s} {L_{k,j}\choose A_{k,j}}/(k!)^{A_{k,j}}\;
\end{equation}
where $|P(s)|$ denotes the number of elements in $P(s)$. These
formulas allow for the calculation of the \kahler\ potential. 

The final
information needed to calculate the proper distances using
\eqref{properdistancer} are the boundaries of integration. We start at
the Landau-Ginzburg point and go along trajectories where only one
deformation is turned on. The endpoint of the trajectory is the
conifold. To determine the position of the conifold one has to solve
the transversality conditions for each deformation. As we have
combined several deformations into one, these conditions are highly
non-trivial to solve. This is most easily done by reformulating the
problem as a minimization problem which then can be solved by standard
minimization techniques. As the objective function we choose
\begin{equation}
\Pi=P^2+\sum_i \left({\partial P\over \partial x_i}\right)^2+\Phi\,.
\end{equation} 
The first two terms are the transversality conditions squared. The last term, linear in the deformation parameter, is added to ensure that the global minimum is the smallest value for $\Phi$, as we are interested in the boundary of the conifold which lies in the Landau-Ginzburg phase. 

With the \kahler{} potential from \eqref{Kahlerfinal} we calculate the proper distance between the Landau-Ginzburg point and the conifold. Table \ref{distances101} lists these distances. The periods are evaluated up to twelfth order. 

\begin{table}[ht!]
  \centering
  \begin{tabular}{c c c}
  direction & $\Delta\Theta$ \\
  \hline\hline
   $\Phi_0$ & 0.4656 \\
   $\Phi_1$ & 0.0082  \\
   $\Phi_2$ & 0.0670  \\
   $\Phi_3$ & 0.0585  \\
   $\Phi_4$ & 0.0089 \\ 
  \end{tabular}
  \caption{Proper distances between the conifold and the Landau-Ginzburg point in the complex structure moduli space of the quintic.}
  \label{distances101}
\end{table}
\textit{The distances are all smaller than one, with the distance for
  the fundamental deformation being the largest.} 
This originates mainly from the conifold position, which depends on the amount of deformations included. For $\Phi_0$ the polynomial is
\begin{equation}
P=x_1^5+x_2^5+x_3^5+x_4^5+x_5^5+\Phi_0x_1x_2x_3x_4x_5\,.
\end{equation}
This polynomial fails to be transversal at $\Phi_0=-5$, which is the known result for the quintic at $\psi=1$. If one instead looks at the polynomial
\begin{equation}
P=x_1^5+x_2^5+x_3^5+x_4^5+x_5^5+\Phi_1x_1^2x_2x_3x_4
\end{equation}
the singularity lies at $\Phi_1={-5\over 2^{2/5}}$. If one includes one more deformation, e.g.
\begin{equation}
P=x_1^5+x_2^5+x_3^5+x_4^5+x_5^5+\Phi_1(x_1^2x_2x_3x_4+x_1^2x_2x_3x_5)
\end{equation}
the value changes to $\Phi_1=-5/2^{6/5}$, i.e. it decreases. For more than two deformations this could not be analytically solved. $\Phi_1$ couples in our case to 20 deformations, the numerical procedure described above results in a conifold position of  $\Phi_1=\mathcal{O}(10^{-9})$. So while one includes more moduli in the trajectories, the conifold in these direction is much closer to the Landau-Ginzburg point. The latter effect dominates, resulting in a total decrease of proper distance.

In view of the RSDC this result is very compelling.  It appears that
the radii of the non-geometric phases decrease with the number of
dimensions. In the one-parameter models, the Landau-Ginzburg 
phases have a radius of $\approx 0.4$. In the two parameter models,
the radii of the non-geometric and hybrid phases range between $0.1$
and $0.3$. Finally, in the 101 parameter example of the mirror quintic
the proper distances are 1-2 orders of magnitude lower. 

When the number of dimensions of the moduli space increases, one can
travel through more phases before reaching a logarithmic behavior. But
at the same time, the proper distance per phase decreases. This
suggests that at the boundary to the geometric
phase, eventually the total collected value of $\Theta_0$  does not
scale with the  dimension of the moduli space. This implies that in the non-geometric phases
\eq{
                     \left<{\Theta_0\over  {\rm phase}}\right>\cdot \#({\rm phases})<
                     M_{\rm pl}\,
}
holds. That is, the average $\Theta_0$ per phase times the number of phases the geodesic passes through is smaller than $M_{\rm pl}$. A generic geodesic does not pass through all phases. 
It would be nice to check this explicitly. Sadly, the convergence of the periods and the partition function of the GLSM at the boundaries of the phases is slow. Moreover, finding the global shortest geodesic between two points in a high-dimensional space is a difficult problem. Nevertheless, the fact that the diameters of the non-geometric phases are decreasing is a strong hint that this could be true.

\section{Conclusions}

In this paper we analyzed the full K\"ahler 
moduli space  for  a concrete set of Calabi-Yau manifolds with respect to the Refined
Swampland Distance Conjecture. 
Because in the large radius phases the behavior could be
expected to  be very similar to toroidal models, special emphasis was
put on the non-geometric  phases. We first reviewed the available
two techniques to  compute the K\"ahler potential in these phases
and applied and compared them for  threefolds defined via
hypersurfaces in 
weighted projective spaces with (mostly) small number of complex structure
moduli $h^{21}\le 2$. 

The essential quantity of interest was the
length of trajectories/geodesics in the various phases. It turned out
that in the LG phases
these were all finite and smaller than the Planck-length. 
Had we found numbers that were say an order of magnitude larger
than the Planck-length, the RSDC would have been falsified.
The hybrid phases were also showing a hybrid behavior with
respect to the length of geodesics. For the complex two-dimensional
moduli spaces that we analyzed, there was one direction of finite
size and another one  of infinite size that showed the same
logarithmic scaling behavior as describe by the Swampland Distance
Conjecture. All our findings were consistent with  the RSDC.

We also analyzed the LG phase of the full
101 dimensional K\"ahler moduli space of the mirror quintic.
Here we observed an interesting scaling of the proper field distances 
with the inverse of the number of moduli so that eventually 
the overall collected proper field distance could still remain smaller than one.

Our setting was very close to the original one by Ooguri/Vafa \cite{Ooguri:2006in} , in the sense
that we were considering a flat moduli space, where the axions, the
$B$-fields,  are compact and the only non-compact directions are
saxionic. As has been clarified by recent  
work
\cite{Baume:2016psm,Klaewer:2016kiy,Valenzuela:2016yny,Bielleman:2016olv,Blumenhagen:2017cxt,Hebecker:2017lxm}, 
for application to 
axion monodromy inflation,  one has to formulate an axionic version of this conjecture.
Due to flux, the axions receive a well controlled potential that makes
them non-compact so that now the former shift symmetry  acts not only
on the axions themselves but on the combination of fluxes and axions.
In this case, one is not interested in a RSDC along flat directions but
along axionic directions with a potential. The backreaction of such an
axion excursion onto the saxionic moduli was essential for  finding the same
logarithmic scaling of the proper field distance. 
Thus, there are indications that a RSDC is at work here, too. However, more
work is needed to undoubtedly  establish this and come closer
to a sort of no-go theorem for the existence of  models of
large field inflation  in controllable  effective field theories derived from string theory.


\vskip2em
\noindent
\emph{Acknowledgments:} We are very grateful to 
Anamaria Font, Daniela Herschmann and Eran Palti for helpful discussions.
We also thank Sebastian Greiner for enlightening conversations
about periods. R.B. would like to thank the University of Bonn
for hospitality. F.W. is grateful to the Deutscher Akademischer Austausch Dienst (DAAD) for support and the Instituto de F\'isica Te\'orica (IFT) in Madrid for hospitality.


\clearpage
\appendix

\section{Periods of \texorpdfstring{$\mathbb P^4_{11169} [18]$}{P11169}}
\label{app_11169}

For completeness we record the periods for the mirror to $\mathbb P^4_{11169} [18]$, in forms suitable for analytic continuation into the hybrid regions of large $\phi$ or large $\psi$.

The periods in the Landau-Ginzburg phase are given by
\begin{equation}
\begin{aligned}
  \omega_j(\psi,\phi)&=-\frac16 \sum\limits_{n=1}^{\infty}\frac{\Gamma(n/6)(-2\cdot 3^{11/6}\psi)^n e^{2 \pi i n j/18}}{\Gamma(n)\Gamma(1-n/3)\Gamma(1-n/2)}U_{-\frac{n}{6}}\left(e^{2\pi i j/3}\phi\right)\;,\\
  U_\nu(\phi)&=3^{-\frac32-\nu} \frac{2\pi e^{2 \pi i\nu/6}}{\Gamma(-\nu)}\sum\limits_{m=0}^{\infty}\frac{\Gamma\left(\frac{m-\nu}{3}\right)\left(e^{2 \pi i/3}\phi\right)^m}{\Gamma^2\left(1-\frac{m-\nu}{3}\right)\prod_{i=1}^{3}\Gamma\left(\frac{i+m}{3}\right)}\;,
\end{aligned}
\end{equation}
which converge for $|\phi|<1$ and $|2^2 3^8 \psi^6|<|\phi - \alpha^{-6\tau}|$, where $\alpha$ is an 18th root of unity and $\tau=0,\dots,2$.
The $U$-function can be rewritten in terms of generalized hypergeometric functions as follows
\begin{equation}
  \begin{aligned}
    U_\nu(\phi)&=\frac{3^{-1-\nu}}{2}\frac{e^{2 \pi i\nu/6}}{\Gamma(-\nu)}\left(  2 \frac{\Gamma\left(\frac{-\nu}{3}\right)}{\Gamma^2\left(\frac{3+\nu}{3}\right)}{}_3F_2\left(\frac{-\nu}{3},\frac{-\nu}{3},\frac{-\nu}{3};\frac13,\frac23;\phi^3\right)\right.\\
    &+9 e^{4 \pi i/3}\phi^2\frac{\Gamma\left(\frac{2-\nu}{3}\right)}{\Gamma^2\left(\frac{1+\nu}{3}\right)}{}_3F_2\left(\frac{2-\nu}{3},\frac{2-\nu}{3},\frac{2-\nu}{3};\frac43,\frac53;\phi^3\right)\\
    &+\left. 6 e^{2 \pi i/3}\phi\frac{\Gamma\left(\frac{1-\nu}{3}\right)}{\Gamma^2\left(\frac{2+\nu}{3}\right)}{}_3F_2\left(\frac{1-\nu}{3},\frac{1-\nu}{3},\frac{1-\nu}{3};\frac23,\frac43;\phi^3\right)\right)\;,
  \end{aligned}
\end{equation}
which allows for analytic continuation to $|\phi|>1$.

Alternatively, if we first sum over powers of $\psi$ we get the representation
\begin{equation}
  \begin{aligned}
    \omega_j(\psi,\phi)&=\frac{-\pi}{3^{5/2}}\sum\limits_{r=1}^{18}e^{2\pi i j r/18}e^{-i\pi r/18}\eta_{j,r}(\psi,\phi)\;,\\
    \eta_{j,r}(\psi,\phi)&=\sum\limits_{m=0}^{\infty} \frac{e^{2 \pi i m (j+1)/3}\phi^m}{\prod_{l=1}^{3}\Gamma\left(\frac{l+m}{3}\right)}V_{m,r}(\psi)\;,\\
    V_{m,r}(\psi)&=(-18\psi)^r\frac{\Gamma\left(\frac{m}{3}+\frac{r}{18}\right){}_7F_6\left(a;b;6^69^9\psi^{18}\right)}{\Gamma^2\left(1-\frac{m}{3}-\frac{r}{18}\right)\Gamma(r)\Gamma(1-\frac{r}{2})\Gamma(1-\frac{r}{3})}\;,\\
    a&=\left(1,\frac{r+6}{18},\frac{r+12}{18},\frac{r+6m}{18},\frac{r+6m}{18},\frac{r+6m}{18},\frac{r}{18}\right)\;,\\
    b&=\left(\frac{r+1}{18},\frac{r+5}{18},\frac{r+7}{18},\frac{r+11}{18},\frac{r+13}{18},\frac{r+17}{18}\right)\;,
  \end{aligned}
\end{equation}
which again converges for $|\phi|<1$ and can be analytically continued to $|2^2 3^8 \psi^6|>|\phi - \alpha^{-6\tau}|$.

\clearpage
\bibliography{references}  
\bibliographystyle{utphys}
\end{document}